%% file: galapy.tex
\documentclass{aa}
\usepackage{graphicx}
\usepackage{txfonts}
\usepackage[colorlinks=true,allcolors={blue}]{hyperref}
\makeatletter
\renewcommand*\aa@pageof{, page \thepage{} of \pageref*{LastPage}}
\makeatother

\usepackage{tabularx}
\usepackage{booktabs}
\usepackage{lscape}
\usepackage{multirow}
\usepackage{makecell} %for line-breaks within table-cells
\usepackage{amsmath}
\usepackage{dsfont}
\usepackage{xspace}
\newcommand{\galapy}{GalaPy\xspace}

\usepackage{ulem}
\newcommand\mysout{\bgroup\markoverwith{\textcolor{red}{\rule[.25ex]{2pt}{2.4pt}}}\ULon}

\begin{document}

   \title{GalaPy, the highly optimised C++/Python\\ spectral modelling tool for galaxies}

   \subtitle{I - Library presentation and photometric fitting}

   \author{T. Ronconi
          \inst{1}\fnmsep\inst{2}\fnmsep\inst{3}\thanks{Corresponding author: tronconi@sissa.it}
          \and
          A. Lapi\inst{1}\fnmsep\inst{2}\fnmsep\inst{4}\fnmsep\inst{5}
          \and
          M. Torsello\inst{1}
          \and
          A. Bressan\inst{1}
          \and
          D. Donevski\inst{6}\fnmsep\inst{1}
          \and
          L. Pantoni\inst{7}
          \and 
          M. Behiri\inst{1}
          \and
          L. Boco\inst{1}
          \and
          A. Cimatti\inst{8}
          \and 
          Q. D'Amato\inst{9} 
          \and 
          L. Danese\inst{1}
          \and
          M. Giulietti\inst{1}
          \and
          F. Perrotta\inst{1}
          \and
          L. Silva\inst{10}
          \and
          M. Talia\inst{8}
          \and
          M. Massardi\inst{4}\fnmsep\inst{1}
          }

   \institute{
   Scuola Internazionale Superiore di Studi Avanzati (SISSA), Via Bonomea 265, IT-34136, Trieste, Italy %1
   \and   
   Institute for Fundamental Physics of the Universe (IFPU), Via Beirut 2, IT-34151, Trieste, Italy %2
   \and
   INAF - Osservatorio di Astrofisica e Scienza dello Spazio (OAS), Via Gobetti 93/3, I-40127 Bologna, Italy %3
   \and
   IRA-INAF, Via Gobetti 101, 40129 Bologna, Italy %4
   \and
   INFN-Sezione di Trieste, via Valerio 2, 34127 Trieste,  Italy %5
   \and
   National Centre for Nuclear Research, Pasteura 7, 02-093 Warsaw, Poland %6
   \and
   Sterrenkundig Observatorium Universiteit Gent, Krijgslaan 281 S9, B-9000 Gent, Belgium %7
   \and
   Dip. Fisica e Astronomia 'Augusto Righi', Univ. Bologna, viale Berti Pichat 6/2, 40127, Bologna, Italy %8
   \and
   INAF - Osservatorio Astrofisico di Arcetri, Largo Enrico Fermi, 5, 50125 Firenze FI, Italy %9
   \and
   INAF-OATS, Via Tiepolo 11, 34143 Trieste, Italy %10
   }

   \date{XXX; XXX}
 
  \abstract{ 
    Fostered by upcoming data from new generation observational campaigns, we are about to enter a new era for the study of how galaxies form and evolve. 
    The unprecedented quantity of data that will be collected, from distances only marginally grasped up to now, will require analysis tools designed to target the specific physical peculiarities of the observed sources and handle extremely large datasets. 
    One powerful method to investigate the complex astrophysical processes that govern the properties of galaxies is to model their observed spectral energy distribution (SED) at different stages of evolution and times throughout the history of the Universe.
    To address these challenges, we have developed \galapy, a new library for modelling and fitting SEDs of galaxies from the X-ray to the radio band, as well as the evolution of their components and dust attenuation/reradiation. On the physical side, \galapy incorporates both empirical and physically-motivated star formation histories, state-of-the-art single stellar population synthesis libraries, a two-component dust model for attenuation, an age-dependent energy conservation algorithm to compute dust reradiation, and additional sources of stellar continuum such as synchrotron, nebular/free-free emission and X-ray radiation from low and high mass binary stars. 
    On the computational side, \galapy implements a hybrid approach that combines the high performance of compiled C++ with the user-friendly flexibility of Python, and exploits an object-oriented design via advanced programming techniques. 
    \galapy is the fastest SED generation tool of its kind, with a peak performance of almost 1000 SEDs per second. 
    The models are generated on the fly without relying on templates, thus minimising memory consumption. 
    It exploits fully Bayesian parameter space sampling, which allows for the inference of parameter posteriors and thus facilitates the study of the correlations between the free parameters and the other physical quantities that can be derived from modelling. 
    The API and functions of \galapy are under continuous development, with planned extensions in the near future.
    In this first work, we introduce the project and showcase the photometric SED fitting tools already available to users. \galapy is available on the Python Package Index (PyPI) and comes with extensive online documentation and tutorials.
}


   \maketitle
%
%--------------------------------------------------------------------
\input{sec/intro}
%--------------------------------------------------------------------
\input{sec/models}
%--------------------------------------------------------------------
\input{sec/sampling}
%--------------------------------------------------------------------
\input{sec/validation}
%--------------------------------------------------------------------
\input{sec/summary}
%--------------------------------------------------------------------
\begin{acknowledgements}
The authors are grateful to the anonymous referee for the detailed and thorough comments that contributed to improve substantially the present work. The authors are grateful to Nicoletta Krachmalnicoff for the extensive and patient discussion on statistical inference and for her useful suggestions that improved the readability of the software-related sections of this work. The authors also thank Steph\`an Charlot for stimulating discussion and for his support to the project. 
This paper is supported by: ``Data Science methods for MultiMessenger Astrophysics \& Multi-Survey Cosmology'' funded by the Italian Ministry of University and Research, Programmazione triennale 2021/2023 (DM n.2503 dd. 9 December 2019), Programma Congiunto Scuole; EU H2020-MSCA-ITN-2019 n. 860744 \textit{BiD4BESt: Big Data applications for black hole Evolution STudies}; Italian Research Center on High Performance Computing Big Data and Quantum Computing (ICSC), project funded by European Union - NextGenerationEU - and National
Recovery and Resilience Plan (NRRP) - Mission 4 Component 2 within the activities of Spoke 3 (Astrophysics and Cosmos Observations); PRIN MUR 2022 project n. 20224JR28W "Charting unexplored avenues in Dark Matter"; INAF Large Grant 2022 funding scheme with the project "MeerKAT and LOFAR Team up: a Unique Radio Window on Galaxy/AGN co-Evolution; INAF GO-GTO Normal 2023 funding scheme with the project "Serendipitous H-ATLAS-fields Observations of Radio Extragalactic Sources (SHORES)".
DD acknowledges support from the National Science Center (NCN) grant SONATA (UMO-2020/39/D/ST9/00720).
\newline
This work made use of the C++ \citep{stroustrup2013c++} and Python \citep{van2007python} programming languages, and of the following software: 
Astropy \citep{astropy2013}, NumPy \citep{harris2020numpy}, SciPy \citep{virtanen2020scipy}, Matplotlib \citep{Hunter2007matplotlib}, Emcee \citep{emcee2013}, dynesty \citep{dynesty2020}, GetDist \citep{lewis2019getdist}, pybind11 \citep{jakob2017pybind11}, HDF5 \citep{folk2011overview}, h5py \citep{collette2021h5py}.
\end{acknowledgements}

%--------------------------------------------------------------------
\bibliographystyle{aa}
\bibliography{galapy}
%--------------------------------------------------------------------
%-------------------------------------------------------------------
\appendix
\input{sec/code_design}
\input{sec/appendix}
\input{sec/demo}
\end{document}

%% file: sec/intro.tex
%-------------------------------------------------------------------

\section{Introduction}
\label{intro}

Galaxies are extremely complex astrophysical objects resulting from the processes affecting baryonic matter after its collapse within dark matter haloes. 
Their formation and evolution strongly depend on the interplay of several factors, including their matter reservoir and accretion history, their environment and possible interactions with neighbours and, ultimately, the large scale structure of the Universe and the physics regulating it on cosmological scales. 
By studying the properties of individual galaxies, such as their luminosity, stellar mass, chemical composition, and star formation history, one can learn how such objects form and evolve over time as well as the cosmological conditions that lead to their assembly.

The broadband Spectral Energy Distribution (SED) of a galaxy describes the distribution of its light across the electromagnetic spectrum, from gamma rays to radio waves, and bears the imprints of the baryonic components and processes determining its evolutionary history.
Galaxy SEDs constitute primary tools of extra-galactic astronomy to constrain models of galaxy formation and evolution, which are an essential part for our understanding of the Universe as a whole.
The majority of commonly used SED fitting tools \citep[e.g.][]{magphys2008,beagle2016,bagpipes2018,cigale2019,prospector2021,beagle2022,lightning2023} have been mainly developed for studies of low redshift objects, thus providing the user with empirical fitting recipes that are (mostly) constrained in the local Universe.
Even though such tools have been extensively used in constraining the physical properties of galaxies, even at high redshift, they lack of a physically-motivated interplay between the recipes they use and the actual evolution of the modelled galaxy SED over cosmic time.
In several studies, this has required some tweaking and hacking, especially when it comes to the high-redshift Universe \citep{Novak2017,Jin2018,wang2019,Gruppioni2020,Pantoni2021,Talia2021,Giulietti2022,Enia2022,Jin2022,Castellano2022,Rodighero2023,Finkelstein2023}. 
Moreover, since the quality and spectral resolution of the SEDs in high-$z$ galaxies are typically much worse than for local objects, a detailed modelling of spectral features can be traded off for a focus on the quantities crucial to derive information about the star formation histories, dust content, and properties of the interstellar medium \citep[see, e.g.,][for two reviews on high redshift galaxies and the evolution of their content]{Forster2020,Tacconi2020}. 

On the theoretical side, investigating the SEDs of high-$z$ galaxies can inform us about the evolution of the overall galaxy population across cosmic times. 
For example, one crucial issue in galaxy evolution concerns the formation of local quiescent galaxies; the issue can in principle be cleared by investigating the SEDs of their high-redshift progenitors, that are thought to be dust-enshrouded star-forming objects forming most of their stars at $z\gtrsim 2$, during the so-called cosmic noon or further back in time during cosmic dawn, at $z \gtrsim 3$ \citep{Shapley2011,Lapi2018,Gruppioni2020,Talia2021}. 
Also, more physical but time-consuming radiative transfer SED models \citep[e.g.][]{grasil1998,Camps&Baes2020} are not suitable to be applied to the large available observational data sets.

In fact, on the observational side, ongoing and upcoming experiments are and will be producing an ever-increasing amount of data from galaxies at high redshifts.
For example, ALMA has opened a window up to redshift $z\sim 8$ in the (sub-)millimetre bands \citep[see e.g.][]{Walter2012,Simpson2014,Brisbin2017,GonzalezLopez2017,Scoville2017,Simpson2017,Franco2018,Bischetti2019,Dudzevetc2020,Simpson2020,Gruppioni2020,Pensabene2020, hodge&dacunha2020,Smail2021,Pensabene2021,rebels2022,Hamed2023}, while JWST is inspecting the Universe in the observed near-IR bands, both in photometry and spectroscopy, out to the Epoch of Reionization (EoR) and beyond \citep[e.g.][]{Castellano2022,Naidu2022,labbe2022,finkelstein2022,Adams2023,Atek2023,harikane2023,Yan2023};
these data complement the already available multi-wavelength data-sets from large high-$z$ observational campaigns such as the Great Observatories Origins Survey \citep[GOODS,][]{GOODS2004}, the \textit{Hubble} Ultra-Deep Field \citep[HUDF,][]{HUDF2006}, COSMOS \citep{COSMOS2007}, as well as data from deep and large-area blind surveys in the infrared domain, like PACS Evolutionary Probe \citep[PEP,][]{PEP2011}, \textit{Herschel} Multi-tired Extra-galactic Survey \citep[Her-MES,][]{HerMES2012}, \textit{Herschel} Astrophysical Terahertz Large Area Survey \citep[H-ATLAS,][]{HATLAS2010}, the \textit{Herschel} Extragalactic Legacy Project \citep[HELP,][]{shirley2019,shirley2021}.
On-going experiments, such as the Evolutionary Map of the Universe \citep[EMU,][]{EMU2021}, performed with ASKAP \citep{ASKAP2007,ASKAP2008,ASKAP2016,ASKAP2021} and the MeerKAT \citep{MeerKAT2012,MeerKAT2016} International GHz Tiered Extragalactic Exploration \citep[MIGHTEE,][]{MIGHTEE2016,MIGHTEE2017}, are tackling sensitivities never achieved before at the longest wavelengths of the extra-galactic emission spectrum.
These latter experiments are nonetheless only pathfinders for the unprecedented amount of data and scientific information that will be collected by the Square Kilometre Array Observatory \citep[SKAO,][]{SKAO2015}, in the same wavelength range.
Complementary, the \textit{Euclid} mission \citep{Euclid2018} with its visible imager \citep[VIS,][]{VISEuclid2016} and near infrared imaging photometer \citep[NIP,][]{NIPEuclid2010}, along with the Vera C. Rubin Observatory and its Legacy Survey of Space and Time \citep[LSST,][]{LSST2009}, as well as the Dark Energy Spectroscopic Instrument \citep[DESI,][]{DESI2016}, will probe the visible and infra-red regions of the spectrum on extremely wide areas and high sensitivities.

In this work, we present \galapy, an extensible API for modelling broadband galaxy SEDs with a particular focus on high-redshift objects\footnote{\galapy can be installed from the Python Package Index: \href{https://pypi.org/project/galapy-fit/}{pypi.org/project/galapy-fit}. The documentation is available at: \href{https://galapy.readthedocs.io/en/latest/index.html}{galapy.readthedocs.io/en/latest/index.html}}. 
It provides an easy-to-use Python user interface while the number-crunching is done with compiled, high-performance, object-oriented C++. 
The development of this tool is an on-going project and the software has been designed to envisage modelling extensions and computational upgrades that are already planned and under development.

In the deepest extra-galactic fields, such as COSMOS, the large amount of high-quality, panchromatic data requires not only the derivation of physical parameters, but also their interpretation. 
One possibility to tackle this point, is to provide informative priors on the model defining parameters, in \galapy this is guaranteed by the implemented Bayesian framework, which provides an interface to sophisticated statistical analysis, not possible with template-fitting codes.
Another possible approach is to directly include physical models (e.g., analytic solutions for galaxy evolution) within the SED modelling and fitting code.
This solution seems particularly important in the era of the large programs outlined above (e.g., synergy between JWST, ALMA, Euclid and LSST) that aim to explore the co-evolution of stars, dust, gas, and metals. 
Indeed, in spite of its potential importance, many previous SED models have not considered the co-evolution of all these components in a physically consistent manner.
To this end, along with more classical empirically motivated models, we have implemented a physically motivated model of star formation history (SFH): the In-Situ model based on works from \cite{Lapi2018}, \cite{Pantoni2019} and \cite{Lapi2020}.
With this model it is possible to get to an analytical estimate of various physical quantities characterising a galaxy, such as its dust and gas content as well as its metallicity.
It is mainly designed to interpret the emission of highly star forming galaxies that end up in local early type galaxies, along all their evolution from the highest to the lowest redshifts, but it also proves effective in modelling local late type galaxies.

As it is being confirmed by JWST since it started taking data, the high redshift Universe is populated by objects that are intensively star-forming and, crucially, highly obscured.
Dust plays a main role in shaping the emission of galaxies, especially in the earliest phases of evolution, but it is not granted that its absorption properties at high redshift can be safely modelled with attenuation laws empirically derived from observations of the low redshift Universe.
The approach we implement in \galapy to model dust is inspired by the one presented in the classical GRASIL code \citep{grasil1998}, with two dust components, one for the age-dependent evolution of molecular clouds around star forming regions and the other for diffuse dust, distributed on larger scales along the galaxy structure.
Differently from GRASIL, we account for the twofold role of dust, which obscures the emission at short wavelengths and re-emits at longer wavelengths, with an age-dependent energy conservation algorithm. 
This approach, while being physically motivated, keeps the execution time extremely contained with respect to radiative transfer algorithms.
With our dust model we can derive non-parametric total attenuation laws, blind to assumptions on the grain physics and with two components whose contributions to emission blend, shaping the dust emission peak.

In this work we showcase the current status of the project and we demonstrate its power for modelling broadband photometric data-sets.
The structure of the paper is as follows: in Sec.~\ref{sec:demo} we provide a primer on installation and usage of the package; in Sec.~\ref{sec:models} we describe in detail all the physical models currently delivered with \galapy; in Sec.~\ref{sec:inference} we discuss the statistical inference tools used for sampling the parameter space; in Sec.~\ref{sec:validation} we show the results of the thorough validation tests we have performed to verify reliability of the results and demonstrate the potential of our tool; ultimately, in Sec.~\ref{sec:summary} we summarise the key results presented in this manuscript. 
Throughout the work, we adopt the standard, flat $\Lambda$CDM cosmology from \cite{Planck2020} with rounded parameter values: matter density $\Omega_M \approx 0.3$, baryon density $\Omega_b \approx 0.05$, Hubble constant $H_0 = 100\, h$ km s$^{-1}$ Mpc$^{-1}$ with $h\approx 0.7$. A Chabrier \citep{Chabrier2003} initial mass function is assumed.

%% file: sec/models.tex
%--------------------------------------------------------------------
\section{Library Models}
\label{sec:models}

In this Section we introduce the physics modelled by the \galapy library.
All the physical components and processes have been implemented in separate modules, with the requirement of making each component and process self-consistent, meaning that every module (and therefore any physical process) can be imported and used as a stand-alone module of the library.

Conveniently, a master class \texttt{galapy.Galaxy.GXY} wraps-up all of the physical modules described in this Section, dealing with the interplay between different parameters and components.
The latter allows for computing straightforwardly the overall emission and derived quantities for a given set of parameters, enhancing the general user-friendliness of the workflow.
This class is meant to help the user accessing directly all the functionalities of the models already implemented in the library with minimal effort, as well as to ease the correct setting of the parameters that are inter-dependent among the different modular components.
It is nonetheless always possible to customise the workflow by accessing the API, importing functions and classes from the different modules into which the \galapy library is organised\footnote{Detailed tutorials on customisation instructions are (and will be made) available in the project on-line documentation: \href{https://galapy.readthedocs.io/en/latest/index.html}{galapy.readthedocs.io/en/latest/index.html}}.

Along the rest of this Section we provide a detailed description of all the physics currently implemented in \galapy. 
We refer the reader to Appendix~\ref{apx:params} for a complete list of all the possible free parameters that can be selected when fitting observational data. 
Tab.~\ref{tab:tunable_parameters} provides a handy conversion between the symbol uniquely identifying a parameter in the library and the mathematical symbol used in this manuscript, along with a short description and a reference to locate the position in text where the parameter is used.

%%%%%%%%%%%%%%%%%%%%%%%%%%%%%%%%%%%%%%%
\subsection{Star formation history}\label{sec:sfh}

Implemented in the \texttt{galapy.StarFormationHistory} module, the \texttt{SFH} object allows selecting various parametric and non-parametric star-formation history models. 

\begin{figure}
\centering
\includegraphics[width=\hsize]{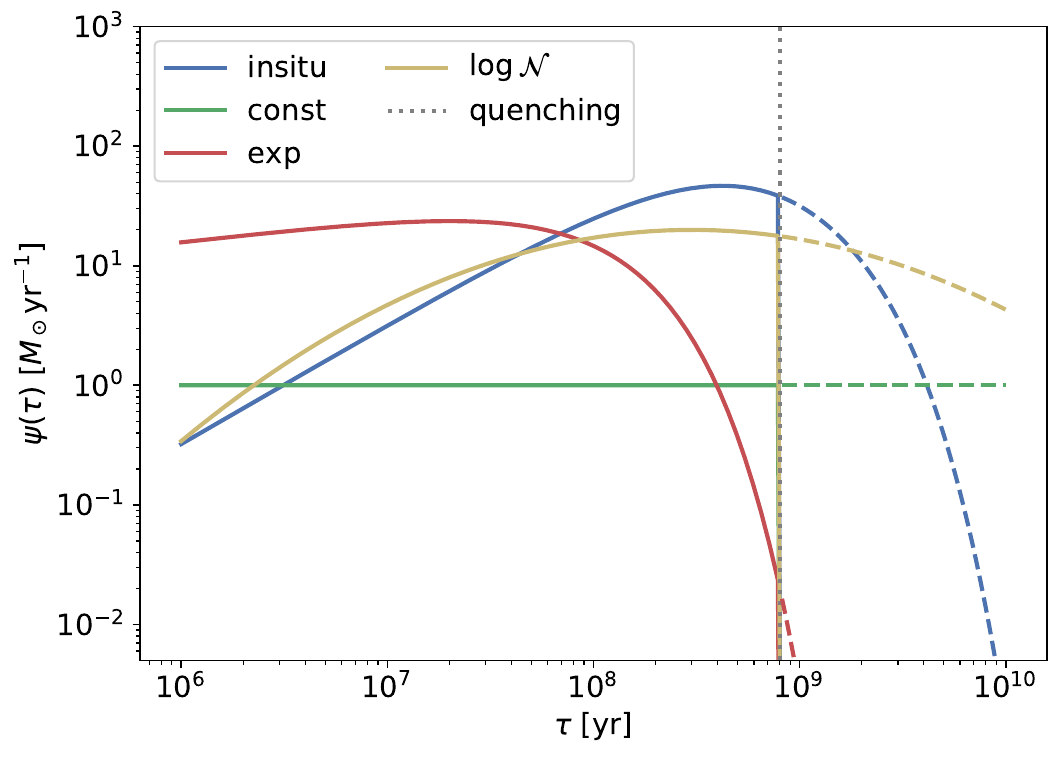}
  \caption{Different models of star formation histories (SFH). Each coloured line shows the star formation rate $\psi$ of a galaxy as a function of its age $\tau$, according to empirical (constant SFR, Eq.~\ref{eq:sfrconst}, green; delayed exponential, Eq.~\ref{eq:sfrgendelexp} with $\kappa=1$, red; log-normal, Eq.~\ref{eq:sfrlognormexp}, yellow) and physically-motivated (In-Situ, Eq.~\ref{eq:sfrinsitu}: blue line) models. The vertical dotted line marks the age $\tau_\text{quench}$ of a possible abrupt quenching event; solid lines refer to the SFH of objects undergoing quenching, while dashed lines to the SFH of objects for which no quenching occurs.}
\label{fig:sfh}
\end{figure}

\galapy offers the possibility to use standard empirical models of SFH (Sec.~\ref{sec:empirical}) as well as non-parametric models (Sec.~\ref{sec:nonparametric}).
However, the default SFH model, i.e. the In-Situ model by \cite{Lapi2018} (Sec.~\ref{sec:insitu}), has proven particularly successful in predicting the evolution of massive proto-elliptical galaxies and it is also very promising in explaining the fast formation of relatively massive galaxies at $z\gtrsim10$, one of the key regimes probed by modern observational campaigns (e.g. JWST, SKAO).
A quick summary of the parameterised SFRs provided by the different models follows.
\begin{itemize}
    \item Constant SFR:
    \begin{equation}
    \label{eq:sfrconst}
    \psi(t) = \psi_0~,
    \end{equation}
    where $\psi_0$ is a constant floating-point value expressed in units of $M_\odot/\text{yr}$.

    \item Generalised version of the delayed exponential SFR:
    \begin{equation}\label{eq:sfrgendelexp}
    \psi(t)\propto \tau^{\kappa}\, \exp{(-\tau/\tau_\star)}~,
    \end{equation}
    where $\tau_\star$ is the characteristic star-formation timescale and $\kappa$ is a shape parameter for the early evolution; $\kappa=0$ corresponds to a pure exponential, while $\kappa=1$ to the standard delayed exponential.

    \item Log-normal SFR:
    \begin{equation}\label{eq:sfrlognormexp}
    \psi(t)\propto \dfrac{1}{\tau}\, \dfrac{1}{\sqrt{2\pi\sigma_\star^2}}\, \exp\left[-\dfrac{\ln^2(\tau/\tau_\star)}{2\,\sigma_\star^2}\right]~;
    \end{equation}
    where $\tau_\star$ and $\sigma_\star$ control the peak age and width.

    \item In-Situ physically motivated model \citep{Lapi2018,Pantoni2019,Lapi2020}:
    \begin{equation}\label{eq:sfrinsitu}
    \psi(t)\propto e^{-x}-e^{-s\gamma\, x}~,
    \end{equation}
    where $x\equiv\tau/s\,\tau_\star$ with $s \approx 3$ a parameter related to gas condensation, while $\gamma$ is a parameter including gas dilution, recycling and the strength of stellar feedback \citep[see][for details]{Lapi2020}, whose value is described in Sec.~\ref{sec:insitu}.
\end{itemize}
We also allow for the existence of an eventual quenching event that stops the star formation.
This is modelled with a heaviside function, multiplying the SFR of choice, which is $1$ before $\tau_\text{quench}$ and $0$ afterwards.
The above rates are plotted for fixed values of the parameters in Fig.~\ref{fig:sfh} where we also show the effect of assuming an abrupt quenching event happening at an age of $\tau\approx10^9$ years. 
Note that, in this first version of the library, we only consider the primary episode of star formation, not secondary bursts that will be included in future updates of the package. 
Nevertheless, pure burst SFHs can be rendered either using the interpolated model or by particular combinations of the free-parameters regulating the shape of the models whose rates are reported above.

In our chemical evolution model, the stellar mass of a galaxy at a given age, $\tau$, is given by the integral
\begin{equation}\label{eq:stellarmass}
M_\star(\tau) = \int_0^{\tau}\text{d}\tau'~[1-\mathcal{R}(\tau-\tau')]\, \psi(\tau')~,
\end{equation}
where $\mathcal{R}$ is the recycled fraction of gas from stellar evolution.
The $\mathcal{R}(\tau)$ factor is given by \citep[see, e.g.,][]{Cimatti2020}
\begin{equation}
    \label{eq:recyclingtrue}
    \mathcal{R}(\tau) = \dfrac{1}{\psi(\tau)} \int_{m_\text{min}(\tau)}^{m_\text{max}} \psi(\tau -\tau_\text{MS}) \phi(m - m_\text{rem}) \text{d}m\,\text{,}
\end{equation}
where $\tau_\text{MS}$ the time spent by a star with mass $m$ in the main sequence, $m_\text{rem}$ is the mass of its remnant, $m_\text{min}(\tau)$ satisfies $\tau_\text{MS}(m_\text{min}) = \tau$ and $\phi(m)$ is the Initial Mass Function (IMF).
For, e.g., a Chabrier IMF \citep{Chabrier2003} it is well approximated by
\begin{equation}\label{eq:sfrrecycled}
\mathcal{R}(\tau) \approx 0.05\,\ln\left(1+\dfrac{\tau}{0.4\, \text{Myr}}\right)~,
\end{equation}
with typical values around $\mathcal{R}\approx 0.4\div0.5$ after $1\div10$ Gyr \citep{Pantoni2019,Lapi2020}; in the instantaneous recycling approximation $\mathcal{R}\approx 0.45$.
Even though, at the current state of development, \galapy implements the aforementioned values only for the case of a Chabrier IMF in the mass range $0.1 \leq M_\star \leq 100$, the library is easily extensible with further models of IMF, including non-standard ones \citep[e.g.][]{Kroupa2013,Fontanot2018}, that will be added in future releases of the library. 

\begin{figure}
\centering
\includegraphics[width=\hsize]{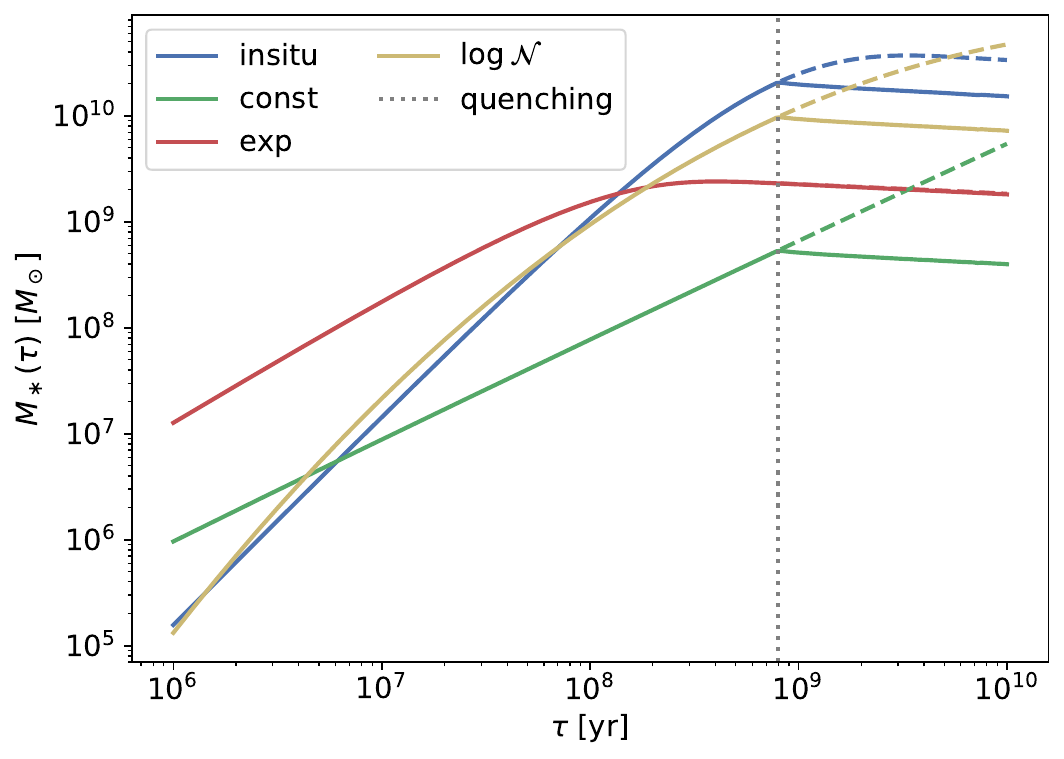}
  \caption{Evolution of the stellar mass $M_\star$ of a galaxy as a function of its age $\tau$, for the SFH models shown in Fig.~\ref{fig:sfh}.}
     \label{fig:Mstar}
\end{figure}

Fig.~\ref{fig:Mstar} shows the stellar mass growth history corresponding to the models of Fig.~\ref{fig:sfh}. 
As it is shown in Fig.~\ref{fig:Mstar}, the overall stellar mass slowly decreases after star formation starts to fade, as a result of the ageing of stellar populations.

%%%%%%%%%%%%%%%%%%%%%%%%%%%%%%%%%%%%%%%
\subsubsection{Empirical models}\label{sec:empirical}

Most of the SED-fitting libraries available in literature are delivered with empirical models of SFH \citep[see, e.g.][for two popular SED-fitting libraries]{magphys2008,cigale2019}.
Such models are primarily motivated by the necessity of reproducing the shape of the cosmic star formation history or, either, to provide a numerically tractable function that returns reasonable values of SFR.
A notable exception from this is Prospector \citep{prospector2021} which provides a step-wise tunable SFH module. 
Even though this approach avoids assumptions on uncertain processes, it ends up in a high dimensional parameter space that slows down inference and reduces accuracy on the parameters estimate.

In tools based on empirical star formation laws, the dust mass $M_\text{dust}$ and the gas/stellar metallicity $Z_\text{gas} = Z_\star$ are typically free parameters, while the gas mass is derived on the basis of a (possibly metallicity-dependent) dust-to-gas mass ratio $D$ gauged on observations. 
A common expression is \citep[e.g.][]{Tacconi2018}:
\begin{equation}
    D_\text{emp}(Z) = \dfrac{M_\text{dust}}{M_\text{gas}} \approx 0.01\, \left(\dfrac{Z_\text{gas}}{Z_\odot}\right)^{-0.85}~.
\end{equation} 
In \galapy, we keep such an approach for backward compatibility and comparison with alternative fitting codes.

%%%%%%%%%%%%%%%%%%%%%%%%%%%%%%%%%%%%%%%
\subsubsection{Interpolated model}\label{sec:nonparametric}

We provide a non-parametric, interpolated, step-wise SFH model with derived components (like dust/gas mass and metallicity) treated as free parameters, as in the previous empirical models presented.
This model is designed for users willing to predict the emission from galaxies for which the stellar mass growth history is available (e.g. obtained from hydro-dynamical simulations or with semi-analytical models) or to test the behaviour of exotic and arbitrarily complex SFH shapes.

\begin{figure}
\centering
\includegraphics[width=\hsize]{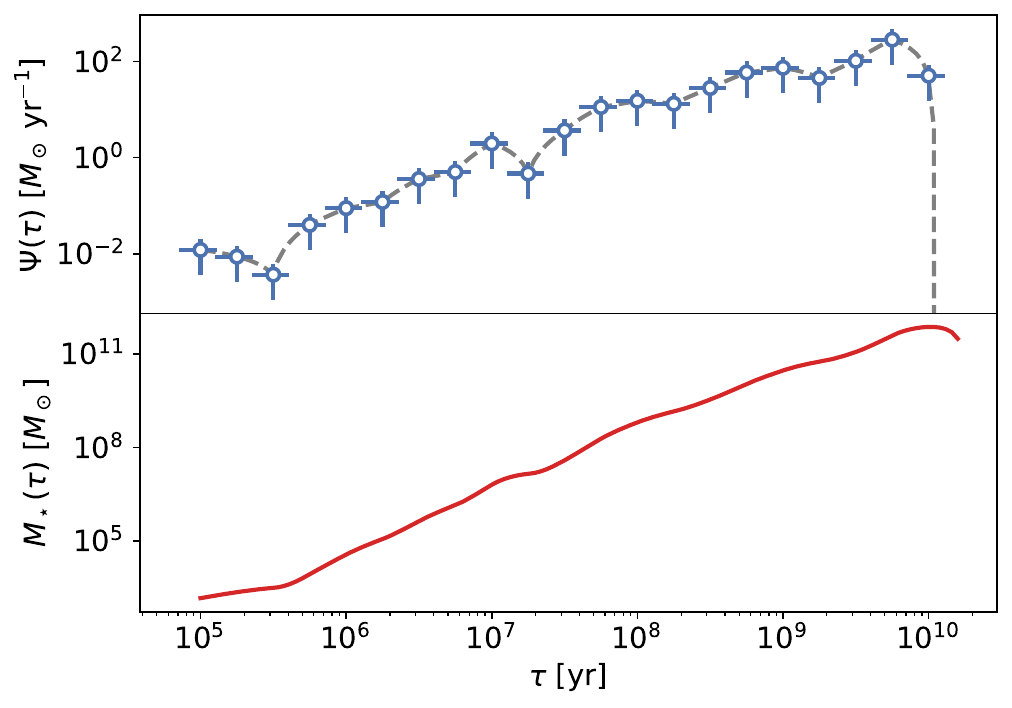}
  \caption{Interpolated SFH model. Upper panel: SFR at different epochs as measured for a simulated galaxy (blue markers with errors) and our interpolated model (dashed grey line). Lower panel: evolution of the integrated stellar mass at different epochs (solid red line) resulting from the SFH interpolated from data in the upper panel.}
     \label{fig:sfh_interp}
\end{figure}
In Fig.~\ref{fig:sfh_interp} we show an example of this non parametric model. 
In the upper panel we plot ``observed'' samplings of a simulated galaxy's SFH (blue markers with error bars) and the up-sampled prediction of our interpolated model (dashed grey line).  
On the lower panel we show the stellar mass growth history resulting from integrating the interpolated SFH along the time-coordinate.

%%%%%%%%%%%%%%%%%%%%%%%%%%%%%%%%%%%%%%%
\subsubsection{In-situ model}\label{sec:insitu}

The In-Situ SFH delivered as default in \galapy implements the (mostly analytic) galaxy formation model first presented in \cite{Lapi2018} for ETGs, further developed in \cite{Pantoni2019} and extended to LTGs in \cite{Lapi2020}.
This model is based on a self-consistent treatment of the black-hole/host-galaxy co-evolution, which captures the fast collapse, with low angular momentum, of the innermost gaseous regions of a galaxy and the resulting stellar feedback.
Such a regime is extremely important when interpreting data-sets of galaxies at considerable redshift (i.e. $z \gtrsim 4$ and beyond).
Furthermore, the model allows for the derivation of age-dependent analytical expressions of the evolution of the gas, metals and dust content in galaxies.

Concerning the SFR, the effects of recycling and stellar feedback are encapsulated in the parameter $\gamma$ that appears in Eq.~\eqref{eq:sfrinsitu} and is defined as
\begin{equation}\label{eq:gammainsitu}
    \gamma \equiv 1 - \mathcal{R} + \epsilon_\text{out}~,
\end{equation}
in terms of the recycled gas fraction $\mathcal{R}$ of Eq.~\eqref{eq:sfrrecycled} and of the mass loading factor of the outflows from stellar feedback, $\epsilon_\text{out}$.
We gauge $\epsilon_\text{out} \approx 3[\psi_\text{max}/M_\odot\text{yr}^{-1}]^{-0.3}$ according to the hydrodynamic simulations of stellar feedback from \cite{Hopkins2012}.
Therefore, the parameter $\gamma$ is completely determined in terms of the free parameter $\psi_\text{max}$ and, eventually, by the age of the galaxy $\tau$, through Eq.~\eqref{eq:sfrrecycled}.

In the In-Situ model, the evolution of the gas/dust masses and of the gas/stellar metallicity can be followed analytically as a function of the galactic age and self-consistently with respect to the evolution of the SFR. 
Specifically, the gas mass is given by 
\begin{equation}\label{eq:insitu_Mgas}
M_\text{gas}(\tau)=\psi(\tau)\, \tau_\star~,
\end{equation}
while the dust mass 
\begin{equation}\label{eq:insitu_Mdust}
M_\text{dust}(\tau)=M_\text{gas}(\tau)\, D_\text{in-situ}(\tau)
\end{equation}
is computed in terms of the gas mass and of the dust-to-gas mass ratio $D_\text{in-situ}$.
As discussed in \cite{Pantoni2019} and \cite{Lapi2020}, for this latter quantity it is possible to derive an analytical expression which writes down
\begin{equation}\label{eq:insitu_dusttogas}
\begin{aligned}
D_\text{in-situ}(\tau) & \approx  \dfrac{s^3\, \epsilon_\text{acc}\,y_D\,y_Z}{[s\gamma-1]\,[s(\gamma+\kappa_\text{SN})-1]\,[s(\gamma+\tilde\epsilon)-1]}\times\\
&\\
& \times \left\{1-\dfrac{(s\, \gamma-1)\, x}{e^{(s\, \gamma-1)\, x}-1}\, \left[1+\dfrac{s\, \gamma-1}{ s\, \tilde\epsilon}\, \left(1-\dfrac{1-e^{-s\,\tilde\epsilon\, x}}{s\,\tilde \epsilon\, x}\right)\right]\right\}~;
\end{aligned}
\end{equation}
where 
\begin{equation}
    \tilde\epsilon\equiv \kappa_\text{SN}+\epsilon_\text{acc}\, s\,y_D/[s(\gamma+\kappa_\text{SN})-1]
\end{equation} 
provides a measure of the efficiency with which dust grains form in terms of the metal coagulation efficiency $\epsilon_\text{acc}\approx 10^6$ onto dust grains, of the dust spallation efficiency $\kappa_\text{SN}\approx 10$  by SN shock-waves, and of the dust production yield $y_D\approx 3.8\times 10^{-4}$.

\begin{figure*}
\resizebox{\hsize}{!}{
\includegraphics{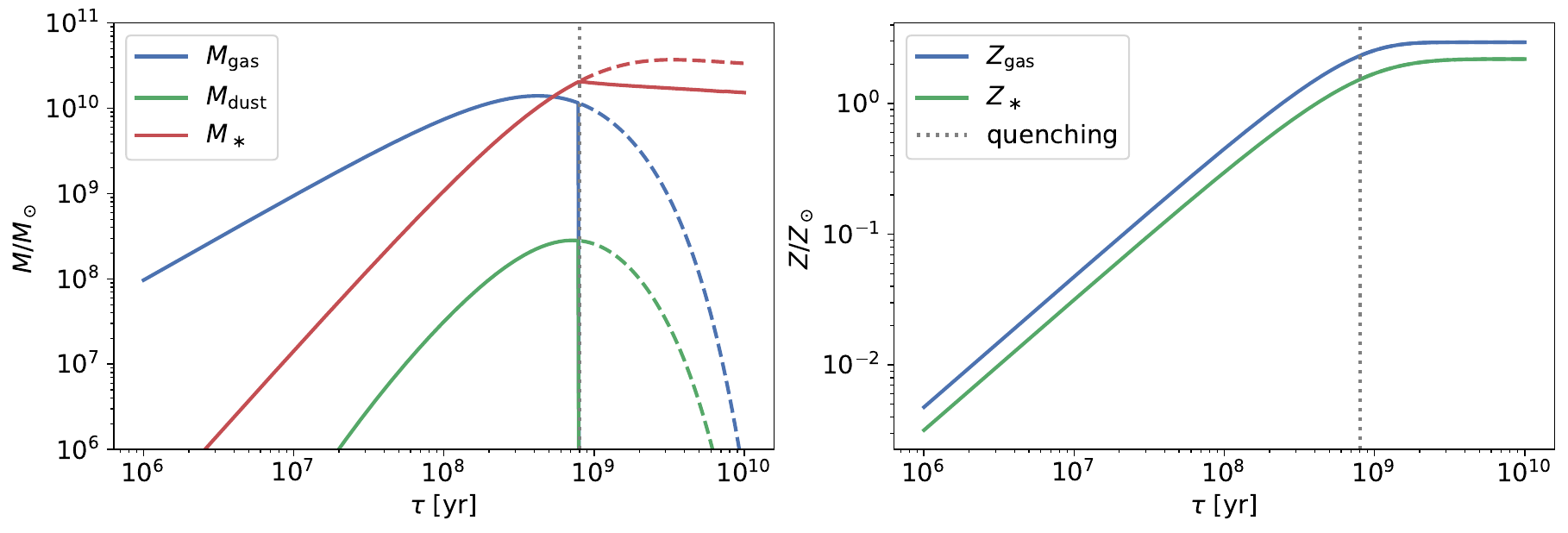}}
  \caption{Evolution of various quantities according to the in-situ SFH model. \textit{Left panel:} stellar mass (Eqs.~\ref{eq:sfrinsitu}-\ref{eq:stellarmass}, red), gas mass (Eq.~\ref{eq:insitu_Mgas}, blue) and dust mass (Eq.~\ref{eq:insitu_Mdust}, green). \textit{Right panel:} gas (blue) and stellar (green) metallicities (Eqs.~\ref{eq:insitu_metals}). Linestyles as in Fig. \ref{fig:sfh}.}
\label{fig:insitu_derived}
\end{figure*}
On the left panel of Fig.~\ref{fig:insitu_derived}, we show the evolution of the gas-mass (blue lines) and of the dust-mass (green lines), for the in-situ SFH model with the same parameters as for the blue line of Fig.~\ref{fig:sfh}.
As it is evident from the figure, the effect of assuming an abrupt quenching event is to wash out the diffuse matter reservoir of the interested galaxy, that therefore ends up loosing its primary source of star formation and starts ageing.

The authors also derive expressions for the gas and stellar metallicity:
\begin{equation}\label{eq:insitu_metals}
\left\{
\begin{aligned}
Z_\text{gas}(\tau) &\approx \cfrac{s\, y_Z}{(s\gamma-1)}\, \left[1-\frac{(s\,\gamma-1)\, x}{e^{(s\, \gamma-1)\, x}-1}\right]~,\\
\\
Z_\star(\tau) &\approx \cfrac{y_Z}{\gamma}\, \left[1-\frac{s\, \gamma}{s\, \gamma-1}\, \frac{e^{-x}-e^{-s\, \gamma\, x}\, [1+(s\, \gamma-1)\, x]}{s\gamma-1+e^{-s\, \gamma\, x}-s\gamma\,e^{-x}}\right]~,
\end{aligned}
\right.
\end{equation}
where again $x\equiv \tau/s\tau_\star$, and $y_Z\approx 0.04$ is the metal production yield (already including recycling) for a Chabrier IMF.
We show the behaviour of the metallicity evolution of the two different components on the right panel of Fig.~\ref{fig:insitu_derived}. 
As the gaseous component is expected to be enriched more readily with respect to stars, $Z_\text{gas}$ is higher than $Z_\star$ consistently along all the evolution history of the galaxy.

Despite being a spatially averaged description of the interplay between the different galaxy components, as well as their evolution, having access to analytical expressions allows to effectively reduce the volume of the parameter space that has to be sampled for fitting an SED.
This not only allows for a faster convergence to an optimal SED, but it also increases the accuracy of our estimates.
Furthermore, it is worth highlighting that such a consistent interplay between components, as well as the age-evolution of these derived quantities, are not commonly present in SED-fitting libraries based on energy conservation. 
In this respect, \galapy constitutes an innovative and powerful tool for providing non-parametric estimates of the components building up galaxies.

%%%%%%%%%%%%%%%%%%%%%%%%%%%%%%%%%%%%%%%
\subsection{Stellar emission}\label{sec:stellar}

Under the assumption of an Universal IMF, the stellar birthrate of a galaxy can be split \citep[see their Eq.s~(1-3)]{Bressan1994} in the product of a mass dependent function (i.e. the IMF) and of a time dependent function (i.e. the SFR). 
In this scenario, the intrinsic luminosity of stars in a galaxy at a given age is the result of the evolution of the several simple stellar populations (SSPs) that have formed and have aged within the structure in all of its history.
Each of the SSPs yields a luminosity $L_\text{SSP}$ that is computed by the convolution of an initial mass function (IMF), $\phi(m_\star)$, with the luminosity of single stars from stellar evolutionary tracks, $L_\text{star}$.
\begin{equation}\label{eq:LSSP}
L_\text{SSP}(\lambda,\tau,Z)=\int\text{d}m_\star\, m_\star\, \phi(m_\star)\, L_\text{star}(m_\star, \lambda, \tau, Z)~,
\end{equation}
where $m_\star$ is the mass of a single star, $\lambda$ is the wavelength, $\tau$ is the age and $Z$ is the metallicity of the given SSP.
%\textcolor{cyan}{The formulation of Eq.~\eqref{eq:LSSP} stands true under the assumption of an Universal IMF, that does not vary with cosmic time \citep[, see their Eq.s~(1-3)]{Bressan1994}.}

\begin{figure*}
\resizebox{\hsize}{!}{
\includegraphics{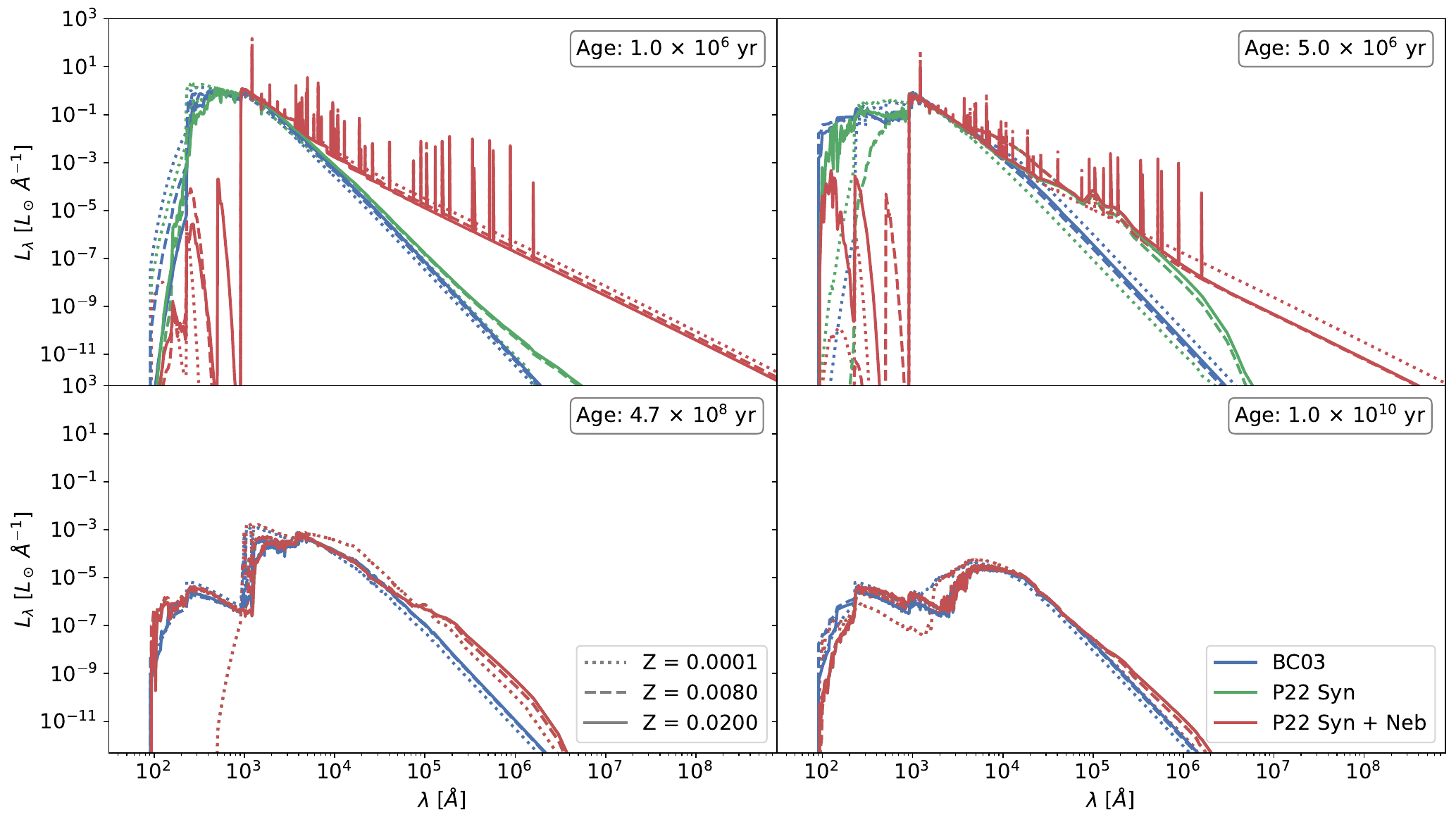}}
  \caption{Comparison among the different simple stellar population libraries included in \galapy: classic BC03 (blue), PARSEC22 without line emission (green), PARSEC22 with line emission (red). Linestyles refer to different metallicity as reported in legend. In all panels the approximate age of the stellar population is reported. In the top two panels we show SSPs from young stellar populations for whom the PARSEC22 models include the contribute of synchrotron (green models) as well as that of nebular emission (red models).}
\label{fig:ssp}
\end{figure*}
In \galapy we use pre-computed SSP libraries in the form of binary files with specific formatting\footnote{Note that we also provide functionalities to convert eventual custom SSP libraries into the accepted format.}. 
In its first release \galapy is distributed with two main libraries. 
\begin{itemize}
    \item The first one is the classic and popular \cite{BruzualCharlot2003} in its updated version (v2016). 
    This library provides the continuum luminosity from SSPs for a set of different IMFs, at varying wavelength, age and metallicity.
    We refer to these set of libraries as BC03. 
    Blue lines in Fig.~\ref{fig:ssp} show SSPs extracted from this set of libraries, for different metallicities (different line-styles) and for different ages (different panels).
    \item As an alternative we have also produced an additional set of SSPs with the PARSEC code \citep{Bressan2012,Chen2014,Chen2015} for a Chabrier IMF and varying ages and metallicities, including emission from dusty AGB stars \citep{Bressan1998}. 
    These libraries come in two flavours, the first one with continuum emission only (green lines in Fig.~\ref{fig:ssp}) and the second also including nebular emission (red lines in Fig.~\ref{fig:ssp}).
    In the former, besides continuum stellar emission, non-thermal synchrotron emission from core-collapse supernovae is also included in each SSP spectrum \citep[see, e.g. ][]{Vega2008}.
    In the latter, on top of the stellar continuum and non-thermal synchrotron, nebular emission is also included, with both free-free continuum and nebular emission \citep[see, e.g.][]{Mayya2004}, calculated with CLOUDY \citep{cloudy90,cloudy13,cloudy17}.
    We refer to these set of libraries as PARSEC22.
\end{itemize}
We highlight that, using the PARSEC22 SSP libraries come with the advantage of reducing the total amount of computations the code has to perform for getting to a final equivalent SED.
Namely, using our custom SSP libraries avoids the need to compute the radio stellar emissions that otherwise would require building the synchrotron and nebular free-free contributions described in Section~\ref{sec:radio}.
Furthermore, nebular line emission is currently only available with the PARSEC22 SSP libraries.
We refer the reader to Appendix~\ref{apx:ssp} for further discussion on the differences between the two SSP libraries distributed with \galapy.

At a given age $\tau$ a galaxy that has followed a SFH $\psi(\tau)$ will host a composite stellar population (CSP) resulting from all the SSPs formed and evolved up to that moment.
The overall unattenuated stellar luminosity of the CSP, $L_\text{CSP}^\text{i}$ (where the superscript ``$\text{i}$'' stands for \textit{intrinsic}), is therefore computed by integrating the contribution of SSPs at different ages and stellar metallicities weighted by the formed stellar mass:
\begin{equation}\label{eq:Lunatt}
L_\text{CSP}^\text{i}(\lambda,\tau) = \int_0^{\tau}\text{d}\tau_\text{SSP}\, L_\text{SSP}\left[\lambda, \tau_\text{SSP},Z_\star(\tau-\tau_\text{SSP})\right]\, \psi(\tau-\tau_\text{SSP})~,
\end{equation}
where $\tau_\text{SSP}$ is the age of the SSP, $L_\text{SSP}$ is the luminosity of the SSP per unit stellar mass defined in Eq.~\eqref{eq:LSSP} and $Z_\star(\tau-\tau_\text{SSP})$ is the metallicity of stars at a given instant in the galactic history of metal enrichment.
Eq.~\eqref{eq:Lunatt} is computed by summing up the light emitted by all the contributing SSPs (as described in Appendix~\ref{apx:register} and Eq.~\eqref{eq:LCSP_numerical}), the resolution used to compute $\psi(\tau-\tau_\text{SSP})$ is fixed at a value $\text{d}\tau = 10^5$ years.
For both the BC03 tables and the PARSEC22 tables, the time domain is sampled on an irregular grid that reaches a maximum accuracy of $\delta\tau = 10^5$ years. 

\begin{figure*}
\centering
\resizebox{\hsize}{!}{
\includegraphics[width=0.99\hsize]{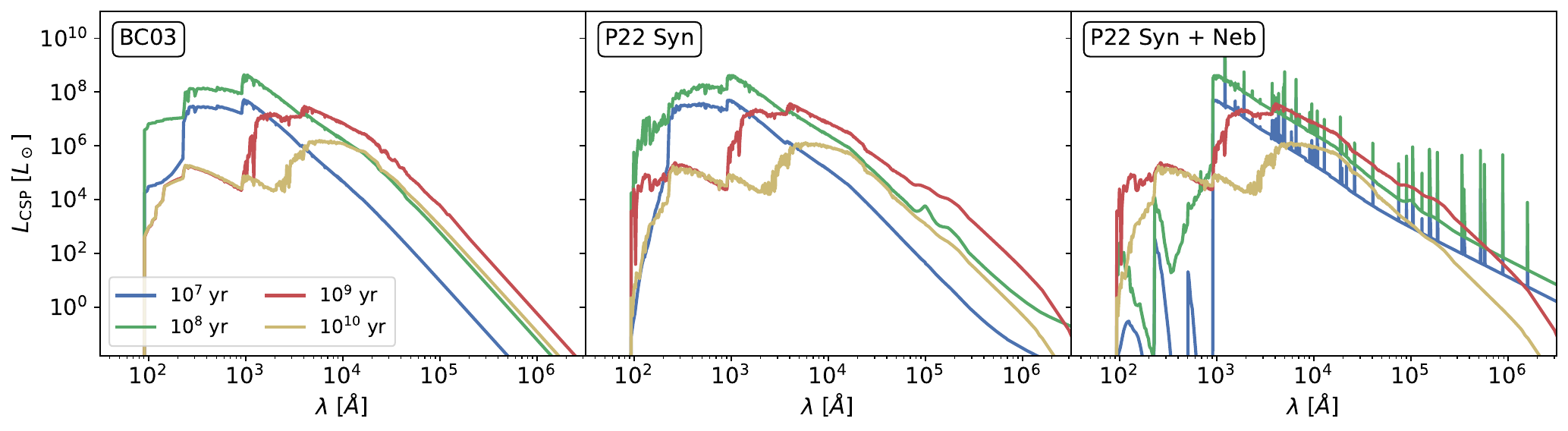}}
  \caption{Composite stellar populations as a function of the age, on assuming the in-situ SFH (blue line of Fig.~\ref{fig:sfh} with quenching at $\tau = 8 \times 10^8$ years) and different single stellar population libraries (left: BC03; centre: PARSEC22 without nebular emission; right: PARSEC22 with nebular emission).}
     \label{fig:csp}
\end{figure*}
In Fig.~\ref{fig:csp} we show the unattenuated stellar emission computed with \galapy and by integrating Eq.~\eqref{eq:Lunatt} up to different galactic ages. 
We use our PARSEC22 SSP libraries with nebular emission and integrate them along an in-situ SFH with parameters set as for the blue line in Fig.~\ref{fig:sfh} with quenching.
Note that the younger CSPs ($\tau = 10^7$ years in blue and $\tau = 10^8$ years in green) also show at the longer wavelengths the radio component resulting from the SN-synchrotron and nebular emission.
It is also worth mentioning that the UV part of the spectrum in the aforementioned CSPs is somewhat depressed as that fraction of the energy budget is absorbed in nebular regions around massive stars and re-emitted by line-transitions.

As a further open question in galaxy evolution is whether the models of IMF developed from studies on the local Universe are representative of stellar populations in the high redshift Universe, we plan to detach from fixed IMF models.
In particular and thanks to its object oriented design, the library is already prepared to work with a parameterised IMF.
This would mean to integrate Eq.~\ref{eq:LSSP} directly instead of getting it from pre-computed libraries.

%%%%%%%%%%%%%%%%%%%%%%%%%%%%%%%%%%%%%%%
\subsection{The age-dependent, two-component dust model}
\label{sec:ism}

Despite the dust mass in galaxies is usually a few orders of magnitude less than other components (behaviour that is captured by our In-Situ SFH model, as it is shown in the left panel of Fig.~\ref{fig:insitu_derived}), it plays a fundamental role on the emitted spectrum \citep{DraineLi2001,Draine2007,draine2011}.
Interstellar dust grains contribute to a galaxy's spectrum by playing a dual role: they absorb and scatter the intrinsic stellar radiation from CSPs, especially in the UV/optical range, while re-radiating the absorbed energy, primarily in the infrared part of the spectrum.

In \galapy we implement an age-dependent, two-component dust model. It comprises a typically hotter molecular cloud phase (in literature also referred to as ``birth clouds'') and a colder diffuse medium (in literature also dubbed ``cirrus''). The fraction of dust that resides in molecular clouds, $f_\text{MC}$ is a free-parameter of the \galapy model. 
This parameter anyways assumes typical values around $0.5$ and is likely larger in more violently star-forming systems and towards high redshift.

The modelling of the two dust components and of their attenuation and re-radiation  has been inspired from previous works,
namely from the radiative transfer code \texttt{GRASIL} \citep{grasil1998}, and from popular SED-fitting libraries such as MAGPHYS \citep{magphys2008}, CIGALE \citep{cigale2019} and Prospector \citep{prospector2021}. 
Like in the latter libraries, \galapy bypasses the computational cost of radiative transfer by exploiting an energy conservation scheme; however, \galapy implements energy conservation in an age-dependent way. 
This means that the attenuation from dust is age-dependent (like in \texttt{GRASIL}) and this in turn determines, via a self-consistent energy conservation calculated time-step by time-step, age-dependent dust temperatures across the galaxy lifetime. 
Details of such modelling are provided in the rest of this Section, that we divide into two parts for clarity, separating between the two main effects that result from the presence of a dust component, namely attenuation of UV-optical radiation and re-emission in the IR and (sub)mm bands.

\subsubsection{Extinction and attenuation}\label{sec:extinctionattenuation}

\begin{figure}
\centering
\includegraphics[width=\hsize]{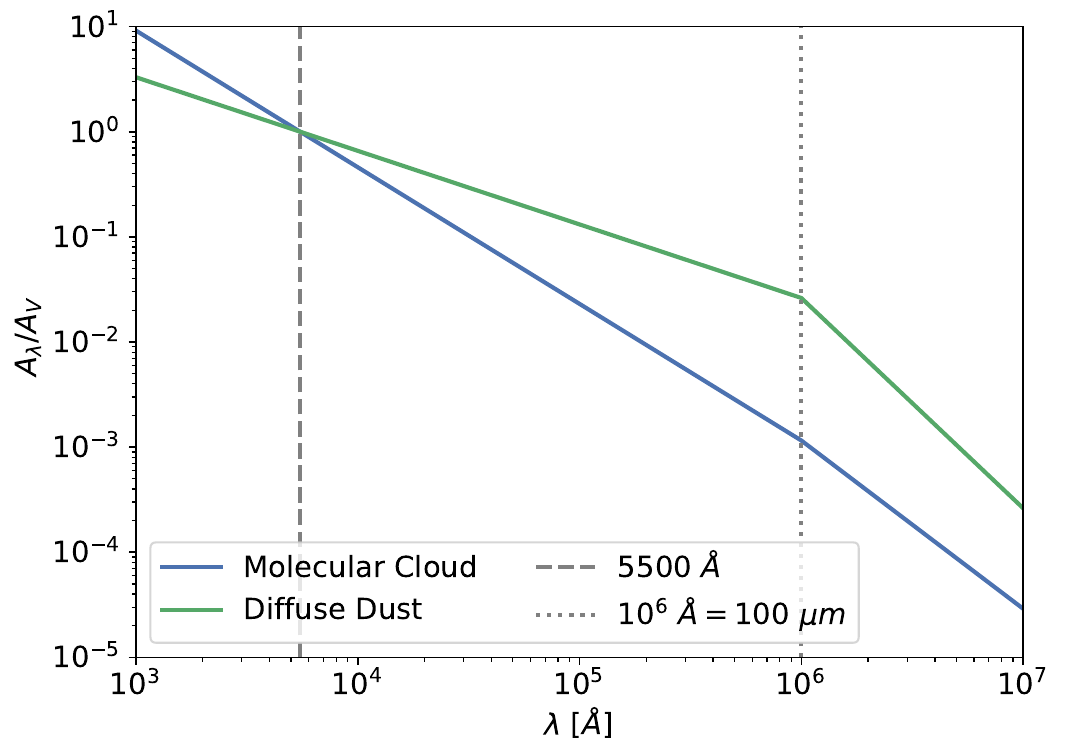}
  \caption{Extinction laws adopted in \galapy, normalised to the value in the $V-$band at $5500\ \text{\text{\AA}}$, for diffuse dust (green) and dust in molecular clouds (blue). The extinction is modelled as a double powerlaw (see Eqs.~\ref{eq:ADD}-\ref{eq:AMC}) with slope changing at around $100\mu$m (marked with a dotted line; the default values of the slopes are adopted in the plot (see Sect. \ref{sec:ism} for details).}
     \label{fig:extinction}
\end{figure}

We assume two different, piece-wise extinction curves for the diffuse dust and molecular cloud components (DD and MC, hereafter).
The behaviour of the two extinctions is shown in Fig.~\ref{fig:extinction} normalised to their value computed in the $V-$band ($\lambda_V\approx 5500$ \AA), $A_V$.
This value is parameterised differently for the two components.

For the DD phase we assume an extinction normalisation scaling as
\begin{equation}\label{eq:AVDD}
A_V^\text{DD} = \mathcal{C}_V^\text{DD}\, \frac{1-f_\text{MC}}{0.5}\, \dfrac{M_\text{dust}}{10^8\, M_\odot}\, \left(\dfrac{R_\text{DD}}{1\, \text{kpc}}\right)^{-2}~,
\end{equation}
here $M_\text{dust}$ is the dust mass in the galaxy (that can be age dependent if the in-situ SFH model is selected), $R_\text{DD}$ is the characteristic radius of the diffuse dust component and $\mathcal{C}_V^\text{DD}$ is a normalisation constant of order unity.
The extinction law for the diffuse component is prescribed to follow the piece-wise power-law behaviour
\begin{equation}\label{eq:ADD}
A_\text{DD}(\lambda) = A_V^\text{DD}\, \left(\dfrac{\lambda}{5500\, {\text{\AA}}}\right)^{-\delta_\text{DD}^\text{l/u}}~,
\end{equation}
with $\delta_\text{DD}^\text{l}$ taking on values of around $\approx 0.7$ for $\lambda\lesssim 100\, \mu$m and $\delta_\text{DD}^\text{u}\approx 2$ for $\lambda\gtrsim 100\,\mu$m; nonetheless \galapy allows to give up on these two reference values by directly fitting the parameters $\delta_\text{DD}^\text{l}$ and $\delta_\text{DD}^\text{u}$.
In Fig.~\ref{fig:extinction} we mark the normalised DD extinction law with reference slopes with a green line.

The piece-wise behaviour imposed by the relation above results from observations of the different cross section of dust grains settling-in at around $\approx 100~\mu$m.
The flatter power-law dependence has been previously assumed by \cite{CharlotFall2000} and \cite{magphys2008} basing on the observed relation between the ratio of far-infrared to UV luminosity and the UV spectral slope for nearby starburst galaxies, while the break to a steeper slope reflects the behaviour of the scattering and absorption cross section of dust grains at longer wavelengths \citep[e.g.][]{grasil1998,DraineLi2001}.

On the other hand, for the MC component we adopt a V-band extinction
\begin{equation}\label{eq:AVMC}
A_V^\text{MC} = \mathcal{C}_V^\text{MC}\, \dfrac{Z_\text{gas}}{Z_\odot}\, \dfrac{f_\text{MC} M_\text{gas}}{N_\text{MC} 10^6\, M_\odot}\, \left(\dfrac{R_\text{MC}}{16\, \text{pc}}\right)^{-2},
\end{equation}
which is dependent on the average gas mass of a molecular cloud $M_\text{gas}^\text{MC}/10^6 M_\odot = (f_\text{MC} M_\text{gas})/(N_\text{MC} 10^6\, M_\odot)$ in units of $10^6~M_\odot$, on the radius of the cloud $R_\text{MC}$, on the total number of MCs in the system and on the gas metallicity $Z_\text{gas}$.
The latter dependence is reasonable since within molecular clouds the growth of dust grains is expected to mainly occur via sticking/accretion of metals onto core grains produced during stellar evolution. 
The normalization $\mathcal{C}_V^\text{MC}$ can be of order several tens to hundreds, since the $V-$band emission in MCs is expected to be completely absorbed. 
For the MCs we also assume a double power-law extinction curve
\begin{equation}\label{eq:AMC}
A_\text{MC}(\lambda) = A_V^\text{MC}\, \left(\dfrac{\lambda}{5500\, {\text{\AA}}}\right)^{-\delta_\text{MC}^\text{l/u}}~,
\end{equation}
with $\delta_\text{MC}^\text{l}\approx 1.3$ for $\lambda\lesssim 100\, \mu$m and $\delta_\text{MC}^\text{u}\approx 1.6$ for $\lambda\gtrsim 100\, \mu$m. 
The former slope corresponds to the middle range of the optical properties of dust grains between the Milky Way, the Large and the Small Magellanic Clouds \citep[see][]{CharlotFall2000,magphys2008}.
The slope at long wavelengths, which is slightly shallower than for the diffuse medium, has been advocated to reproduce the sub-mm emission for ULIRGs like Arp220 where the MC re-radiation dominate over the cirrus' \citep[see][]{grasil1998,Lacey2016}.
MC extinction with reference slopes is shown in Fig.~\ref{fig:extinction} by a blue piece-wise power law. 
Once again, the slopes can be free-parameters of the model to be fitted directly. 

Given the extinction curves of Eq.~\eqref{eq:ADD} and Eq.~\eqref{eq:AMC}, we compute the attenuated galaxy luminosity (i.e. the transmitted one) as
\begin{multline}\label{eq:Latt}
L_\text{CSP}^\text{a}(\lambda, \tau) = \mathcal{A}_\text{DD}(\lambda) \times\\ 
\\
\times \int_0^\tau\text{d}\tau_\text{SSP} \mathcal{A}_\text{MC}(\lambda,\tau_\text{SSP})\,L_\text{SSP}[\lambda,\tau_\text{SSP},Z_\star(\tau-\tau_\text{SSP})]\, \psi(\tau-\tau_\text{SSP})~,
\end{multline}
where $\mathcal{A}_\text{DD}(\lambda)$ and $\mathcal{A}_\text{MC}(\lambda, \tau)$ are the extinction factors due to diffuse and MC dust, respectively, and the superscript ``a'' stands for attenuated.
We assume the attenuation suffered by radiation from stars that have already escaped their birth MCs to be independent from stellar age; thus the DD extinction factor just reads
\begin{equation}\label{eq:attDD}
\mathcal{A}_\text{DD}(\lambda) = 10^{-0.4\, A_\text{DD}(\lambda)}~.
\end{equation}
where $A_\text{DD}(\lambda)$ is given by Eq.~\eqref{eq:ADD}.

On the other hand, since birth clouds tend to be evaporated as the hosted SSPs evolve, the extinction factor due to dust in MCs is defined to be age dependent
\begin{equation}\label{eq:attMC}
\mathcal{A}_\text{MC}(\lambda) = 1-\eta(\tau)+\eta(\tau)\, 10^{-0.4\, A_\text{MC}(\lambda)}~,
\end{equation}
where $\eta(\tau)$ defines the fraction of stars with age $\tau$ still inside their MC.
We define this latter quantity taking up from \texttt{GRASIL} the parametrization
\begin{equation}\label{eq:eta}
\eta(\tau) = \left\{
\begin{aligned}
&\ 1  & \tau\leq \tau_\text{esc}\\
&\ 2-\frac{\tau}{\tau_\text{esc}} & \tau_\text{esc}<\tau\leq 2\,\tau_\text{esc}\\
&\ 0 & \tau>2\,\tau_\text{esc}~,
\end{aligned}
\right.
\end{equation}
with $\tau_\text{esc}$ a free-parameter which defines the typical time required by stars to start escaping MCs; after $2\, \tau_\text{esc}$ all the stars have escaped.

\begin{figure}
\centering
\includegraphics[width=\hsize]{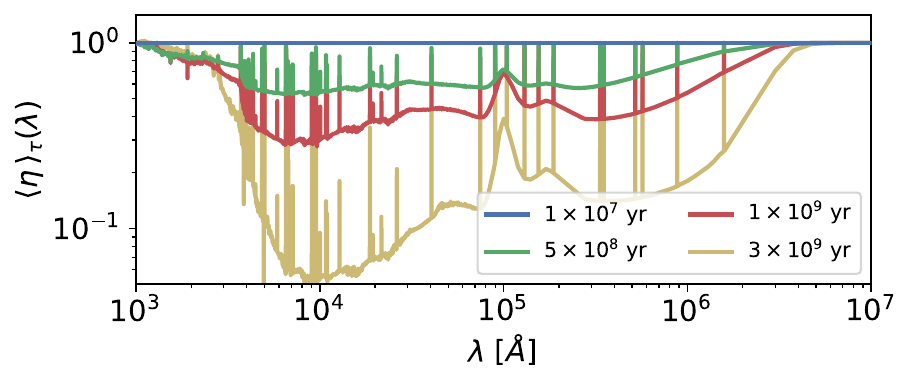}
  \caption{Average fraction of stellar luminosity (see Eq.~\ref{eq:etaavg}) that is absorbed by MCs as a function of wavelength for galaxies of different age (color-coded as in the legend).}
  \label{fig:eta_avg}
\end{figure}

We can take the luminosity-weighted average over stellar ages of the MC extinction law defined in Eq.~\eqref{eq:attMC} and re-write Eq.~\eqref{eq:Latt} in compact form:
\begin{equation}\label{eq:Lattcompact}
L_\text{CSP}^\text{a}(\lambda,\tau) = \mathcal{A}_\text{DD}(\lambda)\, \langle\mathcal{A}_\text{MC}\rangle_\tau(\lambda)\, L_\text{CSP}^\text{i}(\lambda,\tau)~,
\end{equation}
where $L_\text{CSP}^\text{i}(\lambda,\tau)$ is the intrinsic CSP luminosity of Eq.~\eqref{eq:Lunatt} and
\begin{equation}\label{eq:attMCavg}
\langle\mathcal{A}_\text{MC}\rangle_\tau(\lambda) = 1-\langle\eta\rangle_\tau(\lambda)\, \left[1-10^{-0.4\, A_\text{MC}(\lambda)}\right]~,
\end{equation}
with
\begin{multline}\label{eq:etaavg}
\langle\eta\rangle_\tau(\lambda) =\\ 
\\
=\dfrac{\int_{0}^{\tau}\text{d}\tau_\text{SSP}\, \eta(\tau_\text{SSP})\, L_\text{SSP}\left[\lambda,\tau_\text{SSP},Z_\star(\tau-\tau_\text{SSP})\right]\, \psi(\tau-\tau_\text{SSP}) }{\int_{0}^{\tau}\text{d}\tau_\text{SSP}\, L_\text{SSP}\left[\lambda,\tau_\text{SSP},Z_\star(\tau-\tau_\text{SSP})\right]\, \psi(\tau-\tau_\text{SSP})}~.
\end{multline}
We show the behaviour of the latter quantity in Fig.~\ref{fig:eta_avg} for different galactic ages for a fixed value of $\tau_\text{esc} = 5\times10^7$ years.
While almost all the radiation is absorbed in the younger objects (blue and green lines) a considerable part of the stellar radiation escapes when the galaxy ages (red and purple lines).

\begin{figure}
\centering
\includegraphics[width=\hsize]{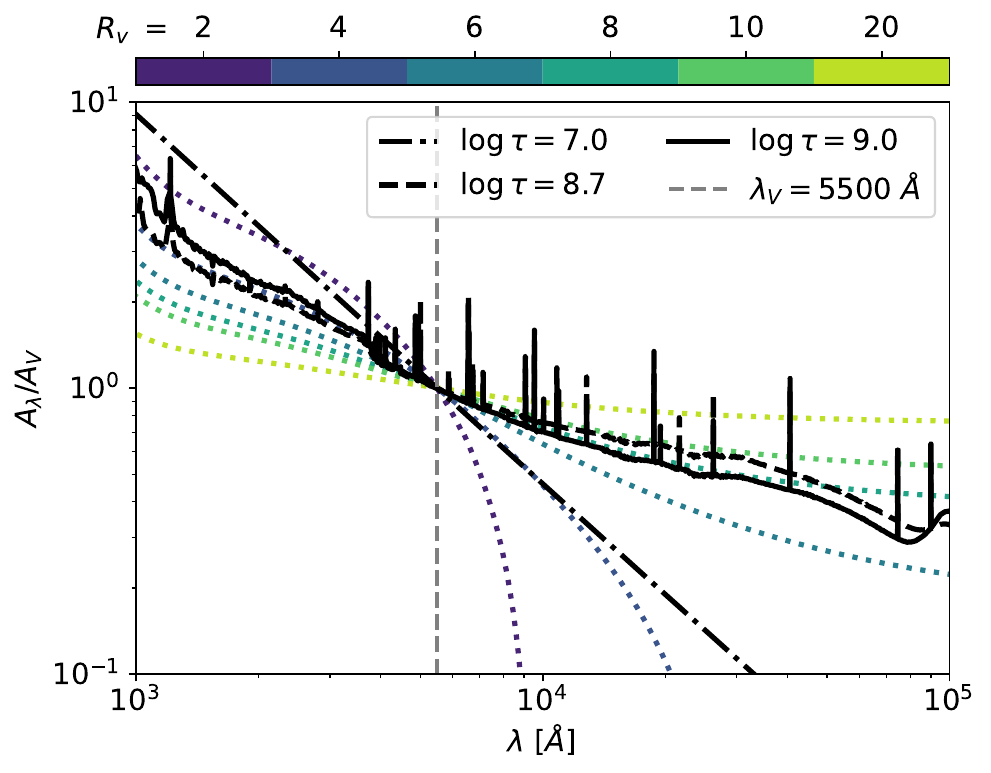}
\caption{Overall attenuation curve at ages of $10^7$ yr (dot-dashed, i.e. $0.5 \tau_\text{esc}$), $5\times10^8$ yr (dashed, i.e. $10 \tau_\text{esc}$), and $10^9$ yr (solid, i.e. $20 \tau_\text{esc}$), we include age effects and plot the value in units of the attenuation in the visible band ($\lambda_V \approx 5500\;\text{\AA}$). Note that these values correspond to the blue, green and red lines of Fig.~\ref{fig:eta_avg}. Our attenuation curves are compared with Calzetti-like attenuation curves (dotted) for different values of the total-to-selective extinction ratio $R_V$ in the $V-$band (colour-coded).}
     \label{fig:extTOT}
\end{figure}
To wrap up, we obtain the wavelength- and age-dependent total galactic attenuation curve, given by
\begin{equation}\label{eq:extTOT}
A_\text{TOT}(\lambda,\tau) = -2.5 \log_{10}\left[\mathcal{A}_\text{DD}(\lambda)\,\langle\mathcal{A}_\text{MC}\rangle_\tau(\lambda)\right]~,
\end{equation}
shown in Fig.~\ref{fig:extTOT} for galaxies with different age, normalised with respect to the global attenuation in the visible band ($\lambda_V\approx 5500 \text{\AA}$).
The attenuation curves are built assuming the in-situ SFH (blue line in Fig.~\ref{fig:sfh}) assuming no quenching and a characteristic escape time from MCs of $\tau_\text{esc} = 5\times10^7$ years\footnote{Note that, even though we limit our discussion to the In-Situ SFH model, all the results can be also obtained using any other SFH model available in the library.}.
In young galaxies (dot-dashed black line) all the stellar populations will still be embedded by their birth MC, therefore their global attenuation curve traces the intrinsic extinction of MCs (blue solid line in Fig.~\ref{fig:extinction}).
Instead, older galaxies (dashed and solid black lines) will host populations of different ages which therefore will provide different contributions to the global attenuation, as some of them will still be embedded in their birth cloud while some others will have partially or completely escaped it.
This evolution with time is reflected on the bending of the global attenuation curve and on the peaks due to the presence of intense emission lines, the latter being the imprint of the youngest stellar populations.

It is also interesting to compare our age-dependent attenuation curves with those predicted by Calzetti-like models \citep[][]{Calzetti2000,dustreview2020}.
In Fig.~\ref{fig:extTOT}, curves with different colours mark different values of the total-to-selective extinction ratio in the $V-$band, $R_V$.
The shape of Calzetti-like attenuation curves is obtained by empirical considerations on the UV/optical photometry of observed galaxies \citep{dustreview2020}.
In our dust model we do not rely on templated nor parametric global attenuation curves and we are independent from dust-grain physics models.
We, nonetheless, manage to derive shapes similar to those found in literature by consistently treating the age-evolution and stellar population dependency of the two different dust phases. 
We derive a model of the global attenuation in a galaxy from data of the transmitted light, by fitting directly the free parameters in Eq.s from \eqref{eq:AVDD} to \eqref{eq:AMC} and Eq.~\eqref{eq:eta}. 

The absorption from the two different components on stellar emission is decomposed in Fig.~\ref{fig:stellar_continuum+ISM} for a $100$ Myr galaxy at the peak of its star formation history and with a characteristic escape time from MCs of $\tau_\text{esc} = 50$ Myr. 
While the dotted black line marks the intrinsic stellar emission, $L_\text{CSP}^\text{i}(\lambda)$, the dashed black line marks the luminosity escaping from molecular clouds, $L_\text{CSP}^\text{a,~MC}(\lambda)$, and the solid black line the obscured stellar emission resulting from Eq.~\eqref{eq:Lattcompact}, $L_\text{CSP}^\text{a,~MC+DD}(\lambda) = L_\text{CSP}^\text{a}(\lambda)$.

We stress that, internally, the overall contribution is computed by integrating numerically Eq.~\ref{eq:Latt}, therefore directly obtaining the SSP luminosity averaged value of Eq.~\ref{eq:extTOT} (shown in Fig.~\ref{fig:extTOT}) by computing the ratio between the attenuated and intrinsic luminosity: $L_\text{CSP}^\text{a}(\lambda,\tau)/L_\text{CSP}^\text{i}(\lambda,\tau)$.
The direct time integration of SSPs attenuation is a peculiarity of our library and is intended to provide blindness to the specific attenuation model.
This choice has been made in order not to rely on parametric total attenuation models and therefore to loose the the assumptions on the dust physics.
We believe that our model will prove relevant, e.g. in studies of primordial galaxies at the highest redshifts, such as those that are currently being probed by JWST (sources at z > 4-6), and, in general, all those cases that lie outside the typical definition range of existing attenuation laws.

Finally, at this stage we do not yet include a treatment of the UV-bump at 2200 \AA~as observed in the total attenuation of some close-by sources \citep[see, e.g.,][]{Noll2009, dustreview2020}, nor its possible relation with the PAH emission, as claimed by some authors.
Note that the object oriented design of the library allows to easily extend the dust modelling to include further components, that we may take in considerations for further developments.

%%%%%%%%%%%%%%%%%%%%%%%%%%%%%%%%%%%%%%%%%%%%%%%%%%%%%%%%%%%%

\subsubsection{Energy conservation and emission}\label{sec:energyconservation}

\begin{figure}
    \centering
    \includegraphics[width=\hsize]{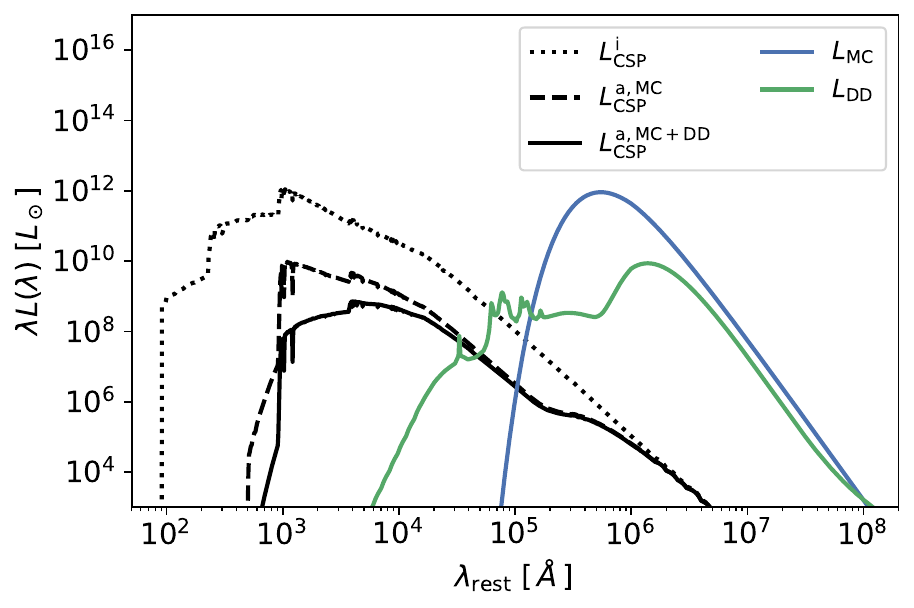}
    \caption{Impact of absorption from the two component dust models (molecular clouds, MC; diffuse dust, DD) implemented in \galapy on the intrinsic stellar emission (dotted black line). Luminosity escaping from MCs is marked by a dashed black line, while the final stellar emission escaping MCs and DD is shown by a solid black line. The luminosity re-emitted in the IR bands is highlighted with a solid blue line for MCs and with a solid green line for the DD(+PAH) component.}
\label{fig:stellar_continuum+ISM}
\end{figure}
By absorbing stellar radiation, dust heats up.
The bolometric intrinsic luminosity coming from the CSP hosted by a galaxy of given age is obtained by integrating Eq.~\eqref{eq:Lunatt} over all the spectrum:
\begin{equation}\label{eq:LbolI}
L_\text{bol}^\text{i}(\tau) = \int_0^{\infty}\text{d}\lambda\,L_\text{CSP}^\text{i}(\lambda,\tau)~.
\end{equation}
In \galapy, such luminosity is absorbed in two stages for stars still embedded within their birth cloud: first it has to pass through the MC phase, and then the remainder of this radiation has then to cross the DD region.
The amount of bolometric luminosity, $L_\text{bol}^\text{i}(\tau)$, which is absorbed by dust in MCs is obtained by filtering the integral above with the age-dependent law defined in Eq.~\eqref{eq:attMCavg}:
\begin{equation}\label{eq:LabsMC}
L_\text{abs}^\text{MC}(\tau) = \int_0^{\infty}\text{d}\lambda\, [1-\langle\mathcal{A}_\text{MC}\rangle_\tau(\lambda)]\, L_\text{CSP}^\text{i}(\lambda,\tau)~.
\end{equation}
The luminosity that has not been transferred to MCs (i.e. $\left[L_\text{CSP}^\text{i} - L_\text{abs}^\text{MC}\right](\lambda,\tau) = \langle\mathcal{A}_\text{MC}\rangle_\tau(\lambda) L_\text{CSP}^\text{i}(\lambda,\tau)$) is then further absorbed by DD:
\begin{equation}\label{eq:LabsDD}
L_\text{abs}^\text{DD}(\tau) = \int_0^{\infty}\text{d}\lambda\, [1-\mathcal{A}_\text{DD}(\lambda)]\, \langle\mathcal{A}_\text{MC}\rangle_\tau(\lambda)\, L_\text{CSP}^\text{i}(\lambda,\tau)~.
\end{equation}

\begin{figure*}
\resizebox{\hsize}{!}{
\includegraphics{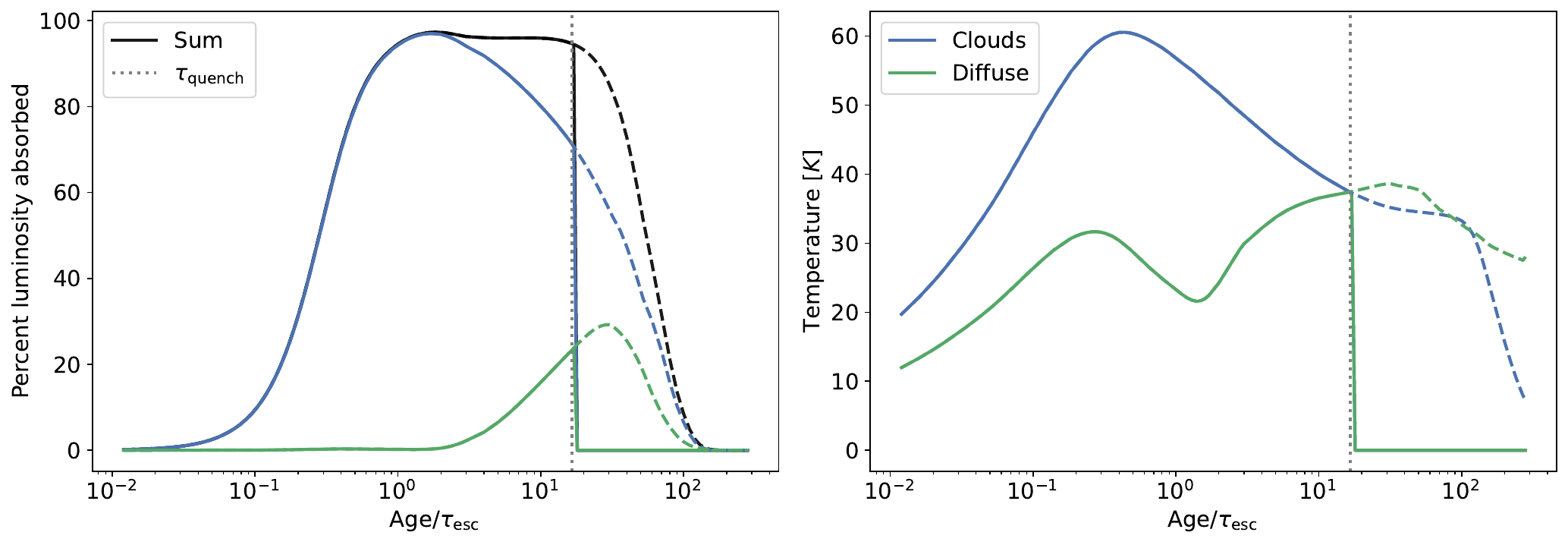}}
  \caption{Behaviour of the two different dust components (molecular clouds, MC; diffuse dust, DD) as a function of galactic age in units of the characteristic escape time from molecular clouds $\tau_\text{esc}$. We assume an in-situ SFH as the blue line in Fig.~\ref{fig:sfh} and set the characteristic escape time to $5\times10^7$ years. \textit{Left panel:} percentage of the intrinsic stellar luminosity absorbed by MC (blue), DD (green), and by both components (black). \textit{Right panel:} temperature of the dust component computed from the age-dependent energy-conservation scheme implemented in \galapy. The vertical dotted line marks the age, $\tau_\text{quench} = 8.8\times10^8$ years, of a possible abrupt quenching event; solid lines refer to the evolution with quenching, dashed lines to that with no quenching.}
\label{fig:ism_components}
\end{figure*}
In the left panel of Fig.~\ref{fig:ism_components}, we show the age-dependency of the percentage of bolometric luminosity that is absorbed by the two different dust phases in units of the typical escape time from MCs, $\tau_\text{esc}$ defined in Eq.~\eqref{eq:eta}.
When the stars begin to radiate energy but the galaxy is still younger than the characteristic escape time $\tau_\text{esc}$, the amount of luminosity absorbed by MCs grows steadily (blue line), while the radiation absorbed by DD is negligible (green line).
At an escape time, the percent of luminosity going to MCs instead starts to decrease, while the relevance of the DD component grows.
All in all though, the combined effect of the two phases is to trap more than $90\%$ of the total energy budget. 
This value decreases only at several tens of escape times, when the SFR decreases (see the in-situ SFH model marked by a blue line in Fig.~\ref{fig:sfh} and compare to Fig.~\ref{fig:extTOT}, commented in Sec.~\ref{sec:extinctionattenuation}) or, either, after an abrupt quenching event that wipes out most of the ISM. 

It has to be noted that, in the earliest stages of evolution (i.e. $\tau \approx 10^{-2}\tau_\text{esc}$) the overall total absorbed fraction of Fig.~\ref{fig:ism_components} is of about $0\%$.
This results from the way we compute the evolution of the absorption curves.
Namely, in our model this process is strictly dependent on the total budget of absorbing medium that is consistently computed from the evolutionary stage of the stars populating a galaxy.
As a consequence, when the galaxy is extremely young (i.e. $\tau \lesssim 10^6$ years $\equiv 0.1\tau_\text{esc}$), stars do not have had time yet to pollute the medium with dust and therefore, no absorption is possible. 

Energy conservation ensures that, having heated up as a consequence of the luminosity absorbed, the two dust components radiate, in good approximation, as two optically thick grey-bodies.
This emission depends on the extinction laws as defined in Eqs.~\eqref{eq:attDD} and \eqref{eq:attMC} for the two respective dust phases.
We define 
\begin{equation}\label{eq:LDD}
L_\text{DD}(\lambda,\tau\,|T_\text{DD}) = \frac{16\,\pi^2}{3}\,R_\text{DD}^2\, [1-10^{-0.4\,A_\text{DD}(\lambda,\tau)}]\, B(\lambda,T_\text{DD})~,
\end{equation}
for the DD phase and 
\begin{equation}\label{eq:LMC}
L_\text{MC}(\lambda,\tau\,|T_\text{MC}) = \frac{16\,\pi^2}{3}\, N_\text{MC}\, R_\text{MC}^2\, [1-10^{-0.4\,A_\text{MC}(\lambda,\tau)}]\, B(\lambda, T_\text{MC})~,
\end{equation}
 for the MC phase.
Notice that, in the limit of an optically-thin emission, the factor $1-10^{-0.4\, A_\text{phase}(\lambda)}\approx \rho_\text{phase}\, k_\lambda\, R_\text{phase}^3$ is approximately the optical depth, thus, it can be written in terms of the dust-phase density $\rho_\text{phase}$  and of the opacity $k_\lambda$. 
Using $\rho_\text{phase}\approx 3\,M_\text{phase}/4\,\pi\, R_\text{phase}^2$, the emitted luminosity can then be recast in the form $L_\lambda\approx 4\,\pi\, M_\text{phase}\, k_\lambda\, B(\lambda,T_\text{phase})$, which frequently occurs in literature \citep[e.g.][]{Lacey2016}.

In both Eq.~\eqref{eq:LDD} and Eq.~\eqref{eq:LMC}, luminosity is given in terms of the black body spectrum
\begin{equation}\label{eq:LBB}
B(\lambda,T) \equiv \frac{2\,h_P c^2}{\lambda^5}\, \frac{1}{ e^{h_P\,\nu_\lambda/k_B\, T}-1}
\end{equation}
where $\nu_\lambda=c/\lambda$ is the frequency corresponding to the wavelength $\lambda$, $c$ is the speed of light, $h_P$ is the Planck constant and $k_B$ is the Boltzmann constant.

At any given age, the temperatures ${T_\text{DD}}$ and ${T_\text{MC}}$ of the diffuse and MC dust component are set by requiring that the total emitted power equals the luminosity absorbed by each of the two phases.
Therefore, in \galapy, the dust temperatures are not free parameters, they are age-dependent outputs obtained by imposing a self-consistent energy conservation.

For the emission coming from MCs we impose 
\begin{equation}\label{eq:EconservationMC}
\int_0^{\infty}\text{d}\lambda\, L_\text{MC}[\lambda, \tau\,|T_\text{MC}(\tau)] = L_\text{abs}^\text{MC}(\tau)~
\end{equation}
where the left hand side is obtained by integrating over the whole spectrum Eq.~\eqref{eq:LMC} and the right hand side has been computed with Eq.~\eqref{eq:LabsMC}.

We make the assumption that the emission from poly-cyclic aromatic hydrocarbons (PAH) is suppressed in MCs, but we include it in the emission coming from the DD phase \citep{Vega2008}.
We define a free parameter $0 \leq f_\text{PAH} \leq 1$ regulating the fraction of absorbed power $L_\text{abs}^\text{DD}(\tau)$ that at some given time is re-radiated by PAH.
Therefore, the temperature of the DD grey-body is computed by imposing
\begin{equation}\label{eq:EconservationDD}
\int_0^{\infty}\text{d}\lambda\, L_\text{DD}[\lambda, \tau\,|T_\text{DD}(\tau)] = (1-f_\text{PAH})\,L_\text{abs}^\text{DD}(\tau)~.
\end{equation}
where the left hand side is obtained by integrating over the whole spectrum Eq.~\eqref{eq:LDD} and the right hand side has been computed with Eq.~\eqref{eq:LabsDD} excluding the fraction of energy that is radiated by PAH.

By solving numerically the two energy conservation equations, Eq.~\ref{eq:EconservationMC} and Eq.~\ref{eq:EconservationDD}, we obtain the temperatures of the two media consistently with their evolution with time.
In the right panel of Fig.~\ref{fig:ism_components} we show the age evolution of the dust phases' temperature obtained above in units of the characteristic escape time from MCs.
Young galaxies display a steady increase in temperature, steeper and some factors larger for the MC component.
When the escape time is reached, the temperatures of both MCs and DD begin to decrease. 
However, while MCs continue to decrease over time, the temperature of DD initially decreases but then begins to increase again.
This instant in the evolution of the galaxy is due to the presence of a large number of stars that have escaped their MC and whose energy therefore contributes only to the heating of the DD phase.
Depending on the value of the model parameters, the DD medium could also become hotter than MCs, when most of them have been evaporated.
It is interesting to notice how in the early stages of evolution, the DD temperature reaches some tens of degrees even though its contribution to absorption in this stage is negligible (cf. left panel of Fig.~\ref{fig:ism_components}).

We adopt the PAH template $L_\text{PAH}(\lambda)$ by \cite{magphys2008} constructed on the behaviour in the photo-dissociation regions of the Milky Way. 
It includes PAH line emission mainly in mid-IR, PAH continuum emission in the near-IR, and mid-IR continuum emission due to very small, hot dust grains. 
All in all, the global emission due to diffuse dust including PAH is given by
\begin{equation}\label{eq:LDD+PAH}
L_\text{DD+PAH}(\lambda,\tau) = L_\text{DD}[\lambda, \tau\,|T_\text{DD}(\tau)] + f_\text{PAH}\, L_\text{abs}^\text{DD}(\tau)\, L_\text{PAH}^\text{norm}(\lambda)~,
\end{equation}
where $L_\text{PAH}^\text{norm}(\lambda) = L_\text{PAH}(\lambda)/\int_0^\infty\text{d}\lambda\, L_\text{PAH}(\lambda)$ is the normalised PAH spectrum.

The total dust bolometric luminosity is given by the all-spectrum integral
\begin{multline}
\label{eq:Ldust}
L_\text{dust}(\tau) = \int_0^\infty\text{d}\lambda\, [L_\text{MC}(\lambda,\tau)+L_\text{DD+PAH}(\lambda,\tau)] =\\= \int_0^\infty\text{d}\lambda\, L_\text{dust}(\lambda,\tau)~.
\end{multline}
Note that other definitions exploited in literature involve this integral over the wavelength range $8-1000\, \mu$m (dubbed FIR for far infrared luminosity) or $3-1100\, \mu$m (dubbed TIR for total infrared luminosity).

The emission from the two different dust components including PAH is shown in Fig.~\ref{fig:stellar_continuum+ISM}.
The blue solid line marks the grey-body emission from molecular clouds, as computed from Eq.~\eqref{eq:LMC}, while the green solid line shows the overall diffuse dust emission, Eq.~\eqref{eq:LDD+PAH}, from both the grey-body of Eq.~\eqref{eq:LDD} and PAH. 

%%%%%%%%%%%%%%%%%%%%%%%%%%%%%%%%%%%%%%%
\subsection{Additional sources of stellar continuum}\label{sec:radio}

The two different SSP libraries delivered with \galapy provide different recipes for the stellar emission.
As already explained in Sec.~\ref{sec:stellar}, while PARSEC22 libraries have been computed either with supernova synchrotron included and with supernova synchrotron and nebular emission included, the BC03 libraries do not include either of these contributes.  
In order to homogenise (and extend to higher energies) the spectral emission due to the stellar component among the different possible SSP library choice, we provide additional (optional) modules for modelling radiative processes that impact mostly on the rest-frame X-ray and radio bands.
With these we extend the spectral coverage of the library to the overall range $1 \text{\AA}\leq \lambda \leq 10^{10} \text{\AA}$, independently on the SSP library of choice.

Including these components is optional, as the photometric system of the data-set in study might not cover such an extended range of wavelengths.
In order to have nebular free-free emission (Sec.~\ref{sec:nff}) and stellar synchrotron (Sec.~\ref{sec:snsyn}) when building models with the helper class \texttt{galapy.Galaxy.GXY}, the user has to require for radio-support (as these components mostly impact on the radio bands).
Note that both nebular free-free and SN-synchrotron might be already present in the SSPs if, as discussed in Sec.~\ref{sec:stellar}, one of the PARSEC22 libraries is chosen.
In this case requiring for radio-support would be redundant and the system will ignore it.\footnote{ 
Specifically, when using PARSEC22 without nebular emission and radio-support is required, the system will automatically include the nebular free-free emission from Sec.~\ref{sec:nff} while, when using PARSEC22 with nebular emission and radio support is required, any source of stellar radio continuum is already present in the SSP and therefore the system will ignore the directive. Further details are given in Appendix~\ref{apx:ssp}.
}
Vice-versa, X-ray binaries (Sec.~\ref{sec:xrb}) are considered by requiring for X-ray support, which will also include high energy emission from an eventual AGN as in Sec.~\ref{sec:agn}, if required.

We underline that, the choice of SSP library should be driven by the data-set studied. In particular, when working with young objects, the attenuation due to line absorption and re-emission has a non-negligible effect also on photometric observations. In such cases is therefore preferable to work with the PARSEC22 library including line-emission.

\begin{figure}
    \centering
    \includegraphics[width=\hsize]{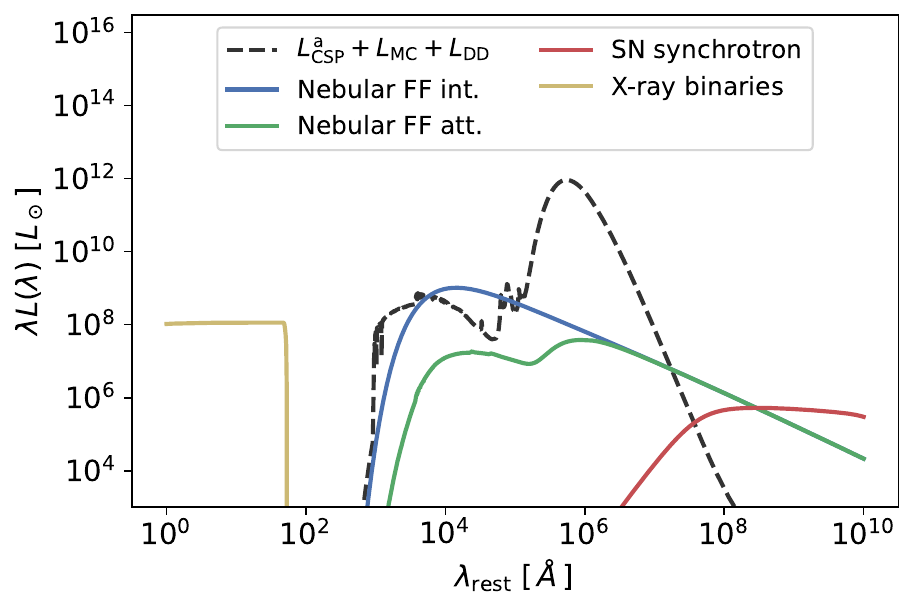}
    \caption{Additional stellar components contributing to the overall continuum emission in \galapy: intrinsic (blue) and attenuated (green) free-free emission from nebular regions, synchrotron emission from SN (red), X-ray emission from high-mass and low-mass binary stars (yellow). For reference, the luminosity from attenuated stellar continuum and dust re-radiation, i.e. the sum of the solid lines in Fig.~\ref{fig:ism_components}, is also reported (dashed black).}
\label{fig:additional_stellar_continuum}
\end{figure}

Fig.~\ref{fig:additional_stellar_continuum} shows the impact of the additional stellar continuum processes, with respect to the continuum coming from stellar atmospheres and dust.  
In the rest of this Section we describe how these components are modelled in \galapy.

%%%%%%%%%%%%%%%%%%%%%%%%%%%%%%%%%%%%%%%
\subsubsection{Nebular free-free}\label{sec:nff}

The nebular free-free (NFF) emission is originated in  HII regions associated to short-lived ionising massive stars (stars with age $\lesssim 10^7$ years in simple stellar evolution models).
For the intrinsic NFF luminosity (blue solid line in Fig.~\ref{fig:additional_stellar_continuum}), we use the expression \citep[see, e.g.][]{bressan2002,Murphy2012,Mancuso2017}
\begin{multline}\label{eq:LNFF}
L_\text{NFF}(\lambda,\tau) \approx 1.8\times 10^{-27}\, \text{erg}\,s^{-1} \text{Hz}^{-1}\cdot\\ \cdot\dfrac{\mathcal{Q}_\text{H}(\tau)}{s^{-1}}\, \left(\dfrac{T_e}{10^4\, K}\right)^{0.3}\, g_\text{NFF}\, \exp\left(\dfrac{-h_P\, \nu_\lambda}{k_B\, T_e}\right)
\end{multline}
in terms of the Gaunt factor \citep[see][]{draine2011}
\begin{equation}\label{eq:gaunt}
g_\text{NFF}=\ln\left\{\exp\left[5.96-\frac{\sqrt{3}}{\pi}\, \ln\left( Z_i\, \frac{\nu_\lambda}{G\text{Hz}}\, \left(\frac{T_e}{10^4\, K}\right)^{-1.5}\right)\right]+\exp(1)\right\}
\end{equation}
and of the Boltzmann correction $\exp(-h_P\, \nu_\lambda/k_B\, T_e)$ due to the thermal nature of the process.
The latter induces a suppression in the high energy part of the spectrum, due to the decreasing number of high energy photons.
In the frequency range satisfying $0.14 < Z_i \nu_9 / T_4^{3/2} < 250$ and $\nu_p < \nu < kT/h$, where $\nu_p$ is the plasma frequency, the free-free emission spectrum is almost flat.
Such frequencies correspond to the microwave and radio part of the NFF spectrum where it can be shown that the emission slowly declines with increasing frequency as $\sim \nu^{-0.12}$ \citep{Vega2008,draine2011,Mancuso2017}.

In Eq.~\eqref{eq:gaunt}, it is often assumed $Z_i=1$, corresponding to a pure hydrogen plasma. 
$T_e$ refers to the electron temperature in HII regions, whose dependence on gas metallicity, $Z_\text{gas}$, can be expressed as \citep{Vega2008}
\begin{equation}\label{eq:ElecTemp}
\log T_e \approx 3.89-0.4802\, \log(Z_\text{gas}/0.02)-0.0205\,[\log(Z_\text{gas}/0.02)]^2~.
\end{equation}
The intrinsic photo-ionisation rate can be computed from the intrinsic stellar luminosity of Eq.~\eqref{eq:Lunatt} by the integral
\begin{equation}\label{eq:PhotIonRate}
\mathcal{Q}_\text{H}(\tau) = \int_0^{\lambda_\text{ion}}\text{d}\lambda\, \frac{L_\text{CSP}^\text{i}(\lambda,\tau)}{h_P\, \nu_\lambda}~,
\end{equation}
with $\lambda_\text{ion}\approx 912\, \text{\AA}$ being the wavelength corresponding to the H ionisation potential.

In Fig.~\ref{fig:additional_stellar_continuum} we also show the attenuation induced by dust in the optical-IR part of the NFF spectrum (green solid line) which is computed as 
\begin{equation}\label{eq:LNFFatt}
L_\text{NFF}^{a}(\tau) = \mathcal{A}_\text{DD}(\lambda)\, \langle\mathcal{A}_\text{MC}\rangle_\tau(\lambda)\, L_\text{NFF}^\text{i}(\lambda,\tau)~.
\end{equation}
where $\mathcal{A}_\text{DD}(\lambda)$ and $\langle\mathcal{A}_\text{MC}\rangle_\tau(\lambda)$ are given by Eq.~\eqref{eq:attDD} and Eq.~\eqref{eq:attMC}, respectively.
When the SSP chosen is not one among the PARSEC22 libraries with nebular emission included (see Sec.~\ref{sec:stellar} and Appendix~\ref{apx:ssp}), this additional source of stellar emission is added a-posteriori to the overall spectrum and, therefore, it is not taken into account automatically in the energy balance described in Section~\ref{sec:energyconservation}.
Given that the amount of energy transferred to dust by this process is negligible (i.e. less than $1\%$ in most of the cases), this choice has not a relevant impact for the majority of sources.
Nevertheless, for particularly young ages, when the contribution from short lived massive stars is dominant and the energy transferred to nebular emission both in terms of continuum and line emission has a relevant impact also in terms of photometric observations, we recommend using the PARSEC22 SSP libraries with nebular emission included (i.e. \texttt{parsec22.ntl} family).
This not only guarantees to account for the presence of emission lines but also guarantees that the nebular emission is accounted for in the energy balance algorithm.

%%%%%%%%%%%%%%%%%%%%%%%%%%%%%%%%%%%%%%%
\subsubsection{Synchrotron from supernovae}\label{sec:snsyn}

The synchrotron (non-thermal) emission is likely originated from relativistic electrons accelerated into the shocked interstellar medium, following
core-collapse SN explosions. 
A possible minor contribution from SN remnants is also possible. 

When not running with the PARSEC22 SSP libraries, we use the equivalent expression \citep{bressan2002,Mancuso2017}
\begin{multline}\label{eq:Lsyn}
L_\text{syn}(\lambda,\tau) \approx 10^{30}\, \text{erg} s^{-1} \text{Hz}^{-1}\,\dfrac{\mathcal{R}_\text{CCSN}(\tau)}{\text{yr}^{-1}}\, \left(\dfrac{\nu_\lambda}{G\text{Hz}}\right)^{-\alpha_\text{syn}}\cdot\\ \cdot\left[1+\left(\dfrac{\nu}{20\, G\text{\text{Hz}}}\right)^{0.5}\,\right]^{-1}\, F[\tau_\text{syn}(\nu_\lambda)]~,
\end{multline}
where $\mathcal{R}_\text{CCSN}$ is the core-collapse SN rate, $\alpha_\text{syn}\approx 0.75$ is the spectral index, the term in square brackets takes into account spectral-ageing effects, and the function $F(x)=(1-e^{-x})/x$ incorporates synchrotron self-absorption in terms of the optical depth $\tau_{\rm sync}\approx (\nu_\lambda/\nu_\text{self})^{-\alpha_\text{syn}-5/2}$ that is thought to become relevant at frequencies $\nu\lesssim \nu_\text{self}\approx 200$ MHz.

The age-evolution of the CCSN rate per unit solar mass of formed stars has been computed from the SSPs and can be rendered in terms of the simple relation 
\begin{equation}\label{eq:rCCSN}
R_{\rm CCSN}(\tau,Z_\star)\simeq R_0(Z_\star)\, \left(\tau/{\rm Myr}\right)^{-R_1(Z_\star)}
\end{equation}
where $R_0$ and $R_1$ are fitting functions dependent on stellar metallicity (some values are tabulated in Tab.~\ref{tab:RCCSN_par}).

The overall rate entering the expression of the synchrotron luminosity is then
\begin{equation}\label{eq:RCCSN}
\mathcal{R}_\text{CCSN}(\tau) = \int_0^{\tau}\text{d}\tau_\text{SSP}\, R[\tau-\tau_\text{SSP}, Z_\star(\tau-\tau_\text{SSP})]\, \psi(\tau-\tau_\text{SSP})~,
\end{equation}
where, as in Sec.~\ref{sec:stellar}, $\tau_\text{SSP}$ is the time passed since some given SSP has formed, $Z_\star(\tau)$ is the metallicity of stars at given galactic age and $\psi(\tau)$ the SFR.

Eq.~\ref{eq:Lsyn} with its normalisation is obtained by requiring that the quantity
\begin{equation}\label{eq:qFIR}
q_\text{FIR} \equiv \log\left(\dfrac{L_\text{dust}}{3.75\times 10^{12}\, \rm W}\right)-\log\left(\frac{L_{\nu_{\lambda}=1.4\, \rm G\text{Hz}}}{\rm W \text{Hz}^{-1}}\right)
\end{equation}
takes on values close to the observed $q_{\rm FIR}\approx 2.35-2.7$ at an age of about $10^8$ yr.

To take into account the lower efficiency in producing synchrotron radiation at small SFRs, we correct the above equation as
\begin{equation}\label{eq|Lsynccorr}
L_\text{syn}^\text{corr}(\lambda,\tau)=\dfrac{L_\text{syn}(\lambda,\tau)}{1+[L_\text{syn}^0/L_\text{syn}(\lambda,\tau)]^\zeta}~,
\end{equation}
with $\zeta\approx 2$ and $L_\text{syn}^0\approx 3\times 10^{28} \text{erg}\, s^{-1}\, \text{Hz}^{-1}$.
The corrected synchrotron emission $L_\text{syn}^\text{corr}(\lambda,\tau)$ is marked by a red solid line in Fig.~\ref{fig:additional_stellar_continuum}.
We stress that  attenuation on the synchrotron emission has been applied but turns out to be irrelevant since the corresponding spectrum is strongly suppressed by self-absorption for wavelengths $\lambda\lesssim 1$ mm.

%%%%%%%%%%%%%%%%%%%%%%%%%%%%%%%%%%%%%%%
\subsubsection{X-ray binaries}\label{sec:xrb}

The X-ray emission associated to star formation comes mainly from high and low mass X-ray binaries. 
For their total output, we use the prescriptions by \cite{Fragos2013} based on stellar population synthesis simulation for a Chabrier IMF. 
Specifically, the contribution to the emission in the $2-10$ keV band from high-mass X-ray binaries can be described via the polynomial expression
\begin{multline}
\label{eq:HMXB}
\log(L_\text{HMXB}/\text{erg}\,s^{-1}) \approx\\ \log({\dot M_\star/M_\odot}\text{yr}^{-1}) + 40.28-62.12\, Z_\star\\+569.44\,Z_\star^2-1883.80\,Z_\star^3+1968.33\, Z_\star^4~,
\end{multline}
while that from low-mass X-ray binaries reads
\begin{multline}\label{eq:LMXB}
\log(L_\text{LMXB}/\text{erg}\,s^{-1}) \approx\\ \log(M_\star/M_\odot) + 40.276-1.503\, \theta-0.423\,\theta^2+0.425\,\theta^3+0.136\, \theta^4~,
\end{multline}
where $\theta\equiv\log(\tau/\text{Gyr})$.

We distribute both emissions according to a power-law with an exponential cutoff
\begin{equation}\label{eq:Xspec}
L_\Gamma^X(\lambda)\propto E^{-\Gamma+3}(\lambda)\, e^{-E(\lambda)/E_\text{cut}}~,
\end{equation}
with $E(\lambda)=h_P\, \nu_\lambda=h_P\, c/\lambda$ the energy of a photon with wavelength $\lambda$. 
The photon index is set to $\Gamma\approx 1.6$ for LMXB and to $\Gamma\approx 2.0$ for HMXB \citep[see][]{Fabbiano2006}; the high-energy cutoff is fixed at $E_\text{cut}\approx 100$ keV, while at the other end the spectrum is extended up to $\lambda\approx 50\,\text{\AA}$.

The resulting total emission from X-ray binaries 
\begin{equation}\label{eq:LXRB}
L_\text{XRB}(\lambda, Z_\star, \tau) = L_\text{HMXB}(Z_\star) L_{\Gamma=2}^X(\lambda) + L_\text{LMXB}(\tau) L_{\Gamma=1.6}^X(\lambda)  
\end{equation}
is marked by a solid yellow line in Fig.~\ref{fig:additional_stellar_continuum}.

%%%%%%%%%%%%%%%%%%%%%%%%%%%%%%%%%%%%%%%

\subsection{Active Galactic Nucleus}\label{sec:agn}

The panchromatic emission from galaxies modelled by \galapy can be enriched by the inclusion of templated spectral models of the emission due to a luminous nuclear component.
Even though \galapy is currently intended for the study of galaxies which are not AGN-dominated, accounting for this component can be important when trying to refine the inference of the hosting galaxy properties from its overall emission, provided that the AGN properties are known.

We adopt the AGN templates $L_\text{AGN}^\text{temp}(\lambda)$ by \citet[][F06 hereafter]{Fritz2006}, which have been computed via a radiative transfer model and take into account three components: the accretion disk around the central supermassive black hole, the scattered emission by a surrounding dusty torus, and the thermal dust emission associated to the heated dust. 

The overall shape of the template depends on $6$ discrete tunable parameters:  the ratio $R_\text{torus}^\text{AGN}$ of the maximum to minimum radii of the dusty torus, the optical depth $\tau_{9.7}^{\rm AGN}$ at $9.7\, \mu$m, the dust density distribution $r^\beta\,e^{-\gamma\,|\cos\theta|}$ in terms of two parameter $\beta$ and $\gamma$, the covering angle $\Theta$ of the torus and the viewing angle $\Psi_\text{los}^\text{AGN}$ between the AGN axis and the line of sight.
The template library we are using, by the variation of these 6 parameters, counts $24000$ spectra among which the user can choose. 
We further vary the overall contribution of the AGN spectrum over the total galactic emission by an additional parameter $f_\text{AGN}$ defined as the contribution of the AGN to the IR emission from interstellar dust.

Specifically, the fraction $f_\text{\rm AGN}$ regulates the fractional intensity of the normalised AGN SED at any given galactic age:
\begin{equation}\label{eq:SEDagn}
L_\text{AGN}({\lambda},\tau) = \dfrac{f_\text{AGN}}{1-f_\text{AGN}}\, \dfrac{L_\text{AGN}^\text{temp}({\lambda})}{\int_0^\infty\text{d}\lambda\, L_\text{AGN}^\text{temp}({\lambda})}\, L_\text{dust}(\tau)~,
\end{equation}
where $L_\text{dust}(\tau)$ is the bolometric dust luminosity at given galactic age as defined in Eq.~\eqref{eq:Ldust}.
This modelling choice is valid for objects where the AGN emission in the IR band is sub-dominant with respect to the inter-stellar dust emission.
For this reason and to guarantee that Eq.~\eqref{eq:SEDagn} is not diverging, \galapy forces $f_\text{AGN} < 1$.

We also model the intrinsic X-ray emission coming from the inner parts of the accretion disk.
This contribution is added on top of the optical/mid-IR template and is modelled by adopting the same shape of Eq.~(\ref{eq:Xspec}), with photon index $\Gamma\approx 1.8$ and high-energy cutoff $E_\text{cut}\approx 300$ keV. 
The normalisation in the $2-10$ keV band is based on the hard X-ray bolometric correction by \cite{Duras2020}
\begin{equation}\label{eq:kxbol}
\cfrac{L_\text{AGN}^\text{bol}}{L_\text{AGN}^X} \approx 10.96\, \left[1+\dfrac{\log(L_\text{AGN}^\text{bol}/L_\odot)}{11.93}\right]^{17.79}~.
\end{equation}
We first compute $L_\text{AGN}^\text{bol}(\tau)=\int_{0}^{\infty}\text{d}\lambda\, L_\text{AGN}(\lambda,\tau) = f_\text{AGN}\, L_\text{dust}(\tau)/(1-f_\text{AGN})$ from Eq.~(\ref{eq:SEDagn}) and then use Eq.~(\ref{eq:kxbol}) to obtain the spectrum normalisation $L_\text{AGN}^X(\tau)$ in the hard X-ray band $2-10$ keV.

F06 templates are excellent for modelling AGNs for which the geometry is well known and/or when the AGN contribution is dominant over the spectrum. 
Nonetheless, due to their high number of dimensions, they prove to add a level of complexity to the model that would require extremely large data-sets to produce significant parameters posteriors if used for fitting. 
Furthermore, our current implementation of the X-Ray emission from AGN does not account for the torus attenuation nor for that due to galactic dust and, therefore, should also be included carefully.
The purpose of the current implementation of the AGN module in \galapy is to add a already known AGN emission to the overall spectrum, when needed.
We plan to extend \galapy with a self consistent parameterised modelling of the AGN accounting for BH-galaxy co-evolution and clumpy torus emission in the nearest future.
This extension will be intended for continuous exploration of the parameter space within the Bayesian framework of the library.

%%%%%%%%%%%%%%%%%%%%%%%%%%%%%%%%%%%%%%%

\subsection{Build a galaxy model}

The overall rest-frame SED is obtained by summing all the contributions from the components building up the galaxy model of choice.
The default minimal set-up provides a total emission given by
\begin{equation}\label{eq:SEDdefault}
L_\text{TOT}^\text{default}(\lambda, \tau) = L_\text{CSP}^\text{a}(\lambda,\tau) + L_\text{MC}(\lambda, \tau) + L_\text{DD+PAH}(\lambda, \tau)~,
\end{equation}
where the different terms are computed via the relevant equations provided in the previous sections.
Including all the optional components to the galaxy model will produce an SED given by
\begin{multline}\label{eq:SEDTOT}
    L_\text{TOT}(\lambda,\tau) = L_\text{TOT}^\text{defaul}(\lambda, \tau) + L_\text{NFF}(\lambda, \tau) + L_\text{syn}^\text{corr}(\lambda,\tau)+
    \\
    \\ + L_\text{AGN}(\lambda,\tau) + L_\text{XRB}(\lambda,\tau)~.
\end{multline}

We provide a convenient class \texttt{GXY} that can be instantiated by importing it from the \texttt{galapy.Galaxy} module.
This class manages the interplay between all the components and provides access to all the free-parameters tuning.
It can both be used for building mock SED observations and for modelling emission. 
A summary of all the tunable parameters available through this interface is reported in Appendix~\ref{apx:params} and in Tab.~\ref{tab:tunable_parameters}.

%%%%%%%%%%%%%%%%%%%%%%%%%%%%%%%%%%%%%%%
\subsubsection{Cosmology and redshifting}\label{sec:cosmo}

When dealing with the emission from galaxies, choosing a cosmological model plays a primary role in deriving the flux received by an observer at a given distance from the source, along with a secondary role in imposing an upper limit to the age the modelled galaxy can have.
Given these considerations, we implemented a \texttt{Cosmology} class with limited functionalities but significant flexibility.
It is built by passing two grids used for interpolation: a redshift-luminosity distance grid and a redshift-cosmic age grid.
Along with the possibility to provide directly grids computed with external libraries \citep[such as with, e.g., \texttt{astropy},][]{astropy2013}, we provide several ready-to-use popular parameterisations in our data-base that can be chosen easily by passing the relevant string (e.g. \texttt{'Planck2018'} for rounded \citealt{Planck2020} parameters\footnote{This choice is driven by the requirement to limit the number of external dependencies of \galapy. See the online documentation for a complete list of available cosmological models and for instructions on how to use custom additional models not present in the database.}).

For a galaxy at redshift $z$, we compute the observed SED in terms of flux density $S_\nu(\lambda_\text{O})$ in mJy at the observation wavelength $\lambda_\text{O} = \lambda_\text{R} (1 + z)$ by redshifting the rest-frame SED, converting it from energy to flux via the cosmological luminosity distance $D_\text{L}(z)$
\begin{equation}\label{eq:SEDflux}
S_\nu(\lambda_\text{O}) \equiv \dfrac{\lambda_\text{R}^2 L_\text{TOT}(\lambda_\text{R}, \tau)}{4 \pi D_\text{L}^2(z)}\times \dfrac{(1+z)}{c}
\end{equation}
expressed as a function of the rest frame wavelength, $\lambda_\text{R}$ and where $L_\text{TOT}(\lambda_\text{R},\tau)$ is given by Eq.~\eqref{eq:SEDTOT}. 

%%%%%%%%%%%%%%%%%%%%%%%%%%%%%%%%%%%%%%%
\subsubsection{Inter-Galactic medium}\label{sec:igm}

The main contribution to the damping of ultra-violet photons due to the absorption from the intergalactic medium (IGM) is ascribed to hydrogen, with a minor contribution from heavier elements. To model this effect we exploit the fitting functions from \cite{Inoue2014}, provided in terms of piece-wise equations for the Lyman-series (LS) and Lyman-continuum (LC) absorption from Damped Lyman-$\alpha$ systems (DLA) and from the Lyman-$\alpha$ Forest (LAF).
The authors approximated model provides a fair trade-off between accuracy and performance.

The optical depth is then given by the sum of the various contributions:
\begin{equation}
    \label{eq:igm_tau}
    \tau_\text{IGM}(\lambda_\text{O}, z) = \tau_\text{LC}^\text{LAF}(\lambda_\text{O}, z) + \tau_\text{LS}^\text{LAF}(\lambda_\text{O}, z) + \tau_\text{LC}^\text{DLA}(\lambda_\text{O}, z) + \tau_\text{LS}^\text{DLA}(\lambda_\text{O}, z)
\end{equation}
where the different terms are computed from Eqs. (20-21-25-26-27-28-29) of \cite{Inoue2014}.

\begin{figure}
\centering
\includegraphics[width=\hsize]{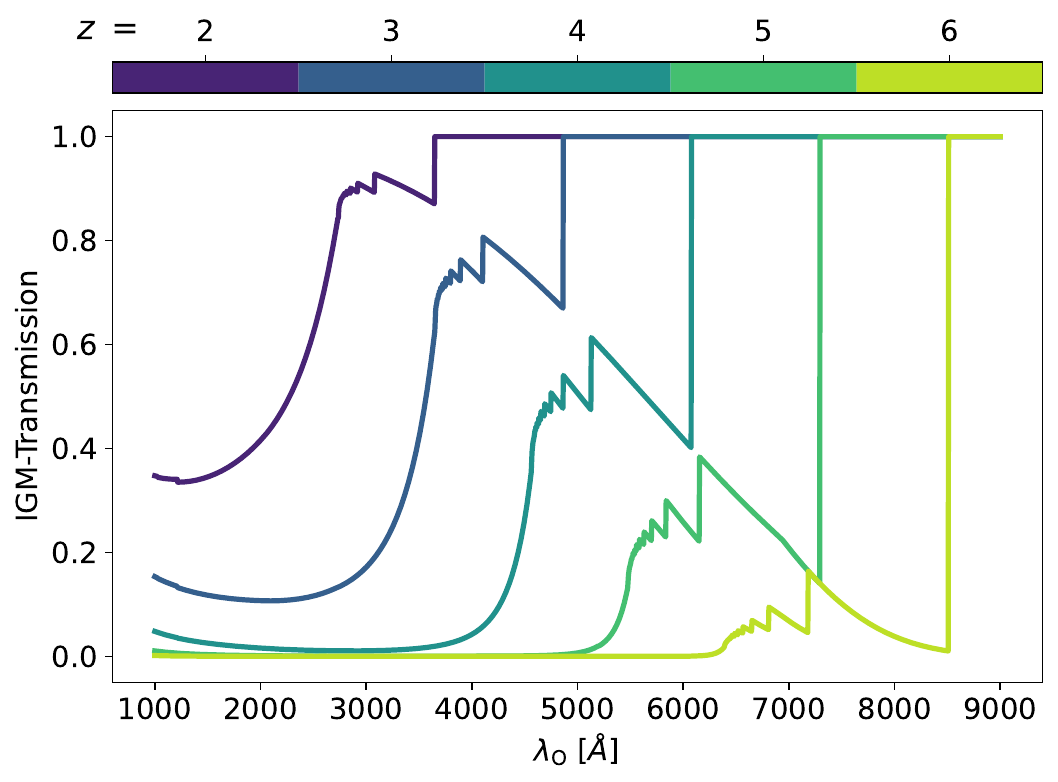}
  \caption{Overall IGM transmission functions predicted by \cite{Inoue2014} for different values of redshift (color-coded).}
     \label{fig:igm}
\end{figure}

As a result, the observed flux of Eq.~\eqref{eq:SEDflux} is multiplied by a transmission term given by $e^{-\tau_\text{IGM}(\lambda_\text{O}, z)}$.
In Fig.~\ref{fig:igm} we show this transmission for different values of redshift above $z = 2$.

%%%%%%%%%%%%%%%%%%%%%%%%%%%%%%%%%%%%%%%
\subsubsection{Photometry and Spectroscopy}\label{sec:photospec}

\galapy is capable of generating synthetic spectra from extra-galactic sources at the native resolution of the input SSP library used.
To compute fluxes in the observational band-pass filters, we use the transmission function $T_i(\lambda)$ in units of photons\footnote{In case of energy counter detectors we make the replacement $$T(\lambda) \rightarrow T(\lambda) / \lambda~\text{.}$$} associated to each instrumental filter $i$. 
We obtain the band-averaged flux as
\begin{equation}\label{eq:bandflux}
    \overline{S}_i = \frac{\int_0^\infty\text{d}\lambda\, T_i(\lambda)\, S(\lambda)/\lambda}{\int_0^\infty\text{d}\lambda\, T_i(\lambda)/\lambda}~,
\end{equation}
where the flux $S(\lambda)$ is computed through Eq.~\eqref{eq:SEDflux}.
Band-emission can be associated to a typical, ``pivot'' wavelength of the filter
\begin{equation}\label{eq:lambdapivot}
\overline{\lambda}_i \equiv \sqrt{\cfrac{\int_0^\infty\text{d}\lambda\, \lambda T_i(\lambda)}{\int_0^\infty\text{d}\lambda\, T_i(\lambda)/\lambda}}~.
\end{equation}

In our data-base, a set of band-pass transmissions from popular experiments is made available (e.g. JWST, SDSS, HST, Spitzer, Herschel, ALMA, VLA), but users can also load their custom transmissions by providing a wavelength-transmission grid over which the system should interpolate the function $T(\lambda)$.

A utility class \texttt{PhotoGXY} can be instantiated by importing the \texttt{galapy.Galaxy} module. 
Objects of \texttt{PhotoGXY} type allow the user to provide a photometric system to simulate and model observations on a set of band-pass transmissions of choice.

At current stage of development, the library does allow for fitting both photometric and spectroscopic data-sets. 
Nonetheless, this requires to modify the default likelihood from the python API (tutorials are available in the documentation).
Automatised sampling of the parameter space with spectroscopic data-sets will be the subject of a focused forthcoming work in the \galapy series.

%% file: sec/sampling.tex
\section{Parameters inference and analysis}\label{sec:inference}

In \galapy we use Monte Carlo techniques in a Bayesian framework in order to sample the posterior probability distributions in the parameter space defining our galaxy models.
A \texttt{sampling} sub-module is accessible from the Python API which provides an interface to two popular pure-Python libraries for parameter-space sampling, \texttt{emcee} \citep{emcee2013} and \texttt{dynesty} \citep{dynesty2020}.
The two libraries offered by \galapy provide distinct techniques and philosophies for addressing the challenge of multi-dimensional sampling. 

We describe here our current implementation and default set-up of the hyper-parameters\footnote{This hyper-parameters, including prior sizes, free model parameters and sampling strategy, can be set through the parameter-file generated calling the \texttt{galapy-genparams} command.}, highlighting though that the sampling module of \galapy will be extended in future developments of the library.
For the latest functionalities of the library, users should always refer to the on-line documentation.
Furthermore, all the discussion below only refers to the out-of-the-box functionalities, accessed through the entry-points described in Appendix~\ref{sec:demo}, that can always be extensively customised by accessing the \galapy Python API.

\subsection{Statistical set-up: likelihood, noise and priors}\label{sec:statistics}

In \galapy, parameter space is sampled with the intent of maximising the log-likelihood 
\begin{equation}\label{eq:loglike}
    \ln \mathcal{L}\left(\,\overline{S}~\big|~\theta\,\right) \equiv -\frac{1}{2}~\chi^2\left(\,\overline{S}~\big|~\theta\,\right)
\end{equation}
where $\overline{S}$ are the observed band-averaged fluxes.
Observations are compared against simulated fluxes obtained by means of Eq.~\eqref{eq:bandflux} with a galaxy model sampled from some position $\theta$ in parameter space.

When observing a source with a given instrument, the signal to noise ratio (S$/$N) might be small due to instrumental or environmental noise of different origins. 
It is not rare that the measurement in some band is not considered a detection, due to the low value of S$/$N.
Even though it should be kept in mind that also small values of the S$/$N are measurements with an associated error, and should therefore be treated as such, we allow for a different treatment of data-points and upper limits.

The $\chi^2$ statistics appearing in Eq.~\ref{eq:loglike} can be expanded as
\begin{equation}\label{eq:modchi2}
    \chi^2\left(\,\overline{S}~\big|~\theta\,\right) \equiv \sum_{i=0}^{N_\text{det}} \left(\frac{\overline{S}_i-\overline{S}_i(\theta)}{\sigma_i}\right)^2 + \sum_{j=0}^{N_\text{up-lims}} f\left[\overline{S}_j, \overline{S}_j(\theta),\sigma_j\right]
\end{equation}
where the first sum is a simple $\chi^2$ among all the $N_\text{det}$ bands where the measurement has been classified as a detection.
The second sum instead runs on the $N_\text{up-lims}$ functions $f$ accounting for the probability that the upper-limit in band $j$ is drawn from a given model $\theta$.

We provide three different possible treatments for upper limits:
\begin{itemize}

\item $\chi^2$ (default): non-detections are treated exactly as detections with a large error;

\item naive: a simple step-wise function setting the log-likelihood to -$\infty$ (i.e., zero probability) when the model predicts a flux larger than observed and to $0$ (i.e., probability equal to one) when the predicted flux is lower than the limit 
\begin{equation}\label{eq:uplims_naive}
f\left[\overline{S}_j, \overline{S}_j(\theta),\sigma_j\right] = \left\{
\begin{aligned}
&\ \text{-}\infty & \overline{S}_j(\theta) > \overline{S}_j\\
&\\
&\ 0& \text{otherwise}~;
\end{aligned}
\right.
\end{equation}

\item \cite{Sawicki2012}: the author proposes a modification of the $\chi^2$ that consists of the integral of the probability of some observation up to the given proposed model. If the errors on data are Gaussian, this integral provides the following analytical expression for the corresponding log-likelihood:
\begin{equation}\label{eq:uplims_sawicki}
    f\left[\overline{S}_j, \overline{S}_j(\theta),\sigma_j\right] = -2 \ln \left\{\sqrt{\frac{\pi}{2}} \sigma_j \left[1 + \text{erf}\left(\frac{\overline{S}_j - \overline{S}_j(\theta)}{\sqrt{2}\sigma_j}\right)\right]\right\}~.
\end{equation}
Even though it can be argued that using the expression above is the most formally correct way of accounting for upper limits when errors are Gaussian, the combination of logarithm and error function is particularly risky in computational terms.
Specifically, it tends to hit the numerical limit of floating point numbers representation accuracy really fast, leading to undefined behaviour. 
These problems, even though negligible in most of the occurrences, might lead to difficulties in the convergence of the posteriors for particularly complex posterior shapes.

\end{itemize}
As already mentioned though, in a Bayesian framework based on direct parameter-space sampling, even in the case of low S$/$N there is no real reason for using a different statistical treatment with respect to detections\footnote{This argument also applies to the extreme case of a negative flux.}.
The large relative error already contains the information necessary to inform the $\chi^2$ about the lack of flux in the specific band.
We therefore set as default behaviour for upper-limits the usage of a standard $\chi^2$.
We strongly recommend to provide the actual measured flux, independently from its S$/$N, to the sampling algorithm.
When such measurement is not available, a safe choice would be to set the flux measurement to the same value of the absolute error.

At the current state, \galapy is thought for the study of individual galaxies and not for the study of the correlation between the parameters in a large sample of objects, therefore, a sophisticated treatment of noise and systematic uncertainties is not necessary \citep[e.g.][]{Kelly2012,Galliano2018}.
Nevertheless, in preparation for future extensions of the library and for completeness, we have implemented a simplistic naive treatment of calibration errors and/or unknown systematic errors that might be present in the modelled data-sets.
To this purpose, we allow for the presence of a nuisance parameter $f_\text{sys}$ that modifies the measured uncertainties as
\begin{equation}
    \label{eq:sigma_sys}
    \Tilde{\sigma_i}^2(\theta, f_\text{sys}) \equiv \sigma_i^2 + f_\text{sys}^2 \overline{S}_i^2(\theta)~.
\end{equation}
This modified error depends on the model parameters $\theta$ through the predicted SED band flux $\overline{S}_i(\theta)$ and on the nuisance parameter $f_\text{sys}$ as well as from the original measured error $\sigma_i$.
By adding a positive value to the observed variance we are making the assumption it had been underestimated by a relative factor $f_\text{sys}$.

We are not accounting for eventual correlations between observational bands thus our Gaussian log-likelihood is simply modified by an additional term accounting for the dependence of the variance on the model parameters.
For the case of detections Eq.~\ref{eq:loglike} becomes
\begin{multline}
    \label{eq:loglikenoise}
    \ln \mathcal{L}\left(\,\overline{S}~\big|~\theta, f_\text{sys}\,\right) \equiv \\
    -\frac{1}{2}~\sum_i\left\{\frac{\left[\overline{S}_i-\overline{S}_i(\theta)\right]^2}{\Tilde{\sigma_i}^2(\theta, f_\text{sys})} + \ln\left[2\,\pi\,\Tilde{\sigma}_i^2(\theta, f_\text{sys})\right]\,\right\}~;
\end{multline}
a similar modification is applied to the case of upper limits.

This simple noise model adds only one parameter to the multi-dimensional space that has to be sampled, therefore it does not particularly burden the sampling procedure.
We have tested on multiple problem set-ups that $f_\text{sys}$ is completely uncorrelated to the other free parameters: the only net effect on the final posterior is to make the constraints less tight, as it would be expected if errors in the observed data-set were larger.
Nonetheless, we have also observed that the addition of this systematic error in some cases help in breaking the degeneracy between parameters, especially in the case of multi-modal posteriors such as, e.g., when estimating photometric redshifts.

In closing this Section, we highlight that the default behaviour of \galapy currently only accounts for uniform uninformative priors, whose limits are set by the user.
This choice is motivated by the argument that each galaxy should be considered as an independent object for which a-priori knowledge of the parameters can be hardly argued.

Accessing the \galapy Python API, the aforementioned behaviours can be easily modified.
Furthermore, more sophisticated statistical tools, such as non-Gaussian errors and non-uniform priors, are planned for future extensions of the library. 
We are already working at the implementation of a hierarchical Bayesian sampling scheme that, along with a more sophisticated treatment of the systematic errors, is intended for the application of \galapy on large samples of galaxies, foreseeing upcoming data from future surveys.

\subsection{Samplers}\label{sec:samplers}

The statistical framework of \galapy comes with a \texttt{Sampler} object that provides a common interface for the parameter-space samplers we rely on, namely, \texttt{emcee} \citep{emcee2013} and \texttt{dynesty} \citep{dynesty2020,koposov2023}.
These two libraries provide different and complementary approaches to Monte Carlo sampling of a multi-dimensional space.
We maintain both tools in order to provide a flexible machinery that can be adapted to different problems.

\begin{itemize}
\item \texttt{emcee}, provides an implementation of the Markov-Chain Monte Carlo (MCMC) technique.
Specifically, it implements an ensemble sampler with affine invariance \citep{GoodmanWeare2010} that, by instantiating many test particles (\textit{walkers}) in the parameter space, builds first order Markov sequences of proposals that are tested against the likelihood.
The dynamics of this system of particles is regulated by the requirement that, at each new step, a better estimate of the parameters is drawn. 
\item \texttt{dynesty} implements Dynamic Nested Sampling \citep{Higson2019}, a generalised version of nested sampling \citep{Skilling2004,Skilling2006} where the number of test particles (here \textit{live points}) is dynamically increased in regions of the posterior where a higher accuracy is required. The parameter space is modelled as a nested set of iso-likelihood regions that are sampled until the overall evidence reaches a stopping criterion set by the user. In our default hyper-parameters set-up we provide an $80\%/20\%$ posterior/evidence split and we model the posterior space with multiple ellipsoids \citep{FerozHobson2009}, as we expect to have multiple peaks and correlations when sampling high dimensional parameter spaces.
We use the default stopping function
$$\mathcal{S}(f_p, s_p, s_{\mathcal{Z}}, n) \equiv 
    f_p \times \frac{\mathcal{S}_p(n)}{s_p} + 
    (1 - f_p) \times \frac{\mathcal{S}_\mathcal{Z}(n)}{s_{\mathcal{Z}}} < 1~\text{,}$$
where $f_p$ is the fractional importance we place on posterior estimation ($20\%$, as mentioned above), $\mathcal{S}_p$ is the posterior stopping function, $\mathcal{S}_\mathcal{Z}$ is the evidence stopping function, $s_p$ is the posterior "error threshold", $s_\mathcal{Z}$ is the evidence error threshold, and $n$ is the total number of Monte Carlo realisations, used to generate the posterior/evidence stopping values.  
\end{itemize}

When sampling high-dimensional large volumes the degeneracy between parameters can easily generate a complex posterior topology, such as multiple peaks on some parameters or non-linear correlations.
Our suggestion for an optimal usage of \galapy is to rely on dynamic nested sampling in this case.
As an empirical rule of thumb, we can recommend to rely on nested sampling when the number of free parameters is larger than $~5$ and when it is not necessary to include extremely complex priors (as this, even though feasible, is not trivial).

On the other hand, MCMC provides a more straightforward interface to the inclusion of sophisticated priors and proves to be efficient and to possibly converge faster when working on smaller and well-behaved volumes, i.e. when multiple peaks and complex correlations among parameters are not to be expected.

\galapy comes with a default set-up for the hyper-parameters determining the behaviour of the two currently available samplers.
The chosen values should work, and have been tested, on several common possible problems.
For both the nested sampler and the MCMC sampler, a drawback of this default set-up is that it might not be the fastest to converge, nonetheless convergence should be guaranteed.
We stress that it is not possible to provide a general set-up of the aforementioned hyper-parameters. 
Experienced users can access and modify the default values to better suit the specific needs of the problem at study.

As already mentioned, we plan to include additional samplers in future extensions of the library.

\subsection{Results}\label{sec:results}

We provide a \texttt{Results} class that collects all the information acquired during the sampling run and computes derived quantities for easy access and analysis.
This includes all the full-SEDs computed for each position in the parameter space, all the derived quantities (masses, metallicities and temperatures) as well as all the coordinates in the parameter space and all the \galapy objects built during the run.

\texttt{Results} objects tend to be particularly heavy in terms of both volatile and non-volatile memory. 
The typical size primarily depends on the number of samples that were needed to obtain a converged posterior and, secondarily, on the number of free-parameters and the other characteristics of the sampling run.
Given the large amount of memory that could be necessary for computation and storage, we offer the possibility to store the results of a sampling run without computing the associated \texttt{Results} object, leaving this process for when the results have to be analysed.

The output formats available in \galapy are
\begin{itemize}
    \item \texttt{pickle}: the standard Python serialisation protocol.
    \texttt{Results} object are computed at the end of a sampling run then serialised and stored in non-volatile memory. 
    The typical size of the output file can reach up to $\sim 1$ GB.
    \item \texttt{hdf5}: the Hierarchical Data Format \citep{folk2011overview}, a widespread method for storing heterogeneous data. 
    When using this format storage in non-volatile memory is possible in two flavours:
    \begin{itemize}
        \item \texttt{light}: store only samples coordinates, likelihood values and weights along with minimal additional information to re-build the models used in the sampling (typical size $~ 10$ MB);
        \item \texttt{heavy}: along with the information available also with the \texttt{light} option, all the additional derived quantities computed when building the \texttt{Results} object are stored (typical size up to $\sim 1$ GB).
    \end{itemize}
\end{itemize}
Once stored, results can be accessed and analysed by users in any moment.
Note that, when choosing the HDF5 format in either its \textit{heavy} or \textit{light} version, results can be accessed even without having to instantiate a \texttt{Results} object and can be loaded in memory as simple dictionaries or accessed as regular HDF5 files.
The drawback of choosing lightweight storage is an additional overhead when instantiating the \texttt{Results} object for the analysis.

By instantiating or de-serialising the \texttt{Results} class several functions for statistical analysis, \TeX~table formatting and plotting are made available.
This should guarantee quick access to data and user-friendliness.
All the plots and tables provided in the following Sec.~\ref{sec:validation} have been produced using these tools.

\subsection{Analysis}\label{sec:analysis}

We distinguish among two broad categories of quantities that are stored and/or that can be computed after a sampling run:
\begin{itemize}
    \item \textit{free parameters}, are all the parameters that define the behaviour of the emission model chosen. 
    These parameters define the size of the parameter-space that is sampled by the Monte Carlo algorithm of choice.
    Parameters of this kind can be, e.g., the age of the galaxy, its redshift, the indexes of the extinction power-law in Eq.s~\eqref{eq:ADD} and \eqref{eq:AMC}.
    A complete list of all the possible free-parameters is provided in Table~\ref{tab:tunable_parameters}.
    \item \textit{derived parameters}, are all those parameters that do not directly define an additional dimension in the parameter-space inspected by the sampler but can be computed by choosing a given position in the parameter space. 
\end{itemize}

\galapy provides several different tools for analysing the results of a sampling run.
These tools are primarily accessible as functions of the \texttt{Results} class described in Sec.~\ref{sec:results} and by importing the sub-package \texttt{galapy.analysis}.
The latter contains two modules: \texttt{plot} and \texttt{funcs}, which respectively provide interfaces for plotting and generating formatted tables of different statistics measured both on the free-parameters, $\mathbf{\theta}$, and on the derived parameters, $\mathbf{\delta}$.

For each sampled position in the free-parameters space we have an associated value of the log-likelihood, $\ln \mathcal{L}(\mathbf{\theta}_i)$, and a weight, $\mathbf{w}_i$.
Furthermore, we pre-compute at the end of the run several derived quantities automatically such as, the full SED in the whole wavelength grid (as it is defined by the SSP library of choice), temperatures of the two ISM components, masses of the different components (stars, dust and gas), metallicities,  star formation rate.

In Bayesian inference we want to get to an estimate of the free-parameters posteriors, $P(\theta|D)$, given a data-set, $D$, a model of the data depending on the free-parameters, $\theta$, and some priors, $P(\theta)$.
From the sampled posterior one can derive an estimate of the true value of each parameter, free $\hat{\theta}$ or derived $\hat{\delta}$, using an estimator (such as, e.g., the weighted mean of samples).
Monte Carlo techniques allow to derive a sample of positions in the parameter space from which we can get to an approximate estimate of the posterior.
It is therefore possible to weight each position in the parameters space by the likelihood and compute weighted summary statistics and estimators.

The two samplers currently available in \galapy provide different philosophies to approximate the posterior. 
The Monte Carlo Markov Chain (MCMC) method implemented in \texttt{emcee} generates samples proportional to the posterior, so that
\begin{equation}
    \label{eq:mcmc_weight}
    w_i \equiv 1\ \forall\ w_i \in \mathbf{w}\text{.}
\end{equation}
On the other hand, the Dynamic Nested Sampling algorithm used in \texttt{dynesty}, generates samples in nested (possibly disjoint) ``shells'' of increasing likelihood.
The associated estimate of the posterior is then obtained by combining the set of samples with weights defined as
\begin{equation}
    \label{eq:nested_weight}
    w_i \equiv \dfrac{1}{2}\left[\mathcal{L}(\mathbf{\theta}_{i-1})+\mathcal{L}(\mathbf{\theta}_{i})\right]\times\left[X_{i-1}-X_i\right]
\end{equation}
where $\mathcal{L}(\mathbf{\theta}_{i})$ is the likelihood of the i-th sample and $X_i$ is its associated volume of the prior\footnote{Note that the prior bounds the algorithm to inspect only a finite region of the multidimensional parameters space, which would otherwise belong to $\mathds{R}^N$, where $N$ is the number of free parameters.} where the likelihood $\mathcal{L}(\theta_i) \geq \lambda$ is above some threshold $\lambda$.

When a sampling run converges, as already mentioned, we provide users with all the samples, their associated log-likelihoods and weights, along with derived-parameter values in all these positions.
In this way users can choose to use their custom estimators to get to an estimate of the true values of these parameters.
Conveniently though, we also provide functions for computing some useful estimators, accessible either from the \texttt{galapy.analysis} sub-package or the \texttt{Results} class.

Along with the weighted average and standard deviation, percentiles and best-fitting value (i.e. the position in the parameter space among all those sampled where the log-likelihood has assumed its maximum value) we also give the possibility to compute credible intervals around a given position of the parameter space.
All of these quantities are weighted with values from Eq.~\eqref{eq:mcmc_weight} and \eqref{eq:nested_weight}.

In particular, we define the central credible interval for a marginalised parameter $\theta$ as that region of the parameter space enclosed in an interval $[\theta_\text{low}, \theta_\text{upp}]$ defined around the best-fitting (i.e. maximum likelihood) value of the parameter, $\theta_\text{best}$.
The limits of this interval are defined by
\begin{equation}
    \label{eq:credible_interval_low}
    \int_{\theta_\text{low}}^{\theta_\text{best}}P(\theta|D)\,\text{d}\theta = \dfrac{\alpha}{2}
\end{equation}
for the lower bound, and
\begin{equation}
\label{eq:credible_interval_upp}
\int_{\theta_\text{best}}^{\theta_\text{upp}}P(\theta|D)\text{d}\theta = \dfrac{\alpha}{2}
\end{equation}
for the upper bound.
A value of, e.g., $\alpha = 0.68$ gives the $68\%$ credible interval.

For highly asymmetric or multi-modal marginalised posteriors, one of the two half-integrals in Eq.s~\eqref{eq:credible_interval_low} and \eqref{eq:credible_interval_upp} might not encompass enough samples to embed the requested probability value.
In these cases only upper/lower limits on the parameter value can be retrieved and the equations become
\begin{equation}
    \label{eq:credible_lower_limit}
    \int_{\theta_\text{low}}^{+\infty}P(\theta|D)\,\text{d}\theta = \alpha
\end{equation}
for lower limits and
\begin{equation}
\label{eq:credible_upper_limit}
\int_{-\infty}^{\theta_\text{upp}}P(\theta|D)\text{d}\theta = \alpha
\end{equation}
for upper limits.

%% file: sec/validation.tex
\section{Validation}\label{sec:validation}

In this Section we present a sanity check for \galapy.
We both verify that all the components of the library are behaving as expected as well as validate the scientific return of the physical models proposed.
Even though we limit our presentation here to the aspects involving the science that can be performed with our tool, we provide some preliminary discussion on the computational side and on software performances in Appendix~\ref{apx:codedesign}.
A thorough discussion on performances and a comparison with other codes goes beyond the scope of this manuscript and is left for the future.

We test our model on both mock and real observations of star-forming and quiescent objects. 
Star-forming objects are complex structures that can host several, if not all, of the different components implemented in \galapy, making them an excellent test-bench for investigating the interplay between the modules building up our library. 
We first test the constraining capabilities of our machinery by building a set of mock observations of simulated galaxies with different physical properties and perform the regression with \galapy (Sec.~\ref{sec:validation_mock}).
In Sec.~\ref{sec:validation_real}, we use the In-Situ SFH model (Sec~\ref{sec:insitu}) along with our dust model (Sec.~\ref{sec:extinctionattenuation}) on a set of real sources, in order to validate the reliability of these models on estimating the astro-physical properties of sources.

\subsection{Validation on mock sources}\label{sec:validation_mock}

In order to verify and prove the efficacy of \galapy we first test it against mock observations generated with the library itself.

As anticipated, we generate a set of mock observations of galaxies simulated using the \galapy modelling framework.
We generate different mock sources by randomly sampling a flat prior space for different models of SFH.
For each of these sources we use the \texttt{parsec22.nt} SSP library (see Sec.~\ref{sec:stellar}; in Appendix~\ref{apx:ssp} we compare the BC03 libraries with the PARSEC22 libraries and show that the results obtained are consistent independently on the choice of SSP library). 

In particular, for the sake of investigating the library reliability on a broad parameter space volume, we generate both actively star-forming and passively evolving galaxies.
\begin{table}
\centering
\caption{\label{tab:mock_summary} Summary of the number of generated sources for the different combinations of SFH models. 
}
\input{tab/mock_sources_summary}
\tablefoot{The X+Y notation marks that the sample is divided in two equivalent sub-samples: for X sources spectroscopic redshift is provided, for Y sources we consider the photometric redshift a free parameter of the model.}
\end{table}
Table~\ref{tab:mock_summary} summarises the number of mock sources generated for this test.
In particular, for each SFH model we generate two sets of objects: 20 sources for which a spectroscopic estimate of redshift exists, and 20 sources for which redshift has to be estimated photo-metrically (i.e. $z_\text{phot}$ is a free-parameter of the model.
Furthermore, for the In-Situ and for Delayed Exponential SFH models, we both generate sources that have still an active star-formation and sources that are passively evolving. 
On the other hand, for the Constant SFH model, we only select actively star forming objects, as this particular model is intended for objects that are undergoing a secular evolution of their stellar content (e.g. late type galaxies).
We therefore run our test on a total of $200$ mock sources, $100$ of which are assumed to not have a spectroscopic determination of redshift.

In practical terms, actively star forming sources are generated by setting the $\tau_\text{quench}$ parameter to an arbitrarily large value, consistent with infinite\footnote{A value $\tau_\text{quench}\ge2\times10^{10}$ Gyrs is enough as it will always prove larger than the Age of the Universe at any epoch.}.
On the other hand, for passively evolving objects, we sample a random value for the quenching time, $\tau_\text{quench}$ and impose that the age should be sampled from an interval of values that is upper-limited by the sampled value of $\tau_\text{quench}$.
\begin{table*}
\centering
\caption{\label{tab:mock_priors} Summary of the free parameters and priors for the mock galaxies of Sec.~\ref{sec:validation_mock}.}
\input{tab/mock_sources_priors}
\tablefoot{For each SFH model and both for actively star forming and passively evolving mock sources, we list the parameter symbols (normalised to their unit) and the prior upper and lower limits. The symbols used are consistent to those used in Sec.~\ref{sec:models} and summarised in Table~\ref{tab:tunable_parameters}.}
\end{table*} 
Each mock source has been generated by sampling uniformly a position in the parameter space defined by the priors summarised in Table~\ref{tab:mock_priors}.

The overall SED of our mock galaxy will therefore consist of:
\begin{itemize}
    \item dust-attenuated emission from stars including nebular thermal emission and non-thermal synchrotron as well as thermal emission coming from the two different components of our dust model, for the case of actively star-forming galaxies;
    \item un-attenuated passively evolving stellar emission, possibly including (if quenching had been particularly recent) some left-over nebular thermal emission and non-thermal synchrotron for source.
    The latter would anyways be extremely sub-dominant and un-investigated given the photometric system shown in Fig.~\ref{fig:pms_mock_passive}.
\end{itemize}

\begin{figure*}
\resizebox{\hsize}{!}{
\includegraphics{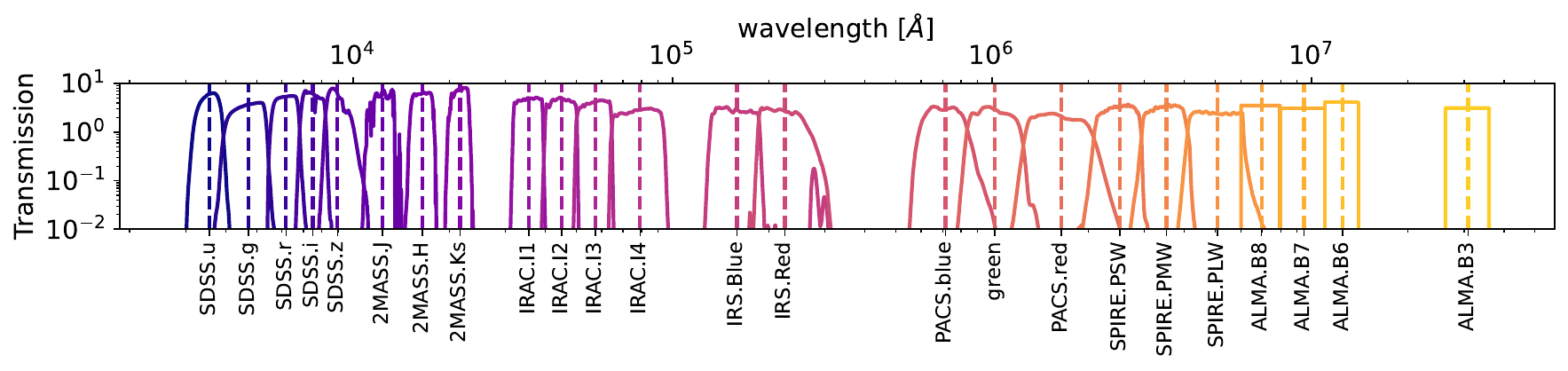}}
  \caption{Photometric system used to generate the mock observation for the actively star forming simulated galaxies of Sec.~\ref{sec:validation_mock}. The lower x-axis shows the keyword name of the band-pass transmission while the upper x-axis shows the corresponding wavelength in angstroms. Transmissions are expressed in terms of photons and the dashed lines mark the position of the pivot wavelength for each band-pass filter. Note that this is just a possible set-up specific to the case of the mock galaxy of Sec.~\ref{sec:validation_mock}. It represents only a sub-set of the band-pass transmissions available in the \galapy data-base.}
\label{fig:pms_mock_active}
\end{figure*}
\begin{figure*}
\resizebox{\hsize}{!}{
\includegraphics{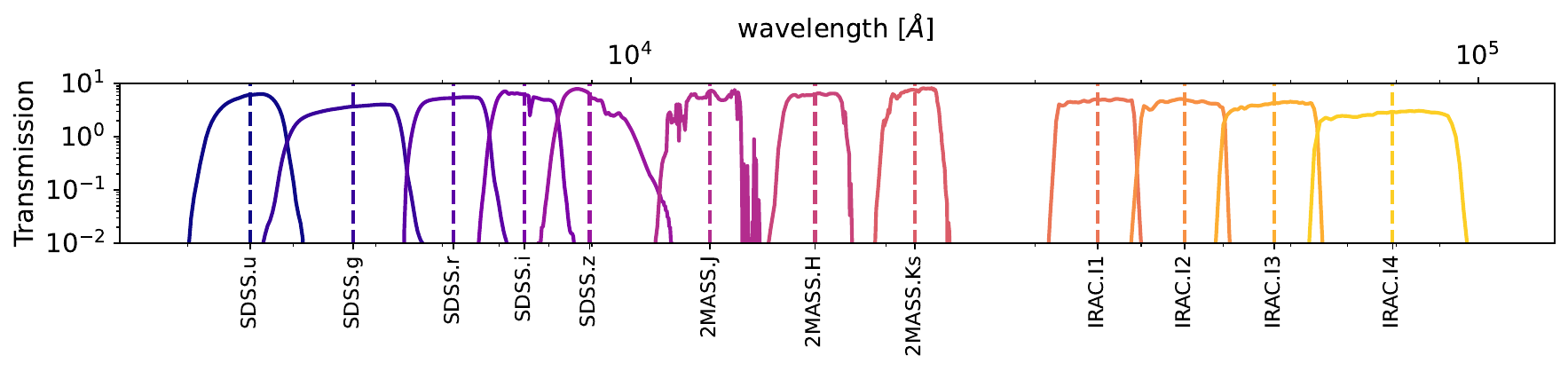}}
  \caption{Same as Fig.~\ref{fig:pms_mock_active} for the mock observation of the passively evolving simulated galaxies of Sec.~\ref{sec:validation_mock}.}
\label{fig:pms_mock_passive}
\end{figure*}
In order to build a mock photometric observation, we need to assume a photometric system. 
This is graphically shown in Figures~\ref{fig:pms_mock_active} and \ref{fig:pms_mock_passive}, where, as a function of wavelength, we show the transmission corresponding to the $24$ band-pass filters we use for actively star-forming mock galaxies and the $12$ used for passively evolving mock sources, respectively.
We have selected filters from different well known experiments, covering a wide range of wavelengths.
While for passive objects we select filters from the UV/Optical bands to the Near Infrared, using transmissions from SDSS, 2MASS and Spitzer, for active objects we extend the spectral coverage up to the sub-mm/mm bands adding also filters from Herschel and ALMA.

To add errors to our mock observation we first associate to each different transmitted flux of each single mock source an error that is randomly chosen to be between $10\%$ and $50\%$ of the flux.
With this value set for all the fluxes, we then generate a random realisation of the mock measurement by extracting it from a Gaussian distribution with mean equal to the real value of the transmitted flux and standard deviation equal to the random error. 

The free-parameters chosen for the sampling runs are the same free parameters we varied in generating the mock sources (Table~\ref{tab:mock_priors}).We allow for $9$ free parameters, including age, SFH and ISM defining parameters, in actively star forming galaxies.
For passively evolving sources we instead vary $4$ free-parameters, including age of the galaxy, age of quenching and SFH defining parameters.
In both cases, for 100 out of the 200 sources, redshift is set to be an additional free-parameter.
Consistently, we model each source with the same SFH model and SSP library used for generating it.

For sampling the free-parameters of our models, we assume a set of uninformative uniform priors whose limits correspond to those listed in Table~\ref{tab:mock_priors}, i.e. the same intervals defining the parameter space volume sampled by the mock sources.
For each source, we run a dynamic nested sampling using \texttt{dynesty} with default \galapy sampling hyper-parameters and stopping criterion\footnote{
Besides the posterior/evidence split and stopping function mentioned in Sec.~\ref{sec:samplers}, we set a higher-bound stopping criterion corresponding to the maximum effective number of likelihood calls $\max(N_\text{eff}) = 5\times10^6$.
Our initial tolerance is set to $\Delta \ln \hat{\mathcal{Z}} \lesssim 0.05$ with an initial maximum number of iterations \texttt{maxiter\_init} $=10^4$.
We then add iteratively $10$ batches of new live-points with a maximum number of iterations per batch corresponding to \texttt{maxiter\_batch} $=10^3$. 
We use the multiple-ellipsoidal decomposition \citep{FerozHobson2009} as bounding criterion.
}.

Sampling runs take, on average, approximately 15 minutes per source to converge on 8 physical cores of an Intel i9-10885H CPU @ 2.40GHz with \texttt{x86\_64} architecture.
The time required for convergence strongly depends on the total number of samples extracted.
When running with dynamic sampling, this number is not known in advance \citep[more details in][]{dynesty2020}.
Our runs typically converge with a total number of valid samples between 10 to 20 thousands, each with a different weight. 
This does not reflect the actual number of likelihood calls that, given an average efficiency between $1\%$ to $5\%$ for these kind of problems, span a range between $5\div10\times10^5$.

\begin{figure}
\centering
\includegraphics[width=\hsize]{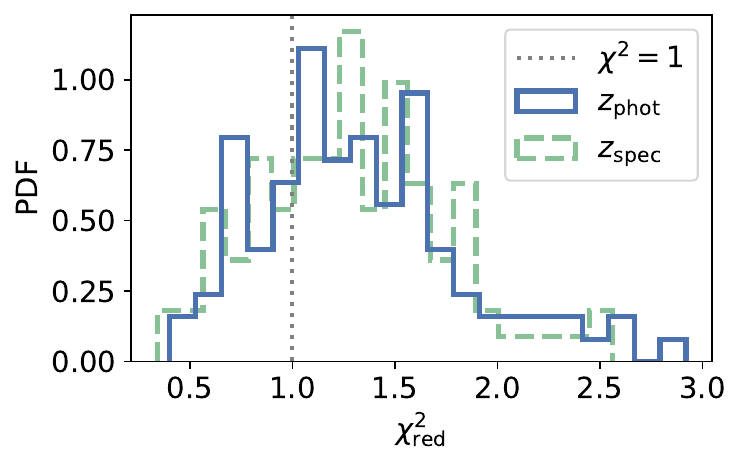}
\caption{Distribution of the reduced $\chi^2$ for the best-fitting parameters of the two sets of sources (with and without spectroscopic redshift in dashed green and solid blue, respectively), as obtained by sampling the free-parameter space with \texttt{dynesty}.}
\label{fig:res_mock_chi2}
\end{figure}
In Fig.~\ref{fig:res_mock_chi2} we show the distribution of the reduced $\chi^2$ values for the best-fitting set of parameters obtained by means of the dynamic nested sampling runs detailed above.
With a dashed green line we mark the distribution of the 100 sources for which we assumed a value of the spectroscopic redshift was available, while the solid blue histogram marks the distribution of the 100 sources for which redshift was a free-parameter.
As a term of comparison, we show as a dotted vertical gray line the value corresponding to the expectanion value $\chi^2=1$.
From the histograms, we can appreciate how, for almost all the sources, $0.5 \le \chi_\text{red}^2 \le 2.5$.

\begin{figure*}
\centering
\includegraphics[width=\hsize]{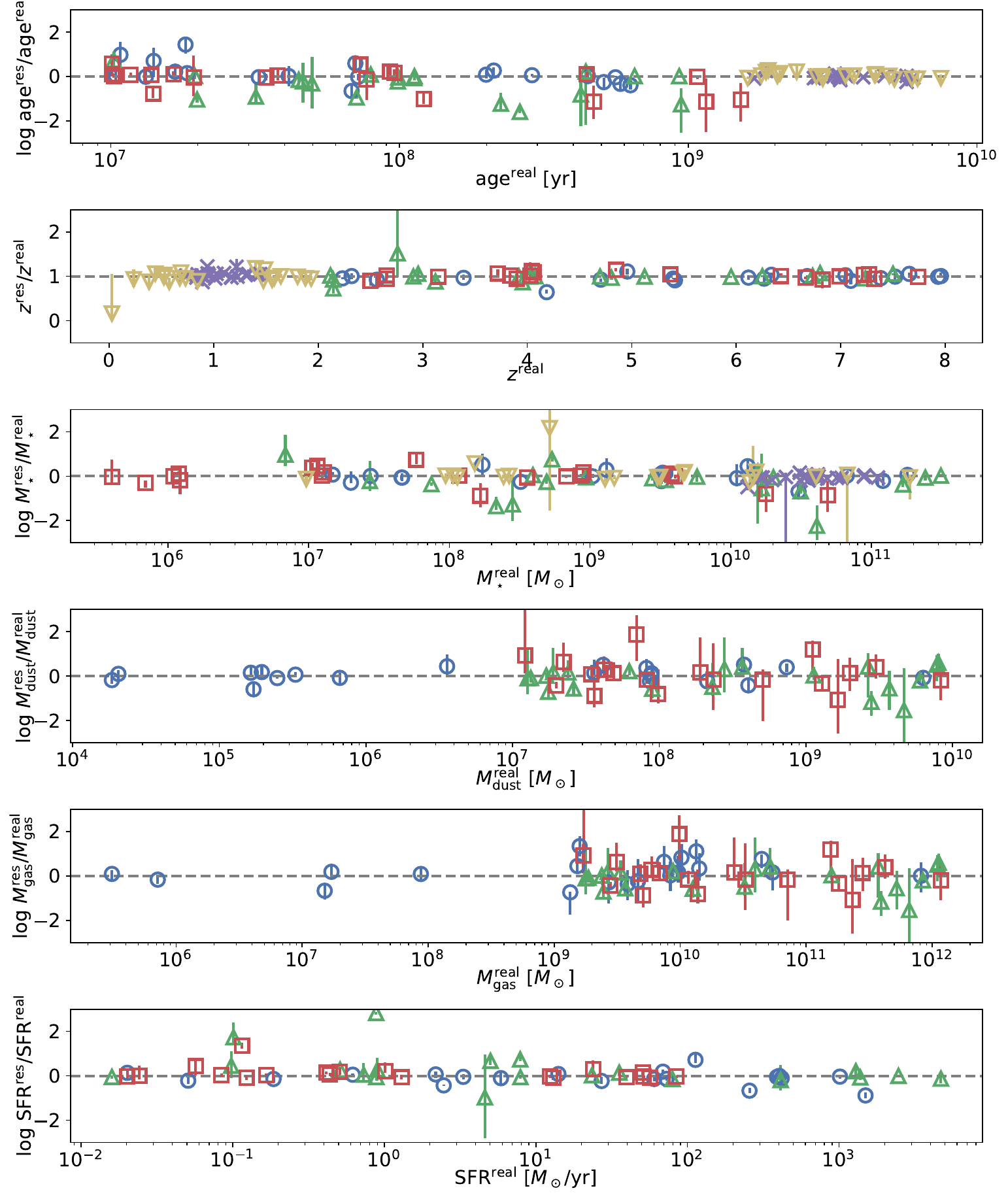}
\caption{Results for some of the free and derived parameters obtained by fitting with \galapy the mock observation set generated without spectroscopic redshift detection (see Sec.~\ref{sec:validation_mock}). The results are presented in terms of the ratio between the median value of the weighted samples collected by running the \texttt{dynesty} sampler. Errors are given in terms of the $16^\text{th}$ and $84^\text{th}$ percentiles, defining a $68\%$ credible interval around the median, as detailed in Sec.~\ref{sec:analysis}.
Blue circles, green upward triangles and red squares are actively star-forming sources modelled with an In-Situ, Delayed Exponential and Constant SFH model, respectively.
Violet crosses and yellow downward triangles are instead passively evolving sources modelled with an In-Situ and Delayed Exponential SFH model, respectively. 
}
\label{fig:res_mock_derived}
\end{figure*}
In Fig.~\ref{fig:res_mock_derived} we show the values for a collection of relevant free and derived parameters obtained by computing the weighted median of the samples with errors given by the $16^\text{th}$ and $84^\text{th}$ percentiles, i.e. the limits of a credible interval embedding a $68\%$ probability.
Different symbols and colours mark different combinations of SFH model and active/passive evolution, as detailed in the caption.
Results are given in terms of the ratio between the measured and real value of the parameter.
The top three panel show quantities available for all sources, actively star forming and passively evolving (i.e. age of the galaxy, redshift and stellar mass), while the lower three panels show quantities that are defined only for the actively star-forming objects (i.e. dust mass, gas mass and current star formation rate).
We highlight that, even though Fig.~\ref{fig:res_mock_derived} collects only posterior values obtained for the 100 sources without spectroscopic redshift, equivalently consistent results have been found for the other set of sources.

The ratios in Fig.~\ref{fig:res_mock_derived} show that the true value of each parameter is within the $68\%$ credible interval for $\gtrsim90\%$ of the mock observations, with this percentage increasing considerably if accounting for a $95\%$ credible interval.
In particular, photometric redshift, stellar mass and star formation rate show an exquisite agreement with the expected value.

It is also interesting to focus on the $M_\text{dust}$ and $M_\text{gas}$ parameters.
As already discussed in Sec.~\ref{sec:sfh}, the method used to estimate these quantities in the In-Situ SFH model is different with respect to other empirical models.
In particular, while for empirical models of SFH $M_\text{dust}$ and $M_\text{gas}$ are free-parameters, the In-Situ model predicts their value analytically, based on the SFH.
It is therefore relevant that the estimates obtained by the In-Situ model (blue circles in Fig.~\ref{fig:res_mock_derived}) show a smaller error and a better agreement with the real value, with respect to the larger error-bars and scatter shown by the Delayed Exponential (green triangles) and Constant (red squares) models. 

We can conclude that the machinery we have built successfully retrieves the correct representation of data.
We highlight that the collection of sources used for testing has been selected randomly from a considerably large parameter-space volume without any prescription for the mock observation to be representative of any real source population.
Nonetheless the agreement of the results is almost perfect in all the dimensions, demonstrating how the tool is not limited to specific populations of objects and does not require a high level of fine tuning to get to a significant result.
This is reflected on the small scatter of the marginalised posteriors of the parameters (as shown in Fig.~\ref{fig:res_mock_derived}). 

\subsubsection{Demonstration on a mock source of the tools available}\label{sec:mock_demo}

Before moving to the analysis of real sources, we select one out of the 200 mock observations generated in the previous Section and we show more in detail the results inferred by the analysis of the posteriors on both the free and derived parameters.
In particular, as it will be subject to a stress test on real sources on the following Section, we pick one of the actively star-forming galaxies generated by sampling the priors defined for an In-Situ model of SFH (Table~\ref{tab:mock_priors}). 

\begin{figure}
\centering
\includegraphics[width=\hsize]{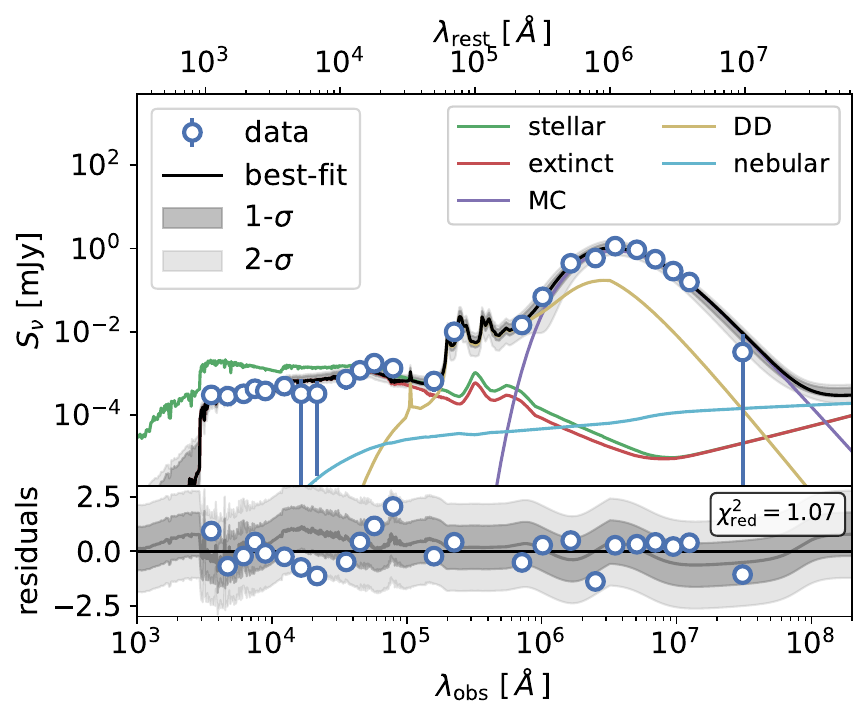}
\caption{Results of the parameters sampling for one of the mock observation of Sec.~\ref{sec:validation_mock}. \textit{Upper panel}: best-fit SED (black solid line) and components (coloured lines) compared to the mock observation data (empty blue markers); 1-$\sigma$ and 2-$\sigma$ confidence around the mean of the samples is also shown with grey shaded areas. \textit{Lower panel}: residuals with respect to the best-fit sample (black); mean and 1-/2-$\sigma$ confidence regions are also shown with a grey solid line and shaded areas, respectively.}
\label{fig:res_mock}
\end{figure}
In Fig.~\ref{fig:res_mock} we compare the original mock observation (blue empty round markers with error-bars) with the model favoured by the free-parameters posterior distribution.
The black solid line marks the best-fit model that results in a reduced $\chi_\text{red}^2 = 1.07$, also reported in the lower panel.
With shades of grey we show the $1$- and $2$-$\sigma$ confidence regions around the mean SED (in solid grey). 
Coloured solid lines show instead the contributions to the best-fitting SED, coming from the different components building up our galaxy model.
It is worth to highlight the different contributions of molecular clouds and diffuse dust to the peak of dust emission, that naturally blend into the final SED to represent a wider distribution of emission in the Mid- to Far-IR. 

By inspecting the standardised residuals (lower panel of Fig.~\ref{fig:res_mock}) 
\begin{equation}\label{eq:stdres}
    \chi_i = \frac{S_\text{O}(\lambda_i)-S_\text{M}(\lambda_i)}{\sigma_\text{O}(\lambda_i)} 
\end{equation}
we see how the best-fit model correctly intercepts the mock-observation while being within $1\text{-}\sigma$ from the mean of the samples.
This agreement is reflected on the marginalised posterior probability of the free-parameters.

\begin{figure}
\centering
\includegraphics[width=\hsize]{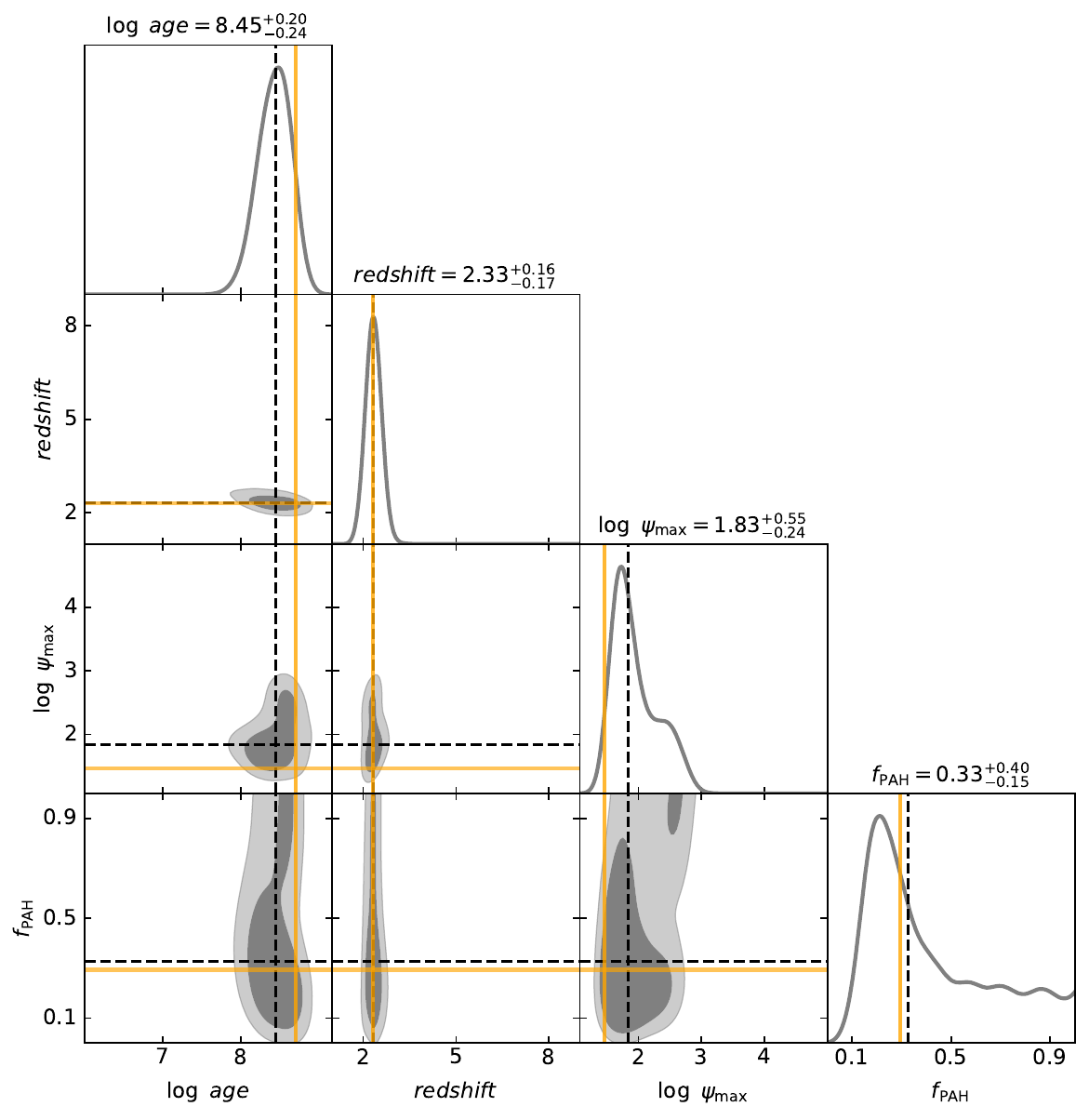}
\caption{Corner plot of the parameter posteriors obtained for the mock galaxy selected in Sec.~\ref{sec:mock_demo}, illustrating the one- and two-dimensional marginalised posteriors for a subset of the parameter space dimensions (for clarity and visualisation reasons). The shaded grey regions highlight $(68\%, 95\%)$ confidence from darker to lighter, corresponding to $(1\text{-}\sigma, 2\text{-}\sigma)$. The dashed black lines mark the position of the weighted median value of parameters while the values above the diagonal panels show the median and $68\%$ percentile around the median (roughly corresponding to $1$-$\sigma$ confidence in a Gaussian approximation). As a term of comparison, we also show mark, with orange solid lines, the real value of each parameter.}
\label{fig:triangle_freepar_mock}
\end{figure}
Figure~\ref{fig:triangle_freepar_mock} shows the triangle plot for a sub-set of the free-parameters posteriors marginalised to 1- and 2-dimensions.
These marginal posterior probabilities are given as histograms on the diagonal and as grey contours, for the 1D and 2D cases, respectively.
We do not show all the parameters in order not to burden the discussion but we limit the demonstration to the overall galaxy parameters (i.e. age and redshift), the normalisation of SFH parameter (i.e. $\psi_\text{max}$) and to the fraction of total emitted diffuse-dust energy contributed by PAH (i.e. $f_\text{PAH}$).
Specifically, the black dashed lines intercept the weighted median value of the samples, darker and lighter grey contours mark the $68\%$ and $95\%$ credible regions (note that, on top of each 1-dimensional marginalised posterior the median value for each parameter with the corresponding $68\%$ credible interval is also reported).
As a term of comparison we also show, with orange solid lines, the fiducial value of each parameter.
All these fiducial values fall within the $68\%$ credible interval.
It is worth to highlight the goodness of our fit for the photometric redshift estimate, as it an extremely sought after quantity and the algorithm was able to correctly infer it with an error of $\sim5\%$.

\begin{figure}
\centering
\includegraphics[width=\hsize]{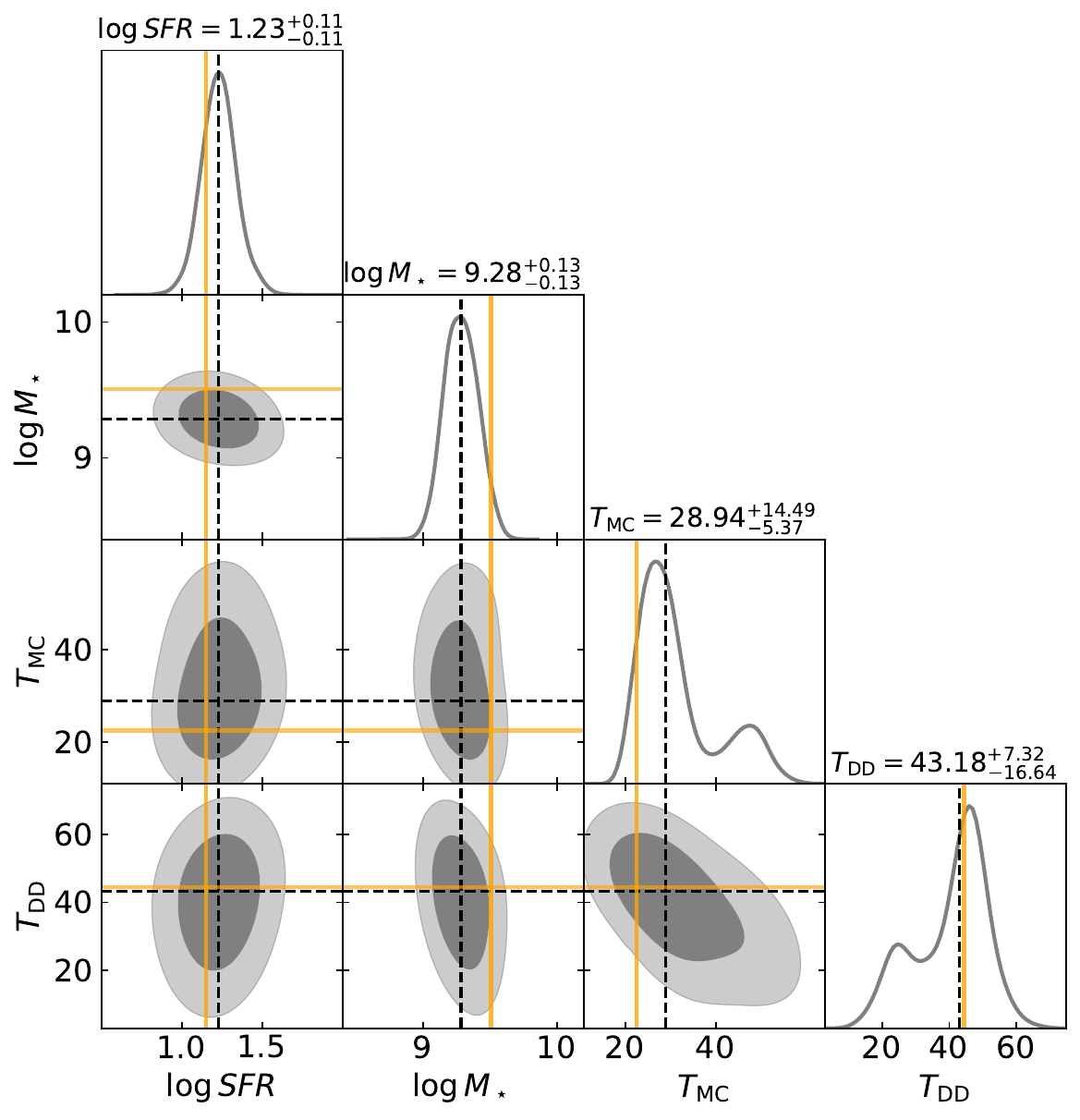}
\caption{Same as Fig.~\ref{fig:triangle_freepar_mock} for a sub-set of the derived parameters reported in Fig.~\ref{fig:res_mock_derived}.}
\label{fig:triangle_derivedpar_mock}
\end{figure}
We can use the results obtained with the sampling algorithm to build probability densities also for the derived parameters, as it is shown in Figure~\ref{fig:triangle_derivedpar_mock}.
We have selected four interesting derived parameters, namely the star formation rate, SFR, stellar mass, $M_\star$ and temperatures of the molecular, $T_\text{MC}$, and diffuse dust, $T_\text{DD}$, components.
The probability densities in 2-dimensional and 1-dimensional space are then built by computing the derived parameter's values in each position of the free-parameters space, as defined by the galaxy emission model used to represent the mock data-set.
Once again, we over-plot with orange solid lines the fiducial value of each parameter which, in all the cases, falls within the region encompassing $68\%$ of the total probability.
It is interesting to observe, on the 1D marginalised probabilities of the two temperatures, how in both cases there is a secondary peak that is symmetric in the two components.
This is not a surprise as the two dust components compete in contributing both to the absorption at short wavelengths and to the re-emission at longer wavelengths.
The modelling we have implemented it's nevertheless successful in distinguishing between the two, therefore favouring one of the two solution over the other.

\subsection{Test on real sources}\label{sec:validation_real}

In order to validate the scientific throughput of the new models introduced in \galapy, we compare their predictions with those performed with other SED-fitting codes and with different models for the SFH and for the dust-model.
To this purpose, we select a small set of sources with interesting properties from several different works.
Most of the objects inspected in this Section are either high-redshift dusty star-forming galaxies or their supposed descendants at low redshift, i.e. quiescent galaxies.

We underline that a validation of other models of SFH has already been performed on mock sources in the previous Section.
Furthermore, as a thorough analysis of the sources would go beyond the scope of this validation Section, we limit the discussion to a simple comparison of the results obtained fitting the SEDs with \galapy with those obtained with other techniques, as they appear in literature. 

\begin{table*}
\centering
\caption{\label{tab:validation1} Table comparing the quantities derived by fitting with \galapy the several sources of Sec.~\ref{sec:validation_real} with the values found in literature (first part).}
\input{tab/real_sources1}

\tablefoot{The values have been obtained with combinations of SED fitting and post-processing. 
SED fits have been performed using MAGPHYS-\textsc{photoz} \citep{magphys2008,Battisti2019} for the median of Sec.~\ref{sec:behiri23} and CIGALE \citep{cigale2019} in all other cases.
A long dash in a cell indicates that the corresponding value is either not available from literature or it does not have meaning in the model set-up used for the corresponding source.}
\tablebib{
(1)~\citet{Pantoni2021}; (2)~\citet{Casasola2020}; (3)~\citet{Giulietti2022}; (4)~\citet{Behiri2023}; (5)~\citet{Donevski2023}.
}
\end{table*}

\begin{table*}
\centering
\caption{\label{tab:validation2} Second part of Table~\ref{tab:validation1}.}
\input{tab/real_sources2}
\tablefoot{The gas metallicity values from literature have all been converted to absolute value from line estimates by means of $\log Z/Z_\odot = 12 + \log(O/H) - \log(O/H)_\odot$ \citep{Nagao2006} assuming $\log(O/H)_\odot = 8.69$ and $Z_\odot = 0.0153$. 
For each source, we provide in the first row the median and $68\%$ credible interval and on the second row, within round brackets, the best fitting value of the given parameter.
A long dash in a cell indicates that the corresponding value is either not available from literature or it does not have meaning in the model set-up used for the corresponding source.}
\tablebib{
(1)~\citet{Pantoni2021}; (2)~\citet{Casasola2020}; (3)~\citet{Giulietti2022}; (4)~\citet{Behiri2023}; (5)~\citet{Donevski2023}.
}
\end{table*}

\subsubsection[Dusty star forming galaxies]{Dusty star forming galaxies from \cite{Pantoni2021}}\label{sec:dsfg}

In the work from \citet[][P21 hereafter]{Pantoni2021}, a set of 11 sources selected from 3 millimetre catalogues (ALMA data from \citealt{Dunlop2017}, reprocessed within the ARI-L project, \citealt{Massardi2021}; LABOCA, \citealt{Yun2012}; AzTEC, \citealt{Targett2013}) in the GOOD-S field and complemented with fluxes from several other bands available in the field.
The selected sources are objects at the peak of cosmic star formation history, strongly attenuated by dust, with redshift spanning between $1.5< z\lesssim2.5$.
The data-set comes with the advantage of having spectroscopic redshifts and a pan-chromatic coverage spanning from visible to radio bands.

In P21 SEDs have been fitted using the CIGALE code \citep{cigale2019}, assuming a delayed exponential SFH and the BC03 SSP library. 
In order to account for the excess in NIR/MIR, the authors also include a power-law component that should ideally model the combination of diffuse dust, PAH and AGN.
For our analysis, we instead use our In-Situ SFH model and the PARSEC22 SSPs including lines from nebular regions. 
Our two-components dust-model automatically accounts for the NIR/MIR excess.

We select 4 sources, among the 11 available, which are not classified as AGN in the NIR-MIR regions of the spectrum. 
All the sources have at least 15 observations spanning from optical to radio. 
We fit the photometries letting 10 model parameters vary, including age, SFH and dust properties.

\begin{figure*}
\resizebox{\hsize}{!}{
\includegraphics{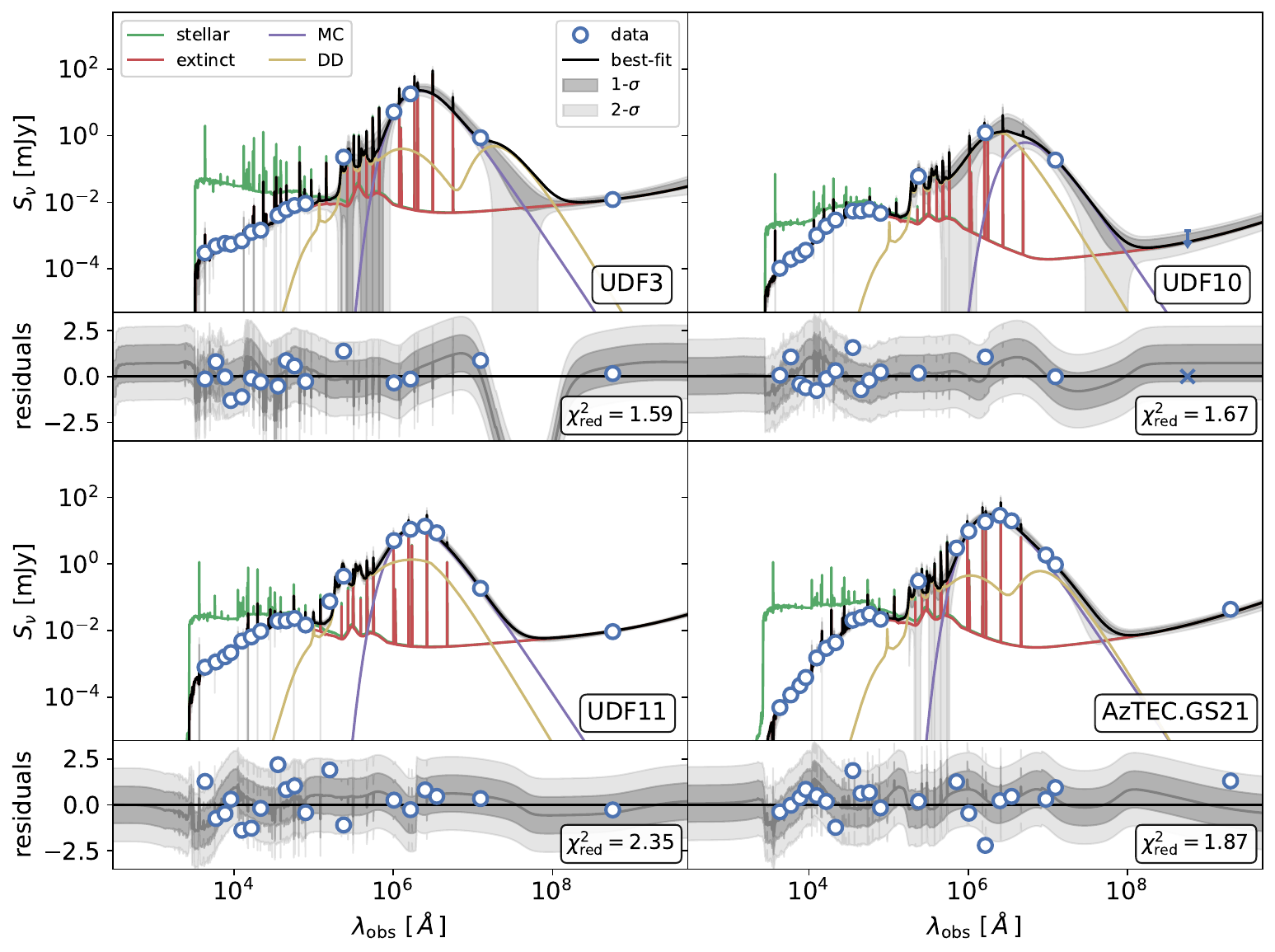}}
  \caption{\textit{Upper panels:} \galapy fits to the photometric data of $4$ galaxies selected from the \cite{Pantoni2021} sample (blue markers with error-bars). Best-fitting model (black solid line), different components (color-coded as in legend), and $1-$ and $2-\sigma$ confidence levels (grey shaded regions) around the mean of the samples are also shown. \textit{Lower panels:} standardised residuals with respect to the best-fitting galaxy models, and $1$-/$2$-$\sigma$ confidence intervals around the mean of the samples (grey shaded regions) are shown; the reduced $\chi^2$ of the fit is reported in each sub-panel.}
\label{fig:pantoni21}
\end{figure*}
The best-fitting SEDs obtained are shown as black solid lines in Fig.~\ref{fig:pantoni21} along with the several components contributing to emission (coloured solid lines) and $1$-/$2$-$\sigma$ confidence around the mean.
Band fluxes for each source are marked with empty blue circles.
In UDF10, the flux measured with VLA in the radio band is not a detection but was classified as an upper limit, we mark it with a downward arrow in the upper part of the panel.
In the lower part of each panel we also plot the standardised residuals with respect to the best-fitting model and the reduced $\chi^2$ value.
We mark the location of the corresponding radio upper limit of UDF10 with a cross in the residuals plot.

We report the main best-fitting values of free and derived parameters obtained for the 4 galaxies in Table~\ref{tab:validation1} and in Table~\ref{tab:validation2}.
We also include the values for the same set of parameters obtained in the original P21 work as a term of comparison.
Given that we implement models that are significantly different from the models used in P21, the values obtained with \galapy do not match exactly those obtained in the original work.
Nonetheless, the inferred properties (e.g. the dust and gas masses as reported in Table~\ref{tab:validation2}) are in good agreement with the ones obtained in the original work.
It as to be further noted that, in the original work these quantities, along with the gas metallicity, were not derived directly from SED-fitting but using post-processing and ALMA-bands emission line analysis.

Focusing on the temperatures of the two dust components, we can first of all notice that, as expected, in most of the cases the MC component is the one having the highest temperature.
On the other hand, it is interesting to notice how UDF10 shows a higher temperature in the DD component, while it is significantly older than the other three objects (i.e. age$\sim10^{9.5}$ years).
If we compare it to the right panel of Fig.~\ref{fig:ism_components}, this result suggests that the object could be older than its characteristic escape time from molecular clouds and therefore be in a late stage of evolution.
We infer that galaxy is significantly older than its characteristic escape time from molecular clouds, whose predicted value is $\log\tau_\text{esc}^\text{UDF10}/\text{yr} = 6.34_{-0.25}^{+0.80}$, suggesting that it might be in a late stage of evolution.
If we trust the In-Situ evolution scenario for the formation of ETG galaxies, about to approach a quiescent phase of evolution.
From both the plots in Fig.~\ref{fig:pantoni21} and the values reported in Tables~\ref{tab:validation1} and \ref{tab:validation2}, it is clear that the assumption of a two-component dust model ensures that the dust peak is better modelled with respect to the case of a single component. 
In particular, the possibility of having two peaks along with PAH emission, allows the overall MIR/FIR model to have more freedom to adapt to the data-set.
This is evident from the flex appearing at $\lambda\approx10^7$ {\AA} in UDF3 and from the broadening of the peak in UDF10, both effects due to the blending of the two grey-bodies. 
We can identify a common trend for all the 4 galaxies in the sample as in the single-component estimate of the temperature from the original work, the resulting measurement is systematically over-estimated with respect to the higher-temperature dust component in the two-component model used in this work.

With \galapy we can also easily derive the characteristic attenuation curves of the modelled galaxies (as detailed in Sec.~\ref{sec:extinctionattenuation}).
\begin{figure*}
\resizebox{\hsize}{!}{
\includegraphics{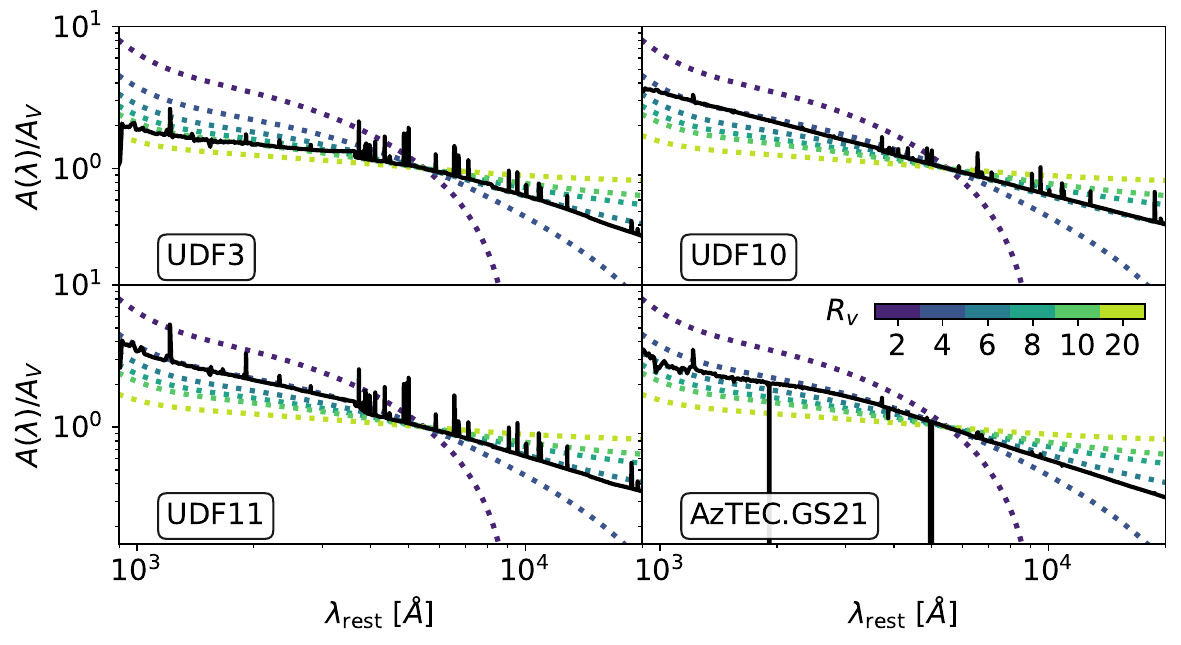}}
  \caption{Average attenuation curves (black solid line) predicted by \galapy in the best-fitting models for the 4 galaxies of P21, normalised to the average value of attenuation in the $V-$band ($\approx 5500$ \AA). For reference, we also show Calzetti-like empirical attenuation curves for different values of the $R_v$ parameter (color-coded).}
\label{fig:att_pantoni21}
\end{figure*}
The average attenuation curve for the 4 galaxies is shown in Fig.~\ref{fig:att_pantoni21} with solid black lines. 
We also plot for reference Calzetti-like \citep{Calzetti2000} attenuation at varying value of the $R_V$ parameter.
For wavelengths bluer than $\lambda_V\sim 5500$ \AA, our attenuation is consistent with $R_V \sim 4$ for UDF10, UDF11 and AzTEC.GS21, while it is $R_V \gtrsim 10$ for UDF3.
For wavelengths redder than $\lambda_V$ all the galaxies have an attenuation that could be represented with $4<R_V<6$ Calzetti-shapes.

\subsubsection[Local late type galaxies]{Local late type galaxies from \cite{Casasola2020}}\label{sec:casasola20}

We extend the validation of our library to a small sub-sample of 4 local ($z < 0.01$) late type galaxies from \cite{Casasola2020}, extracted from the DustPedia database\footnote{Available at \href{http://dustpedia.astro.noa.gr/}{dustpedia.astro.noa.gr}. DustPedia is a collaborative focused research project supported by the European Union under the Seventh Framework Programme (2007-2013) call (proposal no. 606847). The participating institutions are: Cardiff University, UK; National Observatory of Athens, Greece; Ghent University, Belgium; Université Paris Sud, France; National Institute for Astrophysics, Italy and CEA, France.}\citep[see also][]{DeVis2019}.
The archive provides access to multi-wavelength imagery and photometry for 875 nearby galaxies as well as physical parameters for each galaxy \citep{Davies2017,Clark2018} derived by means of the CIGALE code.

To further probe the reliability of \galapy's results, we select 4 galaxies that are not undergoing major interactions, do not show any nuclear activity in the X-ray and are not classified as starburst.
We once again select our In-Situ model of SFH and the SSPs from the PARSEC22 library.
We have performed several exploration runs of the sampler on the 4 galaxies, and consequently decided for a model set-up completely equivalent to the one used for the sources of Sec.~\ref{sec:dsfg} for consistency and as the overall results were not differing substantially.
Nonetheless, this is not intended to constitute a thorough analysis but just a sanity-check of \galapy's validity.
In terms of modelling set-up, we also allow for an eventual systematic error parameterised as described in Sec.~\ref{sec:statistics}, marginalising the results over the nuisance parameter $f_\text{sys}$.

\begin{figure*}
\resizebox{\hsize}{!}{
\includegraphics{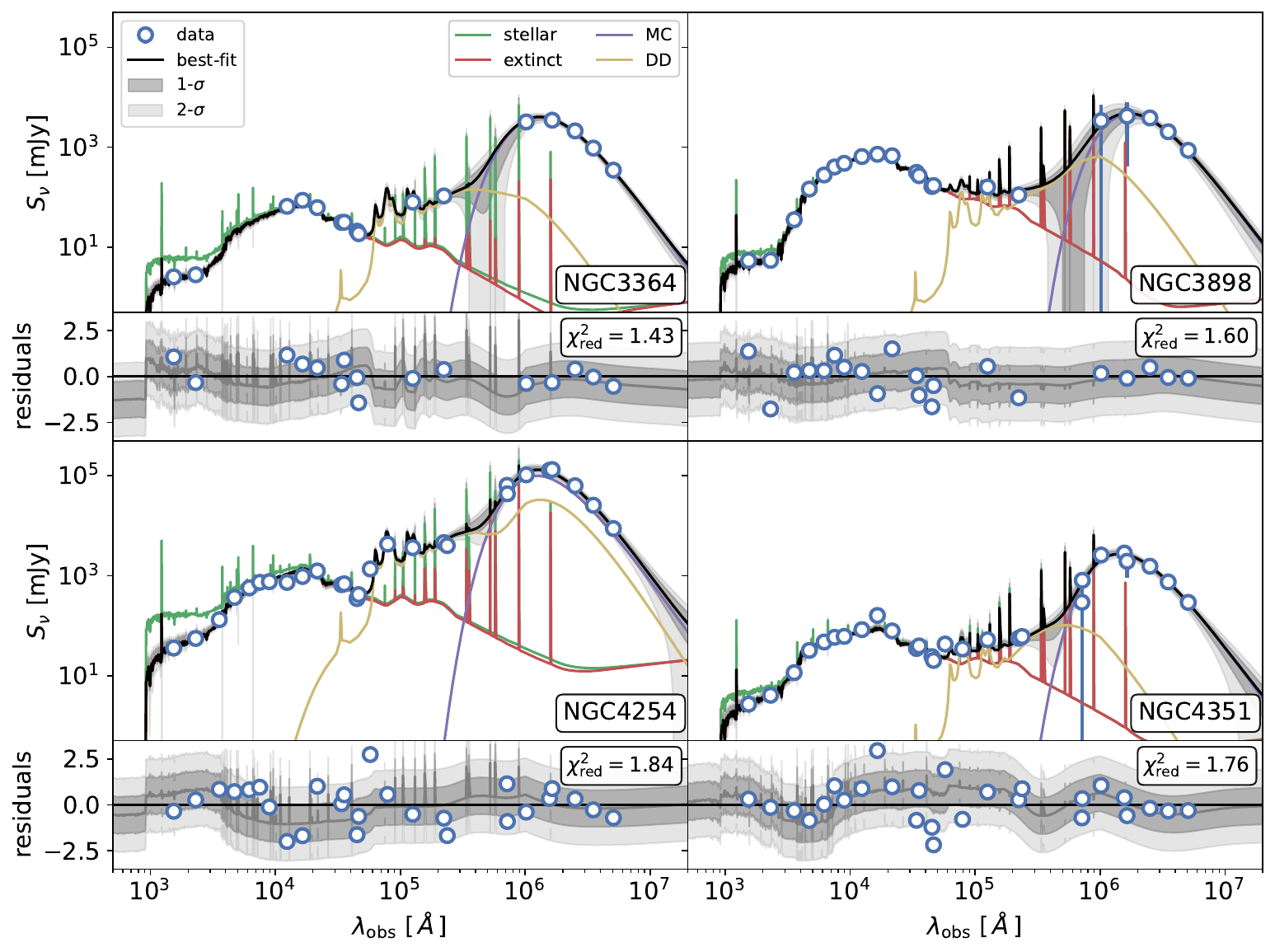}}
  \caption{Same as Fig.~\ref{fig:pantoni21} for the Dustpedia sample of Sec.~\ref{sec:casasola20}.}
\label{fig:ltg_sed}
\end{figure*}
The best-fitting model and $1$-/$2\text{-}\sigma$ confidence regions around the mean of the samples is compared against the multi-band photometry of the sources in Fig.~\ref{fig:ltg_sed} where, as usual, we also show the different contribution to the overall emission and the standardised residuals with respect to the best-fitting model and associated reduced $\chi^2$. 
By inspecting the residuals in Fig.~\ref{fig:ltg_sed} we can see that the estimated best-fit model correctly intercepts the data-points even though, the extremely small errors on the optical flux measurements tend to make the nuisance systematic error parameter converge around values $f_\text{sys} \lesssim 0.1$.

A comparison of the derived astrophysical properties of the galaxies in the sample are provided in Table~\ref{tab:validation1} and Table~\ref{tab:validation2}.
In particular, we show, for each source the median values with associated $68\%$ credible interval, measured from the weighted posteriors, and the best fitting value (between parenthesis below each row of values).
As expected, given the different shape of the delayed truncated SFH model used in the original work \citep{Bianchi2018}, the values for the SFR are in slight disagreement\footnote{A truncated SFH that does not drop to zero after truncation allows to both assume an early stage bulk of star formation and a late time constant stellar mass growth. This would in turn be similar to the approach we propose for ETGs in Section~\ref{sec:donevski23}, if after truncation a null SFR is assumed.}.
While the dust temperatures (Table~\ref{tab:validation1}) do agree within the error-bars, quantities that are more strictly related to the SFH model chosen (i.e. the age and SFR in Table~\ref{tab:validation1}, and the dust and gas masses in Table~\ref{tab:validation2}) deviate by more than 2-$\sigma$.
On the other hand, for the stellar masses (also shown in Table~\ref{tab:validation2}) the agreement between the results is restored, even though this quantity also depends on the SFH and galaxy age.
This last observation motivates us to advocate that \galapy has found a different solution for the most probable properties of the objects.

A special mention must be made for NGC4254, where we measure the largest discrepancy with literature.
In particular, while dust temperatures are still in agreement with the literature result, we measure values consistently higher by a factor $\approx5\div10$ for the other examined quantities.
In \cite{Hunt2019} the authors analyse a sample of objects, including NGC4254, with different SED fitting codes (i.e. MAGPHYS, CIGALE and GRASIL).
With a photometric system similar to the one used here, the authors find values slightly larger to the ones reported here in the \textit{Literature} columns of Tables~\ref{tab:validation1} and \ref{tab:validation2}.
Nonetheless, these are still in disagreement with our estimates for the same parameters.
We have nonetheless checked that the SFR $\approx20~M_\odot/$yr we find is still allowed by the upper limits imposed with empirical relations \citep{Lapi2011} connecting the object's flux in different bands with the SFR, with which we find $\text{SFR}_\text{NGC4254}\lesssim50~M_\odot/$yr.

Finally, we have also tried to run the analysis assuming a constant SFH model. 
The main differences worth to report are higher values for both the SFR and age, as a result of a SFH that, by construction, is more diluted in time.
Nevertheless, the constant model is statistically disfavoured with respect to the In-Situ model as, with a larger number of parameters, produces values of the likelihood that are consistently lower for each of the four sources.

\subsubsection[Lensed NIR-dark galaxy with upper limits]{Lensed NIR-dark galaxy with upper limits from \cite{Giulietti2022}}\label{sec:giulietti22}

As a further test-bench for \galapy, we run the photometric analysis on a lensed, NIR-dark galaxy studied in \cite{Giulietti2022}. 
HATLASJ113526.2-01460 (J1135 hereafter) was selected by \cite{Negrello2016} as a candidate lensed galaxy at redshift $z\sim3.1$ \citep{Harris2012} in the $12^\text{th}$ Gama field of the Herschel-ATLAS survey, and then confirmed as a lensed NIR-dark galaxy by \cite{Giulietti2022}.

The interest in testing \galapy on this object resides on the large number of photometric points that are flagged as upper limits.
As already mentioned in Sec.~\ref{sec:inference}, the method of choice for treating upper limits in \galapy is to consider them as regular points entering the same $\chi^2$ likelihood used for detections.
We assign to each of these points, marked as non-detections, a flux equal to zero and an error equal to the noise value measured in the broadband photometry.
We select the same hyper-parameters chosen for sources in Sec.~\ref{sec:dsfg}: \texttt{parsec22.ntl} SSPs and the In-Situ SFH model, with 10 free parameters.

\begin{figure}
\centering
\includegraphics[width=\hsize]{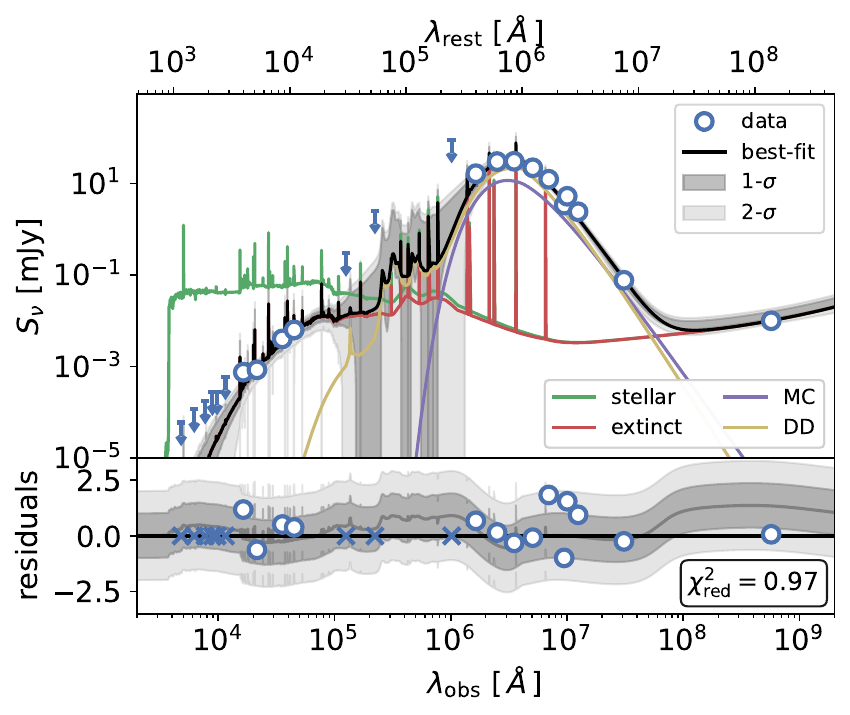}
\caption{Same as Fig.~\ref{fig:pantoni21} for the J1135 lensed galaxy of Sec.~\ref{sec:giulietti22}.}
\label{fig:j1135}
\end{figure}
We show the fitting results in Fig.~\ref{fig:j1135}, upper limits are marked as circles with arrows while detections are round markers with error bars.
In the lower panel we show standardised residuals and the $\chi^2$ associated to the best-fitting model.
As we have associated a flux value of zero to non-detections, we mark their corresponding positions with downward arrows at the 1-$\sigma$ value measured for noise in the upper panel and as crosses in the lower panell.

Besides the solid value of the reduced $\chi^2$ statistics, we can notice from the grey 2-$\sigma$ confidence contour in the upper panel how the NIR-MIR region of the spectrum is just upper-limited, as a result of having just observed upper limits in that part.
This is of course expected, as one of the free parameters of the model (i.e. the fraction of diffuse dust emission that is in PAH, $f_\text{PAH}$) is completely and only determined by measurements in the NIR to MIR.

We conclude this Section by highlighting the excellent agreement (Table~\ref{tab:validation1} and Table~\ref{tab:validation2}) between the parameter values derived with \galapy with respect to the values obtained in \cite{Giulietti2022}, where the analysis has been performed with different, but compatible, methods (i.e. by analysing the line emission properties of the source in the ALMA bands).
In particular, all the quantities do agree within the 1-$\sigma$ credible interval to the measurements obtained in \cite{Giulietti2022}, even though we find both a median and a best-fitting value for the SFR smaller than what found originally by the authors.
We manage to also obtain a more precise measure the age of the object and for the stellar mass, consistent with the upper limit imposed in the original work (i.e. $\log M_\star^\text{GalaPy}/M_\odot = 11.2_{-0.6}^{+0.4} < 11.7 = \log M_\star^\text{Giulietti}/M_\odot$).

Even though not shown in the present manuscript, we have tested the reliability of our treatment of upper limits by both modifying the values used to mark non-detections and by using the other methods presented in Sec.~\ref{sec:statistics} to include them in the likelihood.
Concerning the former test, we have assumed both fluxes equal to the noise measurement and equal to three times this measurement (what is usually referred to as 1- and 3-$\sigma$ upper limits).
We observe that the result tends to be biased towards higher values for the predicted fluxes in the regions where we only have non-detections (i.e. optical and NIR/MIR).
On the other hand, both running with the Sawicki method and with the naive method for the treatment of upper-limits guarantee convergence of the results but, besides requiring more samples (and therefore more time) to converge, the result tends to be statistically less solid, with values of the $\chi^2$ statistics consistently higher.
It is worth to mention that these considerations are not valid as a general test of the different possible approaches and serve solely has motivation for our final choice.
Users of the library should tune their choices on the specific problem at hand.

\subsubsection[Stacked NIR-dark radio selected galaxy with no spectroscopic redshift]{Stacked NIR-dark radio selected galaxy with no spectroscopic redshift from \cite{Behiri2023}}\label{sec:behiri23}

We validate the library on a case that, to some extent, represents a more extreme case of dust obscured photometry with non-detections.
In \cite{Behiri2023}, the authors study a radio-selected sample from the VLA-COSMOS 3 GHz Large Project \citep{Smolcic2017} catalogue, based on a survey covering $2.6\ \text{deg}^2$ in the COSMOS field.
Thanks to the extremely small value of its limiting flux density ($12.6\ \mu\text{Jy beam}^{-1}$ at $5.5\ \sigma$), the survey has delivered one of the deepest samples ever obtained.
Therefore, this data-set proves an ideal laboratory for estimating the contribution of galaxies at $z > 3$ to the cosmic SFRD.

\begin{figure}
\centering
\includegraphics[width=\hsize]{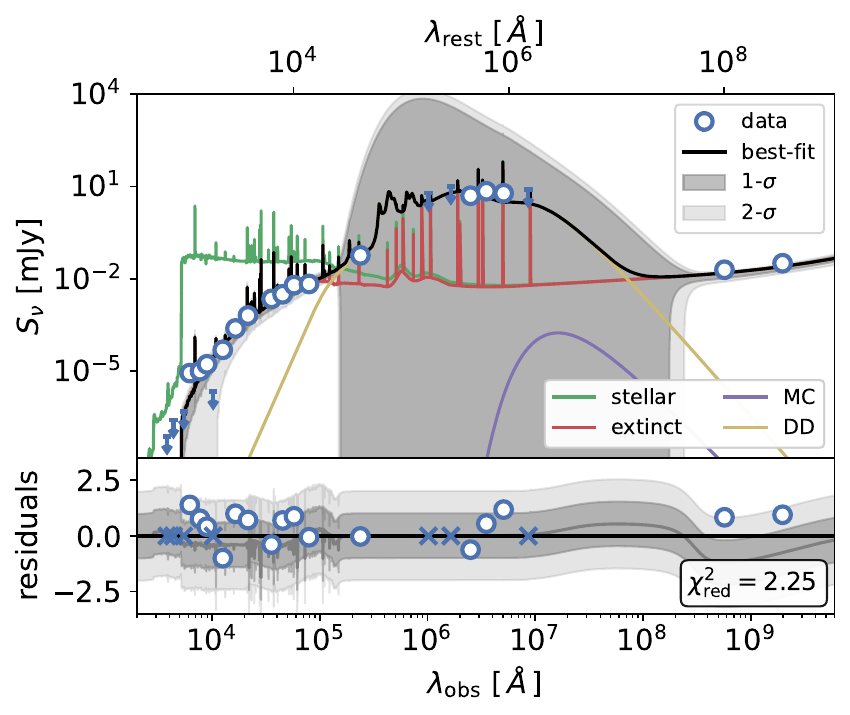}
\caption{Same as Fig.~\ref{fig:pantoni21} for the median of Sec.~\ref{sec:behiri23}.}
\label{fig:behiri23}
\end{figure}

In their work, the authors produce an ensemble analysis of the average sample properties, by performing SED fitting on the \textit{median} photometric properties of the sample.
The median SED (shown as round blue markers and downward arrows in Fig.~\ref{fig:behiri23}), is obtained by stacking the individual sources maps for all the bands where a source is detected, and applying survival analysis \citep{IsobeFeigelson1986} on all the bands where the presence of eventual upper limits has to be taken into account.
This procedure results in 16 bands flagged as detections (empty blue markers with error bar) and 7 bands flagged as upper limits (downward arrows), on an overall wavelength range spanning from $\sim 5\times10^3$ \AA\ to $2\times10^9$ \AA.
The redshift of this artificial median source is unknown, but is expected to be representative of the median redshift of the sample\footnote{The same argument also applies to the other physical properties.}.
The resulting photometry is then fitted using the MAGPHYS+\textsc{photo-z} code \citep{magphys2008, Battisti2019}, obtaining the results reported on the right half of Table~\ref{tab:validation1} and Table~\ref{tab:validation2}.

In running our fit with \galapy, we select the usual set-up with In-Situ SFH and PARSEC22 SSPs including nebular lines, free-free and synchrotron and we let 12 physical parameters vary, including redshift, and an additional systematic error parameter to account for eventual errors that might arise from the stacking procedure.
The best fitting SED with $1$- and $2$-$\sigma$ confidence regions around the mean of the samples are pictured in Fig.~\ref{fig:behiri23} with a black solid line and shaded grey regions, respectively.
The best-fit SED is decomposed into its different contributions, reported as coloured solid lines as listed in the legend.
The reduced value of the $\chi^2_\text{red} = 2.25$ is reported, along with the standardised residuals with respect to the best-fit model, in the lower panel of Fig.~\ref{fig:behiri23}.

We observe a large uncertainty in the thermal dust emission peak, which is a symptom of the large number of upper-limits in the corresponding part of the observed spectrum.
This level of uncertainty is reflected on the large uncertainty for the estimated temperatures of the two media, reported in Table~\ref{tab:validation1} which, compared to the value obtained with MAGPHYS, have fairly small values.
The largest difference between our results with those presented in the original work are though in the parameters mostly depending on the CSP, as this part is the most well sampled by the data-set at hand.
In particular, we measure an age that is less than two times smaller than the one obtained with MAGPHYS, even though the two estimates are in agreement within the $68\%$ credible interval.
The combined effect of this small value of the age and our weighted median photometric redshift estimate of $z_\text{\galapy} = 4.41_{-0.423}^{+0.34}$, determine the extremely large value obtained for the SFR (i.e. $>1400\ M_\odot/$yr, see Table~\ref{tab:validation1}).
Given the combination of age and redshift, this value has to be large to explain the $10$ mJy flux measurement at the IR peak.
As a term of comparison, the photometric redshift estimate obtained with MAGPHYS is $z_\text{MAGPHYS} = 3.25_{-0.11}^{+0.09}$.
Both values are consistent with the photometric redshift distribution of the sample from which the median fluxes have been obtained.
Note though that, even though the median value of this redshift distribution would be more consistent with the MAGPHYS estimate on the median fluxes, also the photometric redshift estimate of the individual sources in the original work have been obtained using MAGPHYS.
It would be interesting to apply our library on the whole sample but this, obviously, goes beyond the scope of this work.

Concerning the component masses instead, we are consistent with the values obtained by the authors of the original work.
Both the total mass in dust and the total stellar mass are consistent within the $68\%$ credible interval, even though the best-fitting values are different by some factors. 

In closing this section, we stress that even though obtained from stacking the fluxes of different sources, this semi-mock observation should embed the characteristics of the NIR-dark radio-selected population of galaxies at high redshift, from which it has been built.
Nonetheless note that, as already mentioned, a thorough comparison of the properties of the median fluxes with the median properties obtained from studying singular objects would require obtaining such individual properties with the same SED-fitting tool, which goes way beyond the objectives of this validation section.

\subsubsection[Quiescent galaxies]{Quiescent galaxies from \cite{Donevski2023}}\label{sec:donevski23}

In order to provide a consistency check that the In-Situ model delivered with \galapy correctly describes the SFH of ETG progenitors and their evolution towards becoming quiescent, we ultimately validate our SED fitting tool on three quiescent galaxies extracted from the parent sample of spectroscopically selected massive quiescent galaxies in the COSMOS field, presented in \cite{Donevski2023}.
Here we fit the deep optical-to-NIR fluxes of representative sources at different redshifts, spanning in the range from $z = 0.1$ to $z = 0.6$.
Testing on quiescent objects is a good test to check on the behaviour of the In-Situ model when the object is evolved, as this model has been derived to describe the progenitors of early type galaxies in all their evolution towards becoming quiescent. 

Along with the In-Situ model of SFH, we once again select the PARSEC22 SSP libraries and allow for 4 free parameters: the galaxy age, the two free parameters of the SFH model and an age corresponding to an eventual abrupt quenching, in order to simulate some violent feedback event cleaning up the galaxy from all its diffuse medium.

\begin{figure*}
\resizebox{\hsize}{!}{
\includegraphics{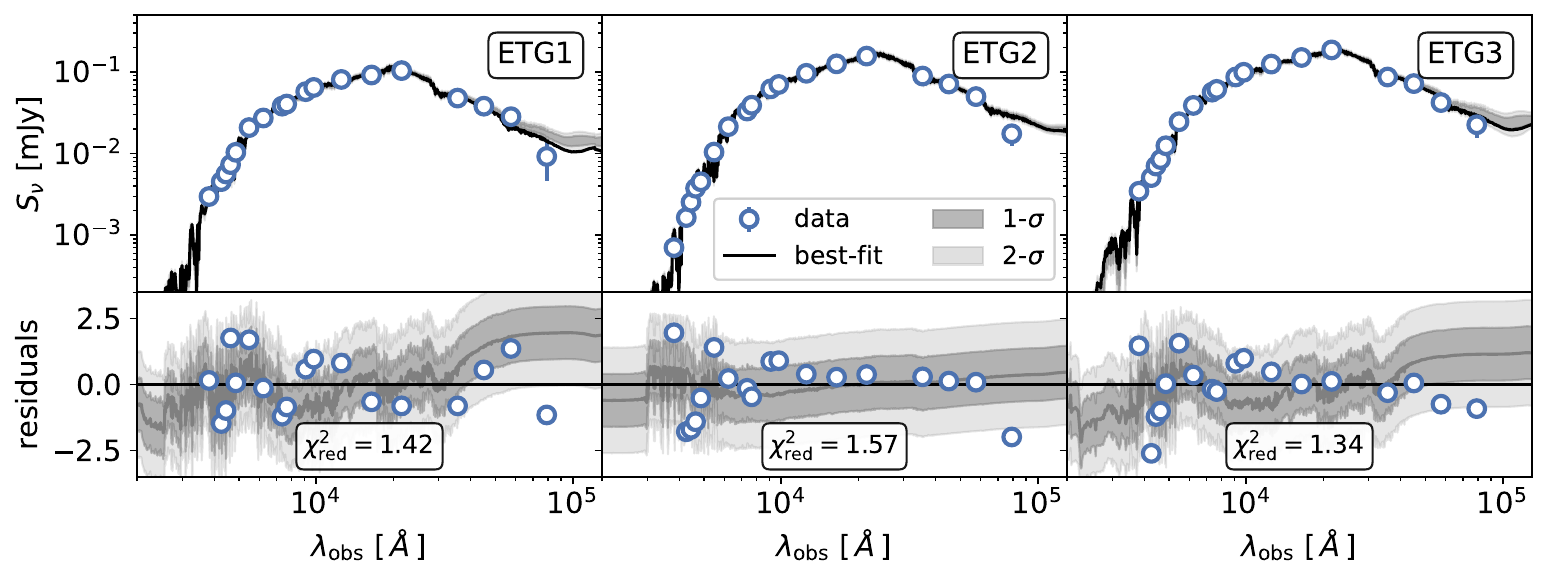}}
  \caption{Same as Fig.~\ref{fig:pantoni21} for the quiescent sample of Sec.~\ref{sec:donevski23}.}
\label{fig:etg_sed}
\end{figure*}
The best-fitting SEDs are shown in Fig.~\ref{fig:etg_sed} as well as $1$- and $2$-$\sigma$ confidence regions around the mean. 
In the lower panels we show the standardised residuals with respect to the best-fitting model and associated reduced $\chi^2$. 

By inspecting the residuals in Fig.~\ref{fig:etg_sed} we can see that the estimated best-fit model is correctly representing the data-set with $\chi_\text{red}^2$ values between $1.3$ and $1.6$, although the extremely small errors on the optical flux measurements tend to make the nuisance systematic error parameter converge around values $f_\text{sys} \lesssim 0.1$.
Also the results for the derived parameters are extremely consistent with those found in literature both in terms of age (Table~\ref{tab:validation1}) and stellar mass content (Table~\ref{tab:validation2}).
In the original work, the authors used a truncated delayed SFH which, in case after truncation the star formation drops to zero, has a functional form which can be easily emulated by our quenched In-Situ shape, as also anticipated in Sec.~\ref{sec:casasola20}. 

\subsubsection{Photometric redshift}\label{sec:photoz}

We perform an final validation test on our machinery by inferring an estimate for the photometric redshift of real sources.
We select the sources from Section~\ref{sec:dsfg}, \ref{sec:donevski23} and \ref{sec:giulietti22}, for which a measurement of the spectroscopic redshift is available.
We then sample again the parameter space by letting the \texttt{redshift} parameter vary along with the other free parameters.

\begin{figure}
\centering
\includegraphics[width=\hsize]{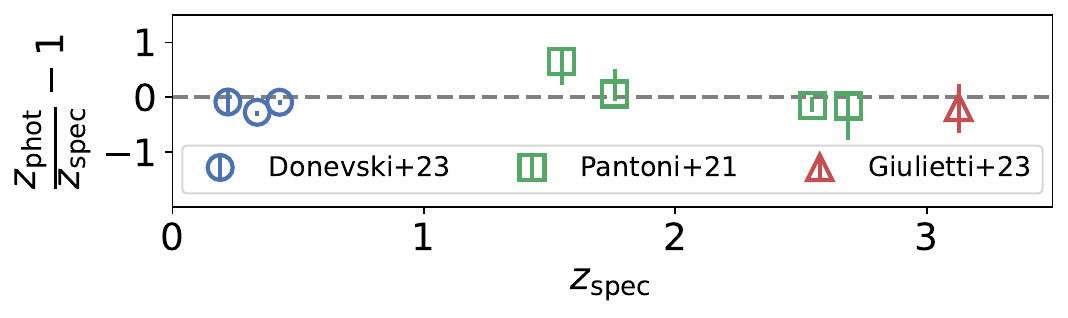}
\caption{Relative difference of photometric redshift estimated with \galapy against the real spectroscopic values for the 4 dusty star-forming galaxies of the \cite{Pantoni2021} sample (green squares), the 3 quiescent galaxies of the \cite{Donevski2023} sample (blue circles) and the lensed galaxy with upper limits from \cite{Giulietti2022} (red triangle). Markers with error bars trace the median and $68\%$ credible interval of the samples.}
\label{fig:zspecVSzphot}
\end{figure}
In Fig.~\ref{fig:zspecVSzphot} we compare the photometric redshift prediction to the real value measured spectroscopically, in terms of the relative redshift difference between estimated ($z_\text{phot}$) and fiducial value ($z_\text{spec}$).
We use coloured markers with error bars for the median value and $68\%$ credible interval of the samples.
The dashed grey line marks the real spectroscopic value.

For all the sources, the median values show at most a 2-$\sigma$ difference with respect to the spectroscopic measurement (i.e. the fiducial value is within the $95\%$ credible interval).
Apart from the low redshift sources and most of the P21 sources, the photometric prediction of the J1135 redshift from \cite{Giulietti2022} is extremely close to the expected value (order of percent relative difference).
This is a remarkable result, considering the large number of fluxes for which only upper limits are available, especially in the UV/optical part of the spectrum.
The reason for this agreement lays on the interplay between the thorough sampling of the dust peak and the precision modelling allowed by our two-component, age-dependent dust model.

These results confirm on real sources the reliability of photometric redshift estimates obtained with \galapy, as already demonstrated on mock sources.
This is an asset that will prove powerful for future observational campaingns targeting distant sources (e.g. JWST).
In the next future, we also plan to further test the photometric redshift determination capabilities of \galapy against large data-sets up to the highest redshifts currently available, e.g., A3COSMOS \citep{A3COSMOS-I-2019,A3COSMOS-II-2019,A3COSMOS-III-2020} and COSMOS-Web \citep{COSMOS-Web2022}.

%% file: tab/mock_sources_summary.tex
\begin{tabular}{l|cc}
\toprule\midrule
SFH model & active & passive \\
\midrule
In-Situ &  20+20 & 20+20 \\
Delayed Exponential & 20+20 & 20+20 \\
Constant & 20+20 & --- \\
\bottomrule\bottomrule
\end{tabular}

%% file: tab/mock_sources_priors.tex
\begin{tabular}{lr|lr|lr}
\toprule\midrule
\multicolumn{2}{c|}{In-Situ} & \multicolumn{2}{c|}{Delayed Exponential} & \multicolumn{2}{c}{Constant} \\\midrule
\multicolumn{6}{c}{Actively star-forming} \\\midrule
Parameter & Prior & Parameter & Prior & Parameter & Prior \\\midrule
$\log$ age/yr                       & $[7,9]$ & $\log$ age/yr                        & $[7,9]$  & $\log$ age/yr                & $[7,9]$ \\
redshift                            & $[2,8]$ & redshift                             & $[2,8]$  & redshift                     & $[2,8]$ \\
$\log~\psi_\text{max}/\dot M_\odot$ & $[0,4]$ & $\log~\psi_\text{norm}/\dot M_\odot$ & $[\text{-}2,2]$ & $\log~\psi/\dot M_\odot$     & $[\text{-}2,2]$ \\
$\log~\tau_\star$/yr                & $[7,9]$ & $\log~M_\text{dust}/M_\odot$         & $[7,10]$ & $\log~M_\text{dust}/M_\odot$ & $[7,9]$ \\
$f_\text{MC}$                       & $[0,1]$ & $f_\text{MC}$                        & $[0,1]$  & $f_\text{MC}$                & $[0,1]$ \\
$\log~N_\text{MC}$                  & $[0,5]$ & $\log~N_\text{MC}$                   & $[0,5]$  & $\log~N_\text{MC}$           & $[0,5]$ \\
$\log~R_\text{MC}$/pc               & $[0,5]$ & $\log~R_\text{MC}$/pc                & $[0,5]$  & $\log~R_\text{MC}$/pc        & $[0,5]$ \\
$\log~\tau_\text{esc}$/yr           & $[4,8]$ & $\log~\tau_\text{esc}$/yr            & $[4,8]$  & $\log~\tau_\text{esc}$/yr    & $[4,8]$ \\
$\log~R_\text{DD}$/pc               & $[0,5]$ & $\log~R_\text{DD}$/pc                & $[0,5]$  & $\log~R_\text{DD}$/pc        & $[0,5]$ \\
$f_\text{PAH}$                      & $[0,1]$ & $f_\text{PAH}$                       & $[0,1]$  & $f_\text{PAH}$               & $[0,1]$ \\ \midrule
\multicolumn{6}{c}{Passively evolving} \\\midrule
Parameter & Prior & Parameter & Prior & \multicolumn{2}{c}{---} \\\midrule
$\log$ age/yr                       & $[9.2,11]$ & $\log$ age/yr                        & $[9.2,11]$ & & \\
redshift                            & $[0,2]$    & redshift                             & $[0,2]$    & & \\
$\log~\tau_\text{quench}$/yr        & $[8,9]$    & $\log~\tau_\text{quench}$/yr         & $[8,9]$    & & \\
$\log~\psi_\text{max}/\dot M_\odot$ & $[2,4]$    & $\log~\psi_\text{norm}/\dot M_\odot$ & $[\text{-}2,2]$   & & \\
$\log~\tau_\star$/yr                & $[7,9]$    & $\log~\tau_\star$/yr                 & $[7,9]$    & & \\
\bottomrule\bottomrule
\end{tabular}

%% file: tab/real_sources1.tex
\begin{tabular}{l | c c c c | c c c}
\toprule
 & \multicolumn{4}{c|}{\galapy} & \multicolumn{3}{c}{Literature}\\
Source & $\log$~age & SFR & $T_\text{MC}$ & $T_\text{DD}$ & $\log$~age & SFR & $T_\text{dust}$\\
& $[\log~\text{yr}]$ & $[M_\odot\,\text{yr}^{-1}]$ & $[\text{K}]$ & $[\text{K}]$ & $[\log~\text{yr}]$ & $[M_\odot\,\text{yr}^{-1}]$ & $[\text{K}]$\\\midrule \midrule 
\multicolumn{8}{l}{Sec.~\ref{sec:dsfg}: dusty star forming (1)}\\ \midrule
UDF3       & $7.58_{-0.51}^{+0.72}$ & $555.80_{-89.86}^{+164.87}$ & $53.44_{-2.76}^{+6.56}$  & $18.49_{-9.03}^{+19.23}$  & $8.37_{-0.10}^{+0.08}$ & $519\pm38$ & $73\pm4$ \\
           & $(7.35)$               & $(702.42)$                  & $(55.47)$                & $(5.54)$                  &                        &            &          \\
UDF10      & $9.41_{-0.08}^{+0.06}$ & $22.88_{-5.96}^{+6.64}$     & $32.04_{-9.95}^{+20.52}$ & $36.37_{-10.11}^{+19.41}$ & $8.96_{-0.07}^{+0.06}$ & $41\pm5$   & $46\pm7$ \\
           & $(9.50)$               & $(18.76)$                   & $(18.69)$                & $(50.12)$                 &                        &            &          \\
UDF11      & $8.85_{-0.19}^{+0.14}$ & $185.70_{-16.94}^{+25.00}$  & $55.10_{-3.49}^{+6.36}$  & $39.53_{-8.57}^{+11.86}$  & $8.58_{-0.11}^{+0.08}$ & $241\pm19$ & $69\pm4$ \\
           & $(8.80)$               & $(182.34)$                  & $(52.61)$                & $(46.84)$                 &                        &            &          \\
AzTEC.GS21 & $8.35_{-0.23}^{+0.20}$ & $335.05_{-58.79}^{+56.67}$  & $52.61_{-3.14}^{+3.68}$  & $16.26_{-4.91}^{+5.16}$   & $8.87_{-0.07}^{+0.06}$ & $360\pm18$ & $63\pm3$ \\  
           & $(7.99)$               & $(241.24)$                  & $(50.22)$                & $(10.55)$                 &                        &            &          \\
\midrule
\multicolumn{8}{l}{Sec.~\ref{sec:casasola20}: local late type (2)}\\ \midrule
NGC3364 & $9.92_{-0.18}^{+0.12}$ & $0.76_{-0.08}^{+0.09}$  & $25.25_{-2.92}^{+13.08}$ & $21.71_{-2.54}^{+33.15}$ & --- & $1.64\pm0.23$ & $22.05\pm1.14$ \\
        & $(10.03)$              & $(0.84)$                & $(33.10)$                & $(36.30)$                & & & \\
NGC3898 & $9.91_{-0.11}^{+0.07}$ & $0.16_{-0.03}^{+0.04}$  & $20.39_{-3.80}^{+16.75}$ & $16.94_{-2.94}^{+14.10}$ & --- & $1.22\pm0.18$ & $16.28\pm4.87$ \\
        & $(9.91)$               & $(0.19)$                & $(18.33)$                & $(37.58)$                & & & \\
NGC4254 & $9.96_{-0.12}^{+0.09}$ & $20.34_{-1.43}^{+1.47}$ & $28.22_{-3.61}^{+13.49}$ & $24.11_{-3.50}^{+8.70}$  & --- & $5.15\pm0.51$ & $24.82\pm0.65$ \\ 
        & $(9.90)$               & $(18.72)$               & $(29.39)$                & $(21.75)$                & & & \\
NGC4351 & $9.35_{-0.13}^{+0.20}$ & $0.48_{-0.07}^{+0.06}$  & $23.29_{-1.48}^{+2.98}$  & $23.11_{-7.08}^{+34.68}$ & --- & $0.11\pm0.02$ & $21.06\pm1.42$ \\
        & $(9.26)$               & $(0.45)$                & $(22.40)$                & $(67.13)$                & & & \\\midrule 
\multicolumn{8}{l}{Sec.~\ref{sec:giulietti22}: lensed NIR-dark with upper-limits (3)}\\ \midrule
J1135 & $8.34_{-0.60}^{+0.43}$ & $815.97_{-159.43}^{+195.03}$ & $51.44_{-22.38}^{+32.67}$ & $44.01_{-6.58}^{+3.62}$ & --- & $933.25_{-157.01}^{+345.77}$ & $37.7\pm1.5$ \\
& $(8.55)$ & $(641.09)$ & $(45.40)$ & $(48.15)$ & $(8.0)$ & & \\\midrule
\multicolumn{8}{l}{Sec~\ref{sec:behiri23}: stacked NIR-dark radio selected (4)}\\ \midrule
median & $7.96_{-0.28}^{+0.39}$ & $1475.12_{-558.84}^{+442.71}$ & $9.51_{-6.58}^{+21.54}$ & $15.29_{-4.65}^{+17.55}$ & $8.38_{-0.22}^{+0.41}$ & $395.37_{-145.91}^{+58.57}$ & $61.35_{-20.4}^{+4.8}$ \\ 
& $(7.94)$ & $(1838.94)$ & $(11.05)$ & $(19.52)$ & $(8.38)$ & $(430)$ & $(61.45)$ \\\midrule
\multicolumn{8}{l}{Sec~\ref{sec:donevski23}: quiescent (5)}\\ \midrule
ETG1 & $9.46_{-0.03}^{+0.36}$ & $0.0$ & --- & --- & $9.45$ & $10^{-7}$ & --- \\ 
& $(9.82)$ & --- & --- & --- & & & \\
ETG2 & $9.62_{-0.07}^{+0.04}$ & $0.0$ & --- & --- & $9.92$ & $10^{-5}$ & --- \\ 
& $(9.60)$ & --- & --- & --- & & & \\
ETG3 & $9.41_{-0.08}^{+0.32}$ & $0.0$ & --- & --- & $9.80$ & $3\cdot10^{-5}$ & --- \\
& $(9.75)$ & --- & --- & --- & & & \\
\bottomrule\bottomrule
\end{tabular}

%% file: tab/real_sources2.tex
\begin{tabular}{l | c c c c c | c c c c}
\toprule
& \multicolumn{5}{c|}{\galapy} & \multicolumn{4}{c}{Literature}\\
Source & $\log~M_\text{dust}$ & $\log~M_\text{gas}$ & $\log~M_\star$ & $Z_\text{gas}$ & $Z_\star$ & $\log~M_\text{dust}$ & $\log~M_\text{gas}$ & $\log~M_\star$ & $Z_\text{gas}$\\
& $[\log~M_\odot]$ & $[\log~M_\odot]$ & $[\log~M_\odot]$ & & & $[\log~M_\odot]$ & $[\log~M_\odot]$ & $[\log~M_\odot]$ & \\\midrule \midrule  
\multicolumn{10}{l}{Sec.~\ref{sec:dsfg}: dusty star forming (1)}\\ \midrule
UDF3       & $8.52_{-0.64}^{+0.60}$ & $10.24_{-1.07}^{+0.91}$ & $10.19_{-0.24}^{+0.50}$ & $0.03_{-0.02}^{+0.04}$ & $0.02_{-0.01}^{+0.02}$ & $8.30_{-0.22}^{+0.15}$ & --- & $10.95_{-0.05}^{+0.05}$ & --- \\
           & $(8.31)$                 & $(9.95)$                  & $(9.98)$                  & $(0.03)$                 & $(0.02)$                 & & $(10.6)$ & & $(0.01)$ \\
UDF10      & $8.66_{-0.27}^{+0.27}$ & $10.28_{-0.34}^{+0.49}$ & $10.60_{-0.07}^{+0.05}$ & $0.03_{-0.01}^{+0.01}$ & $0.02_{-0.01}^{+0.00}$ & $7.98_{-0.53}^{+0.23}$ & --- & $10.40_{-0.06}^{+0.05}$ & --- \\
           & $(8.64)$                 & $(10.24)$                 & $(10.65)$                 & $(0.03)$                 & $(0.02)$                 & & $(10.5)$ & & $(0.01)$ \\
UDF11      & $9.25_{-0.14}^{+0.09}$ & $11.50_{-0.43}^{+0.27}$ & $10.71_{-0.09}^{+0.08}$ & $0.01_{-0.00}^{+0.01}$ & $0.01_{-0.00}^{+0.00}$ & $7.86_{-0.26}^{+0.16}$ & --- & $10.81_{-0.07}^{+0.06}$ & --- \\
           & $(9.19)$                 & $(11.53)$                 & $(10.64)$                 & $(0.01)$                 & $(0.00)$                 & & $(10.3)$ & & $(0.01)$ \\
AzTEC.GS21 & $8.84_{-0.31}^{+0.28}$ & $10.16_{-0.40}^{+0.42}$ & $10.93_{-0.17}^{+0.18}$ & $0.06_{-0.02}^{+0.01}$ & $0.04_{-0.01}^{+0.01}$ & $8.46_{-0.12}^{+0.09}$ & --- & $11.26_{-0.05}^{+0.05}$ & --- \\  
           & $(8.22)$                 & $(9.41)$                  & $(10.81)$                 & $(0.08)$                 & $(0.05)$                 & & $(10.8)$ & & $(0.02)$ \\\midrule
\multicolumn{10}{l}{Sec.~\ref{sec:casasola20}: local late type (2)}\\ \midrule
NGC3364 & $7.17_{-0.22}^{+0.15}$ & $8.97_{-0.27}^{+0.18}$ & $10.03_{-0.08}^{+0.07}$ & $0.02_{-0.00}^{+0.00}$ & $0.02_{-0.00}^{+0.00}$ & $7.10_{-0.10}^{+0.08}$ & --- & $10.24_{-0.22}^{+0.15}$ & --- \\
& $(7.37)$ & $(9.21)$ & $(10.05)$ & $(0.02)$ & $(0.02)$ & & & & \\
NGC3898 & $6.29_{-0.14}^{+0.16}$ & $7.89_{-0.15}^{+0.18}$ & $10.40_{-0.07}^{+0.05}$ & $0.03_{-0.00}^{+0.00}$ & $0.03_{-0.00}^{+0.00}$ & $7.72_{-0.12}^{+0.09}$ & --- & $11.32_{-0.03}^{+0.28}$ & --- \\
& $(6.38)$ & $(7.99)$ & $(10.39)$ & $(0.03)$ & $(0.03)$ & & $(9.14)$ & & \\
NGC4254 & $9.11_{-0.14}^{+0.14}$ & $10.68_{-0.17}^{+0.20}$ & $11.13_{-0.08}^{+0.07}$ & $0.04_{-0.00}^{+0.00}$ & $0.02_{-0.00}^{+0.00}$ & $7.35_{-0.02}^{+0.02}$ & --- & $10.13_{-0.08}^{+0.07}$ & --- \\ 
& $(8.95)$ & $(10.48)$ & $(11.15)$ & $(0.04)$ & $(0.03)$ & & $(9.53)$ & & $(0.01)$ \\
NGC4351 & $6.30_{-0.20}^{+0.25}$ & $7.98_{-0.24}^{+0.28}$ & $9.77_{-0.05}^{+0.09}$ & $0.03_{-0.00}^{+0.00}$ & $0.02_{-0.00}^{+0.00}$ & $5.98_{-0.10}^{+0.08}$ & --- & $9.16_{-0.00}^{+0.00}$ & --- \\ 
& $(6.17)$ & $(7.82)$ & $(9.75)$ & $(0.03)$ & $(0.02)$ & & $(7.92)$ & & $(0.01)$ \\\midrule
\multicolumn{10}{l}{Sec.~\ref{sec:giulietti22}: lensed NIR-dark with upper-limits (3)}\\ \midrule
J1135 & $9.28_{-0.62}^{+0.47}$ & $10.70_{-0.72}^{+0.65}$ & $11.15_{-0.57}^{+0.39}$ & $0.06_{-0.03}^{+0.02}$ & $0.03_{-0.02}^{+0.01}$ & $9.06\pm0.04$ & $11.04\pm0.04$ & $\lesssim 11.73$ & --- \\ 
& $(9.34)$ & $(10.62)$ & $(11.40)$ & $(0.07)$ & $(0.04)$ & & & & \\\midrule
\multicolumn{10}{l}{Sec.~\ref{sec:behiri23}: stacked NIR-dark radio selected (4)}\\ \midrule
median & $9.26_{-0.47}^{+0.39}$ & $11.01_{-0.91}^{+0.59}$ & $10.94_{-0.25}^{+0.24}$ & $0.03_{-0.01}^{+0.04}$ & $0.02_{-0.01}^{+0.02}$ & $8.91_{-0.60}^{+0.08}$ & --- & $11.08_{-0.15}^{+0.19}$ & --- \\ 
& $(9.35)$ & $(11.39)$ & $(10.86)$ & $(0.01)$ & $(0.01)$ & $(8.06)$ & --- & $(11.09)$ & --- \\\midrule
\multicolumn{10}{l}{Sec.~\ref{sec:donevski23}: quiescent (5)}\\ \midrule
ETG1 & --- & --- & $9.97_{-0.07}^{+0.17}$ & --- & $0.04_{-0.02}^{+0.01}$ & --- & --- & $9.90$ & --- \\
 & --- & --- & $(10.15)$ & --- & $(0.02)$ & & & & \\
ETG2 & --- & --- & $10.82_{-0.07}^{+0.10}$ & --- & $0.05_{-0.01}^{+0.00}$ & --- & --- & $10.73$ & --- \\
& --- & --- & $(10.92)$ & --- & $(0.05)$ & & & & \\
ETG3 & --- & --- & $10.48_{-0.94}^{+0.16}$ & --- & $0.04_{-0.02}^{+0.01}$ & --- & --- & $10.85$ & --- \\
& --- & --- & $(9.35)$ & --- & $(0.01)$ & & & & \\
\bottomrule\bottomrule
\end{tabular}

%% file: sec/summary.tex
\section{Summary}\label{sec:summary}

We have presented \galapy, a highly optimised, open-source, hybrid library for parameterised fitting of the Spectral Energy Distribution (SED) of galaxies. 
The tool currently focuses on photometric SED fitting from galaxies, but future versions will extend its functionalities to include spectroscopic fitting at variable resolutions and AGN modelling. 
The API is readily available through terminal entry-points or by importing modules from the \texttt{galapy} package. 
The full documentation, including examples and API usage manual, is available on ReadTheDocs, and the code is available on GitHub.

In Sec.~\ref{sec:models}, we provided a detailed description of the physical models implemented in \galapy, with a particular focus on the In-Situ Star Formation Histories and the two-component age-dependent dust model. 
The former provides a model for the evolution of the extended structure components of a galaxy that depends on both the infall of material in the DM halo and on the evolution of the nuclear regions, driven by the central black hole \citep{Lapi2018,Pantoni2019,Lapi2020}.
The latter, provides a physically motivated model of the time evolution of dust, with overall attenuation directly derived by the contribution of each single simple stellar population hosted in the galaxy.
Additionally, \galapy uses an age-dependent, energy-conservation scheme to derive the evolution of dust temperatures in an analytic way.

In Sec.~\ref{sec:inference}, we described the statistical tools used to obtain parameter posteriors. 
The parameter space sampling is based on Bayesian inference methods, and we provide interfaces to two samplers, \texttt{emcee} and \texttt{dynesty}.

In Sec.~\ref{sec:validation}, we demonstrated the efficacy of \galapy by testing it on various cases, including dusty star-forming galaxies at high redshift, local late-type and early-type galaxies, and a NIR-dark, lensed high-redshift galaxy with mostly upper limits.
We showed that \galapy can be used to study the main physical characteristics of galaxies, such as their star formation histories, matter content, and physical parameters.

Future extensions of \galapy include spectroscopic fitting and Hamiltonian parameter space sampling, as well as a hierarchical Bayesian scheme for modelling data-sets from large catalogues with correlated systematic errors \citep[see, e.g.,][]{Kelly2012,Galliano2018}. 
Additionally, a consistent modelling of the AGN within the BH-galaxy co-evolution In-Situ scenario will be introduced soon. 
Finally, we plan to accelerate posterior inference using active learning techniques.

In conclusion, \galapy is a timely and valuable tool for the astrophysical community that offers a powerful, self-consistent framework for modelling the SED of galaxies, based on physically-motivated models and a Bayesian statistical approach (in Appendix~\ref{apx:logo} we provide recommendations on how to properly acknowledge usage of the library). 
The physical models implemented in \galapy, together with the optimisations made to the fitting algorithms, enable the tool to provide robust and accurate parameter estimates for a wide range of astrophysical applications. 

The main characterising features of \galapy are
\begin{itemize}
    \item self-consistent modelling of the SFH and derived physical properties that not only reduces the size of the parameter space, but it also allows for a straightforward derivation of the physical properties characterising the galaxy and is specifically designed to follow the evolution of high redshift progenitors up to their quiescence, leading to the formation of local early type galaxies;
    \item two component time-dependent energy-conserving treatment of dust attenuation and re-radiation that allows for both a physical treatment of the process without assuming unknown physics of the dust-grain and for a computationally-efficient balancing of energy;
    \item high resolution integration of stellar populations for the study of primordial galaxies which does not burden the computation thanks to a memory-efficient caching of the SSP grid (thoroughly treated in Appendix~\ref{apx:register});
    \item easily extensible database of cosmological models, SSP libraries and photometric band-pass filters;
    \item user-friendly API and extensive documentation, allowing for high level of customisation;
    \item state-of-the-art hybrid C++/Python implementation, reaching high performances with minimal memory consumption;
    \item Bayesian framework for the inference of posteriors in the parameter space.
\end{itemize}
As current and upcoming observational campaigns (e.g. JWST, LSST, SKAO) continue to generate ever-increasing amounts of data, the capabilities of \galapy will become increasingly important for understanding the physical properties of galaxies, especially in the high redshift Universe, and their evolution over cosmic time.

%% file: sec/code_design.tex
\section{Code design}\label{apx:codedesign}

In this Section we provide insights on the design choices made both for optimising the performances of the library and with the intent of keeping the structure modular and user friendly.
The bulk of the library resides in the computation of the parameterised models described in Sec.~\ref{sec:models}. 
Given that the main (but not only) purpose of \galapy is to provide a lightning fast tool for parameter inference, these functions have been implemented to reach high performances on a single core.
We reach this requirement by exploiting different advanced programming techniques, from register proximity, minimisation of operations and interpolation tactics. 
A description of the chosen strategies is provided in Sec.~\ref{apx:strategy}, Sec.~\ref{apx:api} showcases briefly the main modules and subpackages building up the library, while in Sec.~\ref{apx:performances} we show some loose performance measurements of the main functionalities deployed in \galapy.
All the performance measurements presented in this Section have been obtained by running on a Intel i9-10885H CPU @ 2.40GHz personal computer with a \texttt{x86\_64} architecture.
The cache available to CPUs is of average size for modern machines, it has 8 private instances of L1 with 32 KB of memory per instance, 8 private instances of L2 with 256 KB per instance and 1 shared instance of L3 with 2 MB of memory.

\subsection{Implementation strategy}\label{apx:strategy}

\galapy has a hybrid implementation which allows us to exploit both the performance efficiency of a compiled language (C++) as well as the flexibility of an interpreted language (Python).
The Bushido of \galapy software development can be summarised in few points:
\begin{itemize}
    \item compiled Object-Oriented \textbf{C++} crunches all the modelling framework, constituted of complex mathematical relations that burden the computation.
    The physical components described in Sec.~\ref{sec:models} are implemented as independent classes, all of which share a common interface for parameter setting and computation of the eventual emission as a function of wavelength, age and metallicity.
    At construction time, all the quantities that do not depend on the free parameters of the given component are computed in advance and cached, therefore minimising the amount of operations the machine has to perform.
    Modelling though is still extremely light on the RAM, as the volatile memory occupied by this cached information does not go beyond few tens of MB, mostly depending on the size of the SSP library chosen.
    This choice represents a compromise between the acceleration provided by SED grid interpolation and the flexibility of on-the-fly model computation.
    All of these objects can be serialised (i.e. converted in a sequential string of bits), allowing them to be picklable, therefore completely Python-compliant.
    Besides from physical models, the compiled sector of the library also implements a set of data-structures and algorithms for speeding up operations (in Appendix~\ref{apx:interp} we describe the linear interpolation scheme we use in some parts of the library).
    Finally, the compiled sector also manages loading the SSP libraries used to compute stellar emission, this is done to favour CPU cache management as described in Appendix~\ref{apx:register}.
    The only C++ library used is the STL, therefore minimising the problems that might arise in the installation of the package on different systems.
    
    \item \textbf{Python} deals with the interplay between all the components and modules, internal and external, that build up the library. 
    It also provides the user-interface and an extensive documentation.
    Lastly, the terminal commands allowing for quick-access parameterised SED-fitting that come out-of-the-box with library installation (e.g. the \texttt{galapy-fit} command mentioned in Sec.~\ref{sec:demo}) are implemented as Python entry-points.
    By importing the \texttt{galapy} package and sub-packages the \galapy API is exposed, allowing for complete customisation of the algorithms as well as providing the tools for astrophysical modelling and analysis of the sampling results.
    
    \item \textbf{pybind11} is a library to generate Python bindings of compiled C++ code.
    We bind compiled classes and expose our optimised C++ implementation to the Python interface providing access to our functions to users.
    All the functions that can be applied to arrays of values are vectorised, providing a straightforward integration with the most common python packages for scientific computing (e.g. NumPy and SciPy) and therefore allow for array programming.
    We have chosen this strategy because, compared to a CPython wrapping layer, it delivers bindings with negligible latency while providing a more intuitive interface. 
    
\end{itemize}

The primary purpose of \galapy is to derive the parameters that can be inferred from the spectral properties of galaxies.
Our code aims at delivering a high performance serial implementation of parametric SEDs, so that parallelism is not necessary in model generation (some performance testing is shown in Appendix~\ref{apx:performances}).
In this way, the only bottleneck of the work-flow is parameter-space sampling.

Both \texttt{emcee} and \texttt{dynesty} allow for passing a pool of workers to the functions running the sampling.
In \galapy we generate pools exploiting the \texttt{multiprocessing} package of the Python standard library.
Because of the structural limits of Python (i.e. the existence of a Global Interpreter Lock that guarantees parallel threads are not modifying concurrently the reference count in the Python interpreter), allocating a pool of parallel workers, with the intent of speeding-up CPU-bound workflows, requires to generate a copy of the environment.
Copying the whole environment though, results in the necessity of generating deep copies of all the variables that can be referenced in a given scope.
This not only means a larger memory usage, but it also reduces the effectiveness of shared memory parallelism, as passing around chunks of memory slows down severely the computation.

In the entry-point provided for fitting SEDs (i.e. \texttt{galapy-fit}) the default behaviour tries to reach a compromise between memory usage and parallelism. 
The variables that require the larger memory budget (e.g. the SSP libraries and the parametric models) are made global, therefore accessible for all the workers in the pool.
In the meantime, we spawn as many workers as possible to squeeze all the computing power from the architecture.

In future extensions of the library we will investigate more in parallelism and speed-up of the sampling.
We are also considering to implement our own specialised sampler and to test compiled sampling interfaces, that could possibly provide more control on the memory management as well as on the parallel exploitation of CPUs.

\subsubsection{Ordering of the SSP tables and computation of the intrinsic stellar luminosity}\label{apx:register}

A frequently overlooked aspect in scientific software development is the process behind RAM usage and, specifically, the way chunks of data loaded in the volatile memory reach the CPU for usage.
To simplify, sequential data is cached on a hierarchy of memory slots with given size.
The hierarchy ladder is set by the physical proximity of the memory slots to the CPU performing the computation, the closer the higher.
CPUs can access directly only the highest levels of this hierarchy, called registers, which can host a small number of bytes (typically, the amount corresponding to a few floating point numbers).

Registers tend to remain full-filled all the time, meaning that if the CPU needs a number from memory which is not already in one of the registers, this must firstly be emptied and then filled with the number required together with all the numbers which are close to it in memory, until complete occupation.
This process takes also place for the lower levels in the hierarchy ladder, namely, caches.
Tipically, modern computers own three levels of cache, L1 (tens of kBs) and L2 (hundreads of kBs) are private to each CPU in the processor, while L3 (tens of MBs) is shared between all the CPUs.
Since the process of moving cached data from lower to higher levels of cache, up to the registers, is time-consuming, it is desirable that data used in logically sequential operations are also stored sequentially in memory.

This is the reason behind our custom format for storage of SSP libraries, as the operation in \galapy that makes the most massive usage of cached data is integration of SSPs to generate CSPs (Sec.~\ref{sec:stellar}).
Computing Eq.~\eqref{eq:Lunatt} requires to perform, for each wavelength, an integral in time and an interpolation in metallicity.
This can be easily approximated with a linear integration in time and a linear interpolation in metallicity.
The approximated and implemented version of Eq.~\eqref{eq:Lunatt} reads
\begin{multline}\label{eq:LCSP_numerical}
    L_\text{CSP}(\lambda_i, \tau_\text{GXY}) = \sum_{\forall~j~>~0~|~\tau_j~\lesssim~\tau_\text{GXY}} \frac{\tau_j - \tau_{j-1}}{2} \times \\ 
    \times \biggl\{ \psi(\tau_j)~P_{L_\text{SSP}}^{(1)}\left[\lambda_i, \tau_j, Z_\star(\tau_\text{GXY} - \tau_j)\right] +\\
    + \psi(\tau_{j-1})~P_{L_\text{SSP}}^{(1)}\left[\lambda_i, \tau_{j-1}, Z_\star(\tau_\text{GXY} - \tau_{j-1})\right] \biggr\}
\end{multline}
where $\lambda_i$ is the wavelength, $\tau_\text{GXY}$ is the age of the galaxy, $\tau_j$ is the indexed SSP age, $\psi(\tau)$ is the SFR at given time and $P_{L_\text{SSP}}^{(1)}$ is the first order polynomial interpolating linearly the SSP emission between its two tabulated metallicities $Z_k \leq Z_\star(\tau_\text{GXY}-\tau_j) \leq Z_{k+1}$.

\begin{figure*}
\resizebox{\hsize}{!}{
\includegraphics{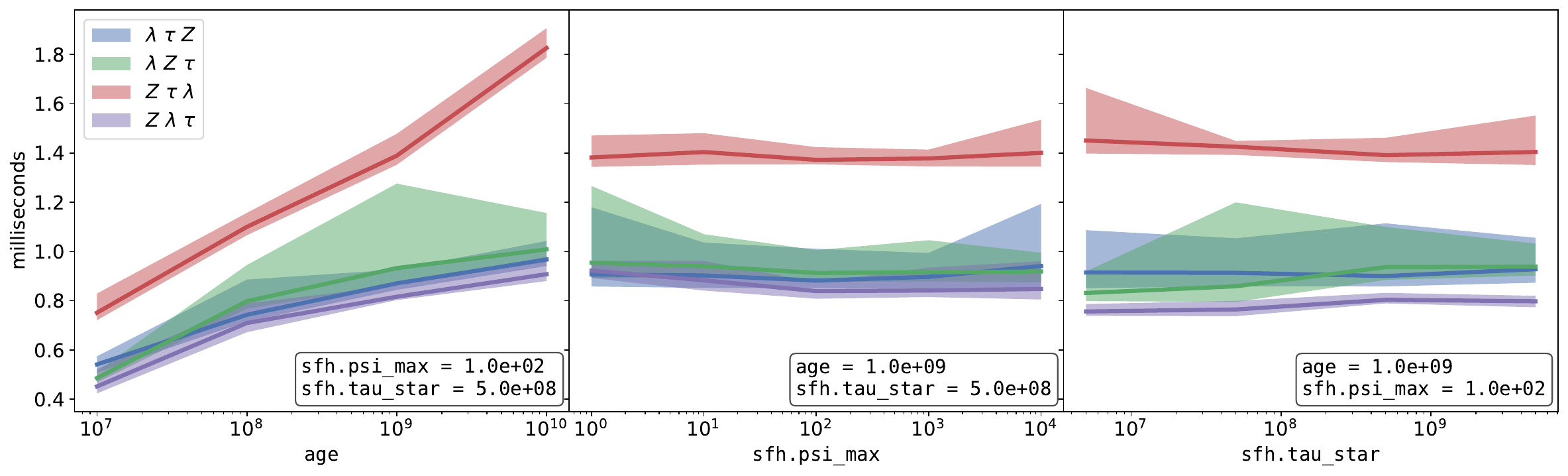}}
  \caption{Dependency of SSP integration performances on the 3-dimensional matrix ordering. Different colours mark different orderings, as reported in the legend. Each panel shows how the integration time changes as a function of one of the three model parameters on which Eq.~\eqref{eq:LCSP_numerical} depends.}
\label{fig:registers}
\end{figure*}

As made evident from Eq.~\eqref{eq:LCSP_numerical}, for each wavelength we first perform an interpolation between 2 metallicities, then sum along the time dimension. 
Even though it might seem that the most logical dimension to keep closest in memory is metallicity, by inspecting Fig.~\ref{fig:registers}, we can easily see this is not true.
In each panel, along the x-axis, we vary the value of one of the model parameters that affect the integration of Eq.~\eqref{eq:LCSP_numerical} while, along the y-axis, we show the integration time in milliseconds for the whole $\lambda$-grid. 
Boxes on the lower right show the fixed value of the two non-varying parameters.
Each different colour marks the performance of a different ordering of the 3-dimensional matrix storing the SSP library, as encoded in the Figure's legend, where the shaded regions show fluctuations over ten runs and the solid line marks the mean execution time.
It is clear that the most efficient ordering is $[Z~\lambda~\tau]$ (in purple).
The reason for this is found in the slow variation of the $Z$-dimension as a function of galaxy age, which means that the metallicity of SSPs in the highest cache levels is updated rarely.

SSP tables are objects counting some millions of double precision floating point numbers and their transposition can easily slow down the code.
For this reason having the SSPs directly stored with the $[Z~\lambda~\tau]$ ordering allows to accelerate the process of building objects that depend on them (i.e. the class \texttt{galapy.CompositeStellarPopulation.CSP} provides the most direct user interface to these functionalities).
Nonetheless, we provide functions in \galapy for converting eventual user-defined SSP tables into the format described above, to foster extensibility and customisation.

\subsubsection{Interpolation technique}\label{apx:interp}

Interpolation is used for many different purposes in \galapy: from the computation of SSP emission between the tabulated values of metallicity to the addition of templated emission on the wavelength grid.
While for some of these cases the values over which to interpolate change with the variation of the model free parameters, for the majority of the occurrences, the interpolation grid is fixed for all the parameter-space sampling\footnote{E.g. when assuming an interpolated SFH model, see Sec.~\ref{sec:nonparametric}.}.
We have developed an \texttt{interpolator} object exploiting this condition to speed up the computation.

The interface is optimised for computing interpolated values on 1-dimensional grids with un-evenly spaced values. 
This is achieved with a high level of specialisation for the functionalities, making therefore the software tool not flexible but extremely efficient when used for all the problem sizes coming up in \galapy.  
This results in a smaller efficiency when building the object itself but, since this operation will be done only once for each galaxy object built, we can safely give up on it.

The \texttt{interpolator} object is based on an interval binary search tree (IBST) without overlapping.
This data structure provides access to nodes that perform analytic linear interpolation, integration and derivation on a single interval of the grid with $\log N$ time scaling, where $N$ is the size of the grid. 
The \texttt{find} function of the IBST implemented in the core C++ sector of \galapy is an order of magnitude faster than its C++ STL equivalent (i.e. \texttt{std::map::find}).
The interface to the \texttt{interpolator} available from \galapy's Python API is up to two orders of magnitude faster than NumPy's linear interpolation module (i.e. \texttt{scipy.interpolate.interp1d} class with \texttt{kind='linear'} which is equivalent to the the NumPy function \texttt{numpy.interp}) on problem sizes comparable to those of interest for our library.
It has to be stressed that \texttt{interpolator} objects from \galapy are not universally more efficient than equivalent functions and classes from external, wide spread and powerful packages such as, e.g., SciPy and NumPy. We reach better performances only when the resolution of the interpolation grid (order of $10^3$ points) and the number of interpolated points (order of $<10^2$) is comparable to those arising from the computation of \galapy models.
Our implementation comes with the additional advantage of being available on both the C++ and the Python sectors as well as providing a uniformed interface for interpolation, numerical integration and numerical derivation.

\subsection{Python API structure}\label{apx:api}

A complete description of the classes and functions implemented in the \galapy package is available in the on-line documentation, in the section Python API.
We hereby provide just a short description of the package structure and of the functionalities provided by each module/sub-package.

\galapy contains modules in the top-level package and on sub-packages as well, divided as follows
\begin{verbatim}
galapy
    |-- galapy.analysis
    |-- galapy.configuration
    |-- galapy.internal
    |-- galapy.io
    `-- galapy.sampling
\end{verbatim}
The content of each sub-package provides different functionalities:
\begin{itemize}
    \item \textbf{\texttt{galapy}}: the root package contains all the modules providing access to the models described in Sec.~\ref{sec:models}. 
    Specifically:
    \begin{itemize}
        \item \texttt{galapy.StarFormationHistory}: contains the class \texttt{SFH} that can be used to build either empirical, In-Situ or the interpolated SFH models of Sec.~\ref{sec:sfh}.
        \item \texttt{galapy.CompositeStellarPopulation}: contains functions for listing and loading SSP libraries and the \texttt{CSP} class, used to build composite stellar populations (Sec.~\ref{sec:stellar}).
        \item \texttt{galapy.InterStellarMedium}: provides access to the dust-model described in Sec.~\ref{sec:ism}. It defines several objects: a base \texttt{ismPhase} class, from which two derived classes inherit, \texttt{MC} and \texttt{DD}, modelling the attenuation and emission due to the two separate dust components; additionally, a \texttt{ISM} type is defined, wrapping the other two components and combining their contributions.
        \item \texttt{galapy.NebularFreeFree}, \texttt{galapy.Synchrotron} and  \texttt{galapy.XRayBinaries}: these modules implement (optional) the additional sources of stellar continuum described in Sec.~\ref{sec:radio}.
        They respectively define the classes \texttt{NFF}, \texttt{SNSYN} and \texttt{XRB}.
        \item \texttt{galapy.ActiveGalacticNucleus}: provides an interface for loading the \cite{Fritz2006} templates and consistently adding on top of them an eventual X-Ray contribution, as described in Sec.~\ref{sec:agn}. These functionalities are accessed through the class \texttt{AGN}, defined in this module.
        \item \texttt{galapy.Cosmology} and \texttt{galapy.InterGalacticMedium}, define respectively the classes \texttt{CSM} and \texttt{IGM}, whose implementation provide access to the models of Sec.~\ref{sec:cosmo} and Sec.~\ref{sec:igm}, respectively.
        \item \texttt{galapy.Galaxy} and \texttt{galapy.Handlers} define utility classes and methods designed to ease modelling through the Python API. In particular, the former defines the class \texttt{GXY} which wraps up the models implemented in the other modules, their interplay and parameter settings, optimising the performances through a minimisation of the number of operations.
        By instantiating one of this objects, i.e.
        \begin{verbatim}
        
from galapy.Galaxy import GXY
gxy = GXY( age = 1.e9, redshift = 1.0 )
        \end{verbatim}
        users can easily modify the value of the free-parameters (e.g. \texttt{gxy.set\_parameters(age = 1.e10)}, see Table~\ref{tab:tunable_parameters} for a list of all the tunable parameters), get the emission or flux (e.g. \texttt{flux = gxy.SED()}), or compute derived parameters (e.g. \texttt{Mstar = gxy.sfh.Mstar( 1.e8 )}, for the stellar mass at an age of $\tau = 10^8$ years).
        Note that all these functionalities are obtained by a combination of tools implemented in the modules listed above.
        The latter, \texttt{galapy.Handlers} module, is designed for managing the free-parameters when sampling.
        \item \texttt{galapy.PhotometricSystem}: implements the class \texttt{PMS} that can be used to manage band-pass transmission filters, both loaded from the database or user-defined.
    \end{itemize}
    \item \textbf{\texttt{galapy.sampling}}: contains sub-modules used for sampling the parameter space, i.e. the two sub-modules \texttt{Sampler} and \texttt{Results} which unify the interface to the different sampling algorithms implemented in the library along with their results (note in particular, the \texttt{Results} class described in Sec.~\ref{sec:results}), the \texttt{Observation} module which collects observational datasets and the \texttt{Statistics} sub-module, defining statistical functions such as likelihoods and estimators.
    \item \textbf{\texttt{galapy.analysis}}: provides the two sub-modules \texttt{funcs} and \texttt{plot}, both defining functions that facilitate the analysis of sampling results. While the former mainly produces tables with the estimates of several statistics (most of the tables in Sec.~\ref{sec:validation} have been produced with these functions), the latter produces plots of fitted SEDs, residuals and posteriors (most of the figures in Sec.~\ref{sec:validation} have been produced with these functions).
    \item \textbf{\texttt{galapy.io}}: used to load and store object types defined in the package.
    \item \textbf{\texttt{galapy.configuration}} and \textbf{\texttt{galapy.internal}}: are mainly for internal usage, even though some classes and functions of \texttt{galapy.internal} might be useful in some parts of the analysis. An example are the \texttt{interpolator} objects described in Appendix~\ref{apx:interp}.
\end{itemize}

\subsection{Insights on performances and scaling}\label{apx:performances}

A solid comparison of performances against other libraries would require a thorough analysis that goes beyond the scope of this presentation work.
We just mention that the computation of a single model (including parameters setting, computation of the flux and band-averaging) requires $\sim 10$ milliseconds, depending on the resolution of the wavelength grid over which the flux is computed.

In our Bayesian framework, the time required for convergence of the free-parameters inference algorithms strongly depends on the likelihood estimation time (defined as the time necessary to set a new position in the parameter-space, compute the corresponding flux model, extract the band-averaged fluxes and compute the likelihood).
We therefore measure our performances in terms of likelihood computations per second, even though, for how the code is structured and given the simplicity of the likelihood of Eq.~\eqref{eq:loglike} and Eq.~\eqref{eq:loglikenoise}, this interval of time is obviously dominated by model computation.
\begin{figure}
    \centering
    \includegraphics[width=\hsize]{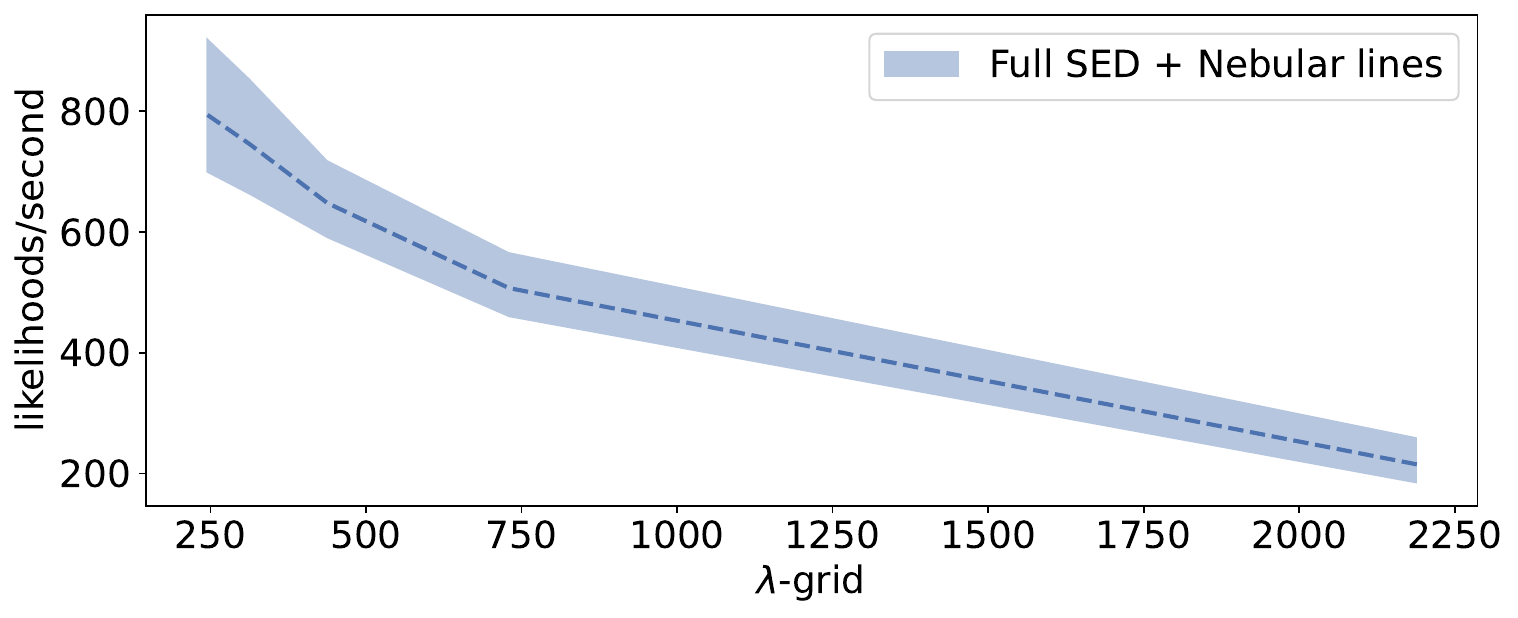}
    \caption{Likelihoods generated per second with \galapy as a function of the wavelength grid resolution. The dashed line marks the average over $1000$ measurements and the shaded region highlights the $1$-$\sigma$ confidence intervals.}
    \label{fig:llike_ngrid}
\end{figure}
This is shown in Fig.~\ref{fig:llike_ngrid} as a function of the wavelength grid thinness for the least optimal model set-up: BC03 SSP libraries with full modelling from X-ray band to radio including nebular and synchrotron emission.
As the Figure shows, the performance of the code decreases with wavelength grid size increasing.
This is expected as, the thinner the wavelength grid, the larger the number of times the code has to compute, e.g., Eq.~\eqref{eq:LCSP_numerical}.

As a term of comparison, on a similar problem set-up the Prospector \citep{prospector2021} documentation\footnote{Specifically at this link: \href{https://prospect.readthedocs.io/en/latest/faq.html}{prospect.readthedocs.io/faq} section ``How long will it take to fit my data'' version \texttt{v1.2.0}.} declares 25 likelihoods per second, to be compared with our $200\div300$ result shown in Fig.~\ref{fig:llike_ngrid}.
Additionally and differently from other libraries, the problem size seems not to affect too much the computation of likelihoods as we do not measure significant variations on performances when increasing the number of photometric bands, or when making the model more complex.
This is mostly due to the highly optimised implementation of \galapy.
The execution time of most of the components is in fact negligible with respect to the computation of Eq.~\eqref{eq:LCSP_numerical} whose scaling also affects how the likelihood-per-second execution time scales.

We point out that the measurements provided in this Section are obtained by running on a single core, as the parallelisation scheme of \galapy is still under development and will be in its final form on future extensions focused on boosting the performances.
At current state, we exploit the parallel strategy already implemented in the samplers available in \galapy (i.e. \texttt{emcee} and \texttt{dynesty}) by passing to the sampling algorithm a pool of processes obtained with the \texttt{multiprocessing.Pool} method of the Python standard library.
This approach can prove to be not optimal in some cases and we will therefore explore different strategies in the future.

%% file: sec/appendix.tex
\section{Additional modelling informations}\label{apx:models}

We provide here an extension on the description of models reviewed in Sec.~\ref{sec:models} with further details that we did not want to face in the main body of this paper in order not to burden excessively the discussion.

\subsection{Difference between CSP emission assuming different SSP libraries}\label{apx:ssp}

Stellar emission, as already explained in Sec.~\ref{sec:stellar}, is computed by assuming a SFH and integrating SSP emission.
In \galapy we provide the tabulated emission from 4 main SSP libraries: the classic \cite{BruzualCharlot2003} libraries in both the low (\texttt{bc03.basel}) and high (\texttt{bc03.stelib}) resolution version, and two libraries, produced specifically for the publication of this package, obtained using the PARSEC code \citep{Bressan2012,Chen2014}.
The latter are delivered in two flavours: 
\begin{itemize}
    \item \texttt{parsec22.nt} contains SSP continuum emission including the non-thermal low energy contribution produced by SN synchrotron;
    \item \texttt{parsec22.ntl} also accounts for the thermal and line emission produced in ionised nebular regions around young massive stars.
\end{itemize}
\begin{table}
\centering
\caption{\label{tab:ssp_lib_summary} Main properties of the different SSP tables delivered with \galapy.}
\input{tab/SSP_lib}
\tablefoot{The first column reports the library name, columns from the second to the fourth list the dimensions of the grids in the wavelength, age and metallicity domains, respectively. The last column provides a list of the tabulated metallicities expressed in absolute units.}
\end{table}
In Tab.~\ref{tab:ssp_lib_summary} we summarise the size of the aforementioned SSP tables along each of the 3 dimensions: wavelength, age and metallicity.
In BC03 tables the time domain spans from $0$ to $2\times10^{10}$ years while PARSEC22 tables go from $0$ to $1.4\times10^{10}$ years.
In both the libraries we have extended the wavelength domain from $1$ to $10^{10}$ \AA. 

\begin{figure*}
\resizebox{\hsize}{!}{
\includegraphics{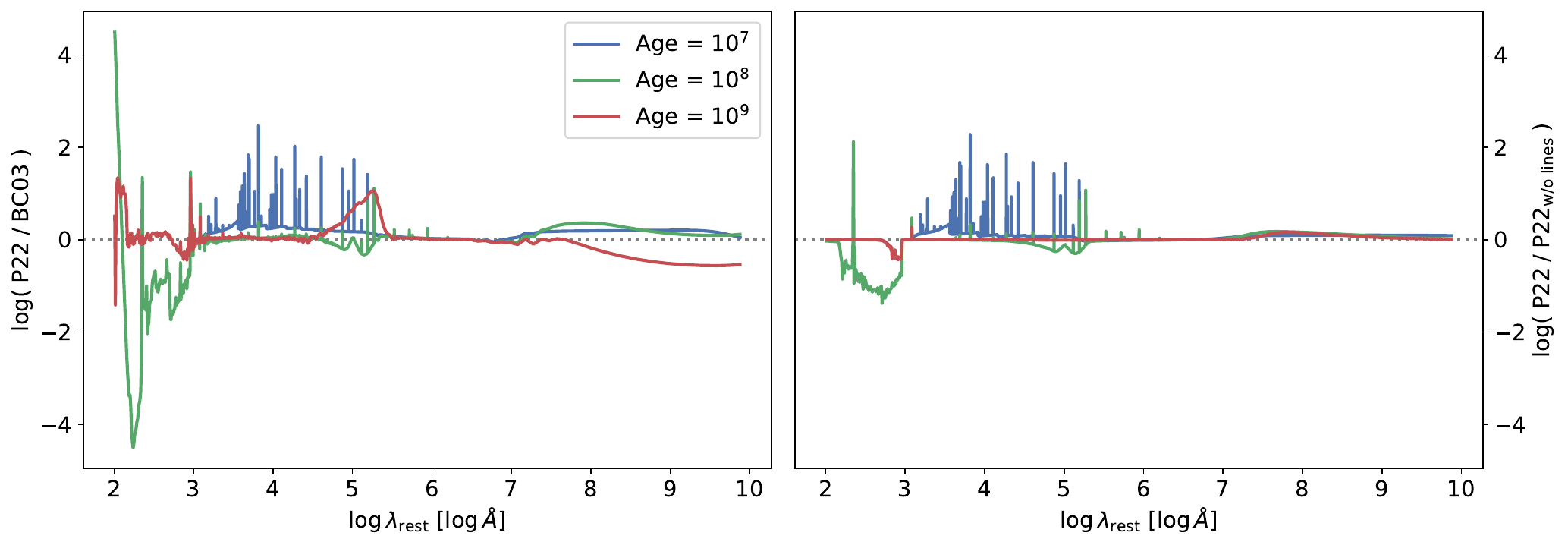}}
  \caption{Logarithm of the ratio between CSPs obtained with different SSP libraries at fixed SFH and age. The \textit{left panel} shows the ratio between PARSEC22 and BC03 while the \textit{right panel} shows the ratio between PARSEC22 with and without lines.}
\label{fig:csp_ratio}
\end{figure*}

The emission predicted by the models tabulated in the 4 libraries do agree in general even though they show minor differences.
In order to highlight how choosing one library over the other contributes differently to the panchromatic emission from a galaxy, we compute the total spectrum due to the composite stellar emission. 
Fig.~\ref{fig:csp_ratio} shows the ratio between CSP emission predicted with the \texttt{bc03.basel} library (left panel) and with the \texttt{parsec22.nt} library (right panel) with respect to the \texttt{parsec22.ntl} library.
In order to have a meaningful comparison, we also include the radio components in the plot, that therefore spans from $100$ \AA~to $10^{10}$ \AA.
This means that, while CSPs built with \texttt{parsec22.ntl} self-consistently contain the contribution from nebular thermal emission and SN synchrotron non thermal emission, in the other two cases the latter components are added to the final CSP emission using models described in Sec.~\ref{sec:nff} and Sec.~\ref{sec:snsyn}.
Fig.~\ref{fig:csp_ratio} shows how the \texttt{parsec22.ntl} library steals energy from the continuum at the shorter wavelengths (Optical to UV) and re-emits it in lines mainly in the IR region of the spectrum.
This effect is more relevant for younger galaxies (blue and green solid lines) while becoming less important to irrelevant for older stellar populations (red solid line).
In the radio bands the emission predicted by the models of Sec.~\ref{sec:nff} and Sec.~\ref{sec:snsyn} are in good agreement with the one obtained using the PARSEC library, even though it seems to deviate more for old stellar populations in the left panel, suggesting that the synchrotron emission might be slightly over-estimated with the model of Sec.~\ref{sec:snsyn}.
In general we suggest, whenever possible, to use the PARSEC22 libraries as it also provide the additional advantage of reducing the size of the parameter space.

For reference, we provide the values, entering Eq.~\eqref{eq:RCCSN}, computed for the 7 metallicities available in the BC03 SSP libraries.
The two metallicity-dependent parameters, $R_0$ and $R_1$, are tabulated in Tab.~\ref{tab:RCCSN_par}.
\begin{table}
\centering
\caption{\label{tab:RCCSN_par} Tabulated values of the two parameters $R_0$ and $R_1$ regulating the rate of CCSN in Eq.~\eqref{eq:RCCSN} with metallicity dependence.}
\input{tab/RCCSN_par}
\tablefoot{Metallicity is given in absolute value, parameter $R_0$ is given in units of $[\text{Gyr}^{-1} M_\odot^{-1}]$ while parameter $R_1$ is adimensional.}
\end{table}

\subsection{Tunable parameters}\label{apx:params}

In Tab.~\ref{tab:tunable_parameters} we provide a complete list of the parameters that can be tuned with \galapy. 
All of the parameters are available from the \texttt{galapy.Galaxy.GXY} object (as well as from derived objects).
We divided the table in sections describing which of the class-objects they model.
The first column contains the API keyword used to access the parameter, the second column contains the symbol used in this manuscript to refer that parameter, in the third column we give a brief description of the parameter and the last column contains the eventual Eq.(s) where the parameter appears.

Each of the tunable parameters can be either fixed or set as free.
In the latter case it will add a dimension to the parameter space explored by the sampler during SED-fitting.
Always remember that, the larger the parameter space (both in terms of prior volume and dimensionality), the longer it will take for the sampler to converge.
It is therefore always crucial to carefully select which parameters to sample and which to keep fixed.
The volume, prior shape and fixed value used for each run depends on several considerations on the data-set that has to be fitted with \galapy.
The choice of these hyper-parameters is left to the user.

The parameters regulating the shape of the AGN template are not described in this manuscript and can be found in \cite{Fritz2006}. 
Since in the current version of \galapy we do not provide a template fitting interface yet, we discourage setting them as free, as it would imply sampling a discrete parameter space.
Nonetheless, this custom behaviour can be achieved by modifying the likelihood in the sampling algorithm through the Python API of \galapy.

\longtab[3]{
    \centering
    \input{tab/tunable_parameters}
}

%% file: tab/SSP_lib.tex
\begin{tabular}{l r r r r}
\toprule
Library & $N_\lambda$ & $N_\tau$ & $N_Z$ & Metallicities \\
\hline\midrule
\multicolumn{5}{c}{\textbf{BC03}}\\ \midrule
Stelib & $7325$ & \multirow{2}{*}{$221$} & \multirow{2}{*}{$7$} & $[0.0001, 0.0004, 0.004,$\\
BaSeL & $2223$ &&& $0.008, 0.02, 0.05, 0.1]$\\ \midrule
\multicolumn{5}{c}{\textbf{PARSEC22}}\\ \midrule
NT & \multirow{2}{*}{$1562$} & \multirow{2}{*}{$146$} & \multirow{2}{*}{$6$} & $[0.0001, 0.0005, 0.001,$\\
NTL &&&& $0.004, 0.008, 0.02]$ \\
\bottomrule\bottomrule
\end{tabular}

%% file: tab/RCCSN_par.tex
\begin{tabular}{l | r r r r r r}
\toprule
$Z_\star$ & $0.02$ & $0.008$ & $0.004$ & $0.001$ & $0.0005$ & $0.0001$ \\
$R_0$ & $1.5141$ & $1.2679$ & $1.1820$ & $1.0924$ & $1.0692$ & $1.0425$ \\
$R_1$ & $0.5146$ & $0.4454$ & $0.4140$ & $0.3789$ & $0.3673$ & $0.3531$ \\
\bottomrule\bottomrule
\end{tabular}

%% file: tab/tunable_parameters.tex
\begin{longtable}{lrcr} %p{10cm}
    \caption{\label{tab:tunable_parameters} Complete list of the tunable parameters available in \galapy. Each of these parameters (except for \texttt{sfh.model}) can be set as an additional free dimension for the sampler to explore.} \\
    \toprule
    \textbf{Keyword} & \textbf{Symbol in text} & \textbf{Description} & \textbf{Section ref.} \\ \hline \midrule
    \endfirsthead
    \caption{continued.} \\
    \toprule
    \textbf{Keyword} & \textbf{Symbol in text} & \textbf{Description} & \textbf{Section ref.} \\ \hline \midrule
    \endhead
    \bottomrule
    \endfoot
    \bottomrule
    \endlastfoot
    \multicolumn{4}{c}{\textbf{Global}}\\ \midrule
    \texttt{age} & $\tau$ & Age of the galaxy & \ref{sec:models} \\ 
    \texttt{redshift} & $z$ & Redshift of the galaxy & \ref{sec:cosmo} \\\midrule
    \multicolumn{4}{c}{\textbf{Star Formation History}} \\\midrule
    \texttt{sfh.model} & --- & SFH model: one among the keywords listed below & \ref{sec:sfh} \\ 
    \texttt{sfh.tau\_quench} & $\tau_\text{quench}$& Age of the abrupt quenching & \ref{sec:sfh} \\ \midrule
    \multicolumn{4}{c}{Constant (keyword: \texttt{constant})} \\ \midrule
    \texttt{sfh.psi} & $\psi_0$ & Value of the constant SFR & \ref{sec:sfh} \\
    \texttt{sfh.Mdust} & $M_\text{dust}$ & Total dust mass in galaxy at the given age & \ref{sec:empirical} \\
    \texttt{sfh.Zgxy} & $Z_\text{gxy}$ & Metallicity of all phases in galaxy at the given age & \ref{sec:empirical} \\ \midrule
    \multicolumn{4}{c}{Delayed Exponential (keyword: \texttt{delayedexp})} \\ \midrule
    \texttt{sfh.psi\_norm} & $\psi_\text{norm}$ & Normalisation & \ref{sec:sfh} \\
    \texttt{sfh.k\_shape} & $\kappa$ & Shape parameter of the early evolution & \ref{sec:sfh} \\
    \texttt{sfh.tau\_star} & $\tau_\star$ & Characteristic timescale & \ref{sec:sfh} \\
    \texttt{sfh.Mdust} & $M_\text{dust}$ & [same as for constant SFH] & \ref{sec:empirical} \\
    \texttt{sfh.Zgxy} & $Z_\text{gxy}$ & [same as for constant SFH] & \ref{sec:empirical} \\ \midrule
    \multicolumn{4}{c}{Log-Normal (keyword: \texttt{lognormal})} \\ \midrule
    \texttt{sfh.psi\_norm} & $\psi_\text{norm}$ & Normalisation & \ref{sec:sfh} \\
    \texttt{sfh.sigma\_star} & $\sigma_\star$ & Characteristic width & \ref{sec:sfh} \\
    \texttt{sfh.tau\_star} & $\tau_\star$ & Peak age & \ref{sec:sfh} \\
    \texttt{sfh.Mdust} & $M_\text{dust}$ & [same as for constant SFH] & \ref{sec:empirical} \\
    \texttt{sfh.Zgxy} & $Z_\text{gxy}$ & [same as for constant SFH] & \ref{sec:empirical} \\ \midrule
    \multicolumn{4}{c}{Interpolated (keyword: \texttt{interpolated})} \\ \midrule
    \texttt{sfh.Mdust} & $M_\text{dust}$ & [same as for constant SFH] & \ref{sec:empirical} \\
    \texttt{sfh.Zgxy} & $Z_\text{gxy}$ & [same as for constant SFH] & \ref{sec:empirical} \\ \midrule
    \multicolumn{4}{c}{In-Situ (keyword: \texttt{insitu})} \\ \midrule
    \texttt{sfh.psi\_max} & $\psi_\text{max}$ & Normalisation & \ref{sec:sfh} \\
    \texttt{sfh.tau\_star} & $\tau_\star$ & Characteristic timescale & \ref{sec:sfh} \\\midrule
    \multicolumn{4}{c}{\textbf{Inter-Stellar Medium}} \\\midrule
    \texttt{ism.f\_MC} & $f_\text{MC}$ & Fraction of dust in the MC phase & \ref{sec:ism} \\ \midrule
    \multicolumn{4}{c}{Molecular Clouds} \\ \midrule
    \texttt{ism.norm\_MC} & $\mathcal{C}_V^\text{MC}$ & Normalisation of the MC extinction in the visible band & \ref{sec:ism} \\ 
    \texttt{ism.N\_MC} & $N_\text{MC}$ & Number of MCs in the galaxy & \ref{sec:ism} \\ 
    \texttt{ism.R\_MC} & $R_\text{MC}$ & Average radius of a MC & \ref{sec:ism} \\ 
    \texttt{ism.tau\_esc} & $\tau_\text{esc}$ & Time required by stars to start escaping their MC & \ref{sec:ism} \\ 
    \texttt{ism.dMClow} & $\delta_\text{MC}^\text{l}$ & Extinction power-law index at wavelength $\lesssim100 \mu m~(10^6 \text{\AA})$ & \ref{sec:ism} \\ 
    \texttt{ism.dMCupp} & $\delta_\text{MC}^\text{u}$ & Extinction power-law index at wavelength $\gtrsim100 \mu m~(10^6 \text{\AA})$ & \ref{sec:ism} \\ \midrule
    \multicolumn{4}{c}{Diffuse Dust} \\ \midrule
    \texttt{ism.norm\_DD} & $\mathcal{C}_V^\text{DD}$ & Normalisation of the DD extinction in the visible band & \ref{sec:ism} \\ 
    \texttt{ism.Rdust} & $R_\text{DD}$ & Radius of the diffuse dust region embedding stars and MCs & \ref{sec:ism} \\ 
    \texttt{ism.f\_PAH} & $f_\text{PAH}$ & Fraction of the total DD luminosity radiated by PAH & \ref{sec:ism} \\ 
    \texttt{ism.dDDlow} & $\delta_\text{DD}^\text{l}$ & Extinction power-law index at wavelength $\lesssim100 \mu m~(10^6 \text{\AA})$ & \ref{sec:ism} \\ 
    \texttt{ism.dDDupp} & $\delta_\text{DD}^\text{u}$ & Extinction power-law index at wavelength $\gtrsim100 \mu m~(10^6 \text{\AA})$ & \ref{sec:ism} \\\midrule
    \multicolumn{4}{c}{\textbf{Nebular Free-Free}} \\\midrule
    \texttt{nff.Zi} & $Z_i$ & Average atomic number of ions & \ref{sec:nff} \\\midrule
    \multicolumn{4}{c}{\textbf{Synchrotron}} \\\midrule
    \texttt{syn.alpha\_syn} & $\alpha_\text{syn}$ & Spectral index & \ref{sec:snsyn} \\ 
    \texttt{syn.nu\_self\_syn} & $\nu_\text{self}$ & Self-absorption frequency & \ref{sec:snsyn} \\\midrule
    \multicolumn{4}{c}{\textbf{Active Galactic Nucleus}} \\\midrule
    \texttt{agn.fAGN} & $f_\text{AGN}$ & AGN fraction & \ref{sec:agn} \\ \midrule
    \multicolumn{4}{c}{Templates} \\\midrule 
    \texttt{agn.ct} & $\Theta$ & Torus half-aperture angle & --- \\
    \texttt{agn.al} & $\alpha$ & Density parameter (exponential part) & --- \\
    \texttt{agn.be} & $\beta$ & Density parameter (power-law part) & --- \\
    \texttt{agn.ta} & $\tau_{9.7}^\text{AGN}$ & Optical depth at $9.7 \mu m$ & --- \\
    \texttt{agn.rm} & $R_\text{torus}^\text{AGN}$ & Radial ratio of the torus & --- \\
    \texttt{agn.ia} & $\Psi_\text{los}^\text{AGN}$ & Inclination angle & --- \\ 
    \bottomrule
\end{longtable}

%% file: sec/demo.tex
\section{Practical information}\label{sec:demo}

\subsection{Installation and post-installation operations}

\galapy is available on the Python Package Index (PyPI) and can be installed by running 
\begin{verbatim}
    $ pip install galapy-fit
\end{verbatim}
on a terminal\footnote{\galapy can also be installed from source by cloning the github repository: \href{https://github.com/TommasoRonconi/galapy}{github.com/TommasoRonconi/galapy}}.
Once the package has been installed, before the first usage run
\begin{verbatim}
    $ galapy-download-database
\end{verbatim}
to download on the file-system the database necessary for running.
The database contains the formatted SSP libraries (Section~\ref{sec:stellar}), a collection of bandpass transmission filters (Sec.~\ref{sec:photospec}), the AGN templates (Sec.~\ref{sec:agn}) and pre-computed tables for cosmological calculations (Sec.~\ref{sec:cosmo}).

These two steps should be sufficient to obtain a working installation of \galapy.
For further details on how to install in developer mode and on the required dependencies of the library, the user can refer to the installation guide available in the documentation (\href{https://galapy.readthedocs.io/en/latest/general/install_guide.html}{galapy.readthedocs.io/en/latest/general/install\_guide.html}).

\subsection{Running the automatised sampling script}

We provide command-line tools for sampling the parameter space with our automatised set-up.
The automatic fitting requires to set-up a parameter file that is generated by calling
\begin{verbatim}
    $ galapy-genparams [-n/--name NAME]
\end{verbatim}
If the optional argument \texttt{NAME} is not provided, the generated file will be assigned the default name \texttt{galapy\_hyper\_parameters.py}.
By modifying this self-explanatory file the user can set all the relevant information (e.g. free-parameters and priors, data-set, input/output files) required for running a sampling.
The file is divided in four main sections:
\begin{itemize}
    \item \textbf{Loading of the data-set}: users can use their preferred method for loading the data-set (band-pass transmission filters, either from the data-base or custom, fluxes, errors and upper-limits). Since the parameter file is effectively a python script, external libraries (such as NumPy or Pandas) can be imported to ease the process. We also provide a function (i.e. \texttt{galapy.internal.utils.cat\_to\_dict}) for the conversion of ASCII catalogues\footnote{Like those compiled with TopCat \href{https://www.star.bris.ac.uk/~mbt/topcat/}{https://www.star.bris.ac.uk/~mbt/topcat/}} to dictionaries. 
    \item \textbf{Galaxy model set-up}: here the user chooses their preferred models among those available in the package. Namely, the SFH model, the SSP library, whether to include an AGN, X-ray emission, radio support, cosmology and an eventual treatment of noise.
    \item \textbf{Sampling parameters}: for choosing priors on the free parameters or fixing part of them to values different from their default values.
    \item \textbf{Sampler and output choices}: choose the sampler and format for the sampling output file(s).
\end{itemize}

Once the parameter file has been set, by calling the command
\begin{verbatim}
    $ galapy-fit galapy_hyper_parameters.py
\end{verbatim}
the sampling starts on all the available parallel CPUs (see the documentation for further details on how to customize the parallel scheme or to run serially on a single CPU).

Besides this terminal entry-points, the \galapy API is made available upon installation of the library.
On a python script, shell or notebook the user can import modules, classes and functions from the \texttt{galapy} Python package.
The entry-point themselves are made available for inspection and customisation on the sub-module \texttt{galapy.sampling.Run}.
For more in-detail description, please inspect the API documentation at the following link: \href{https://galapy.readthedocs.io/en/latest/index.html}{galapy.readthedocs.io/en/latest/index.html}. 

\subsection{Acknowledging usage of the library}\label{apx:logo}

\galapy relies on models and samplers that might require additional references along with this manuscript.
We encourage authors to check the documentation for further instructions on how to acknowledge the relevant works.

\begin{figure}
    \centering
    \includegraphics[width=\hsize]{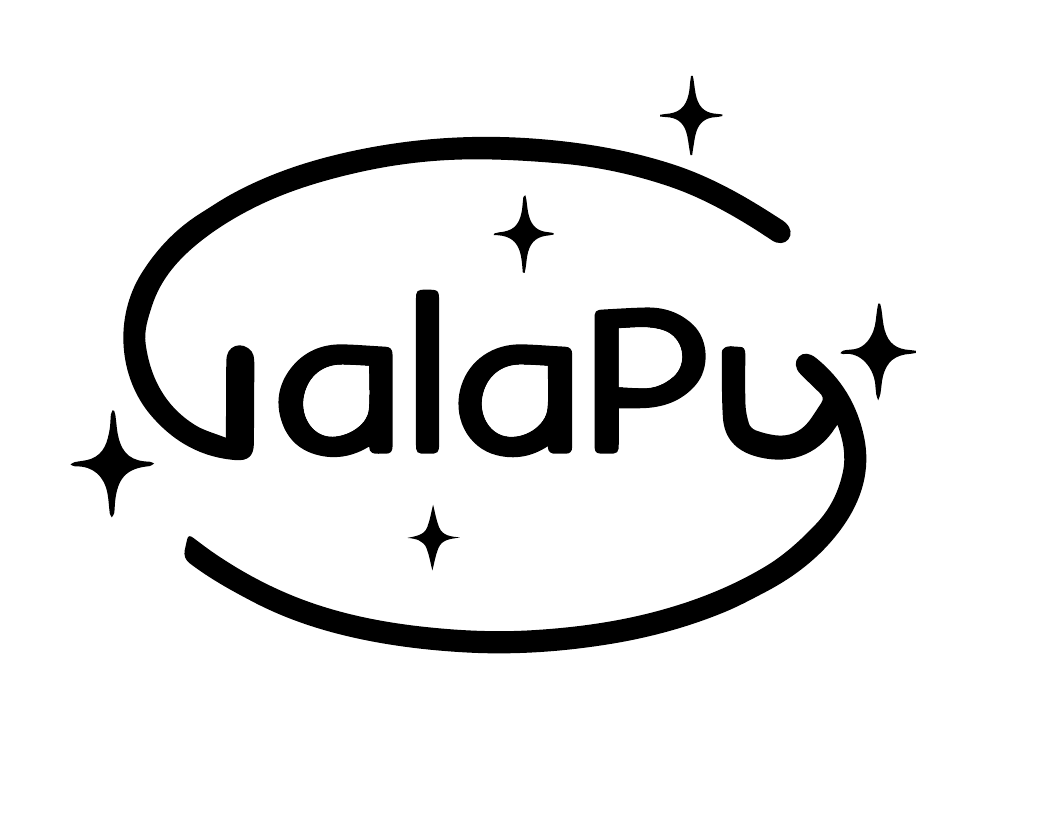}
    \caption{Version \texttt{1.0.0} of the \galapy logo. We encourage authors presenting results obtained with \galapy to add their preferred version of the logo in public presentations. A variety of formats is available on the website and in the data-base.}
    \label{fig:logo}
\end{figure}
Several data-formats for the \galapy logo (Fig.~\ref{fig:logo}) are also available in the documentation.
Even though we do not imprint the logo on the figures produced by our plotting API, we encourage authors in adding it to their presentations if presenting results obtained using \galapy.